\RequirePackage{lineno}
\documentclass[twocolumn,letterpaper,aps,prc,longbibliography,superscriptaddress,nofootinbib,floatfix]{revtex4-1}

\usepackage{graphicx}	
\usepackage{appendix}
\usepackage{float}
\usepackage{bbm}
\usepackage{hyperref}
\usepackage{amsmath}
\usepackage{verbatim}

\hypersetup{
    colorlinks=true,
    linkcolor=blue,
    filecolor=blue,
    urlcolor=blue,
    citecolor=blue,
}

\usepackage{xspace}	

\newcommand{\pt}{\mbox{$p_T$}\xspace}
\newcommand{\pttrig}{\mbox{$p_T^{\rm Trig}$}\xspace}
\newcommand{\ptassoc}{\mbox{$p_T^{\rm Assoc}$}\xspace}

\newcommand{\nch}{\mbox{$N^{\rm ch}$}\xspace}

\newcommand{\sqs}{\mbox{$\sqrt{s}$}\xspace}
\newcommand{\sqsn}{\mbox{$\sqrt{s_{_{NN}}}$}\xspace}
\newcommand{\deta}{\mbox{$\Delta\eta$}\xspace}
\newcommand{\dphi}{\mbox{$\Delta\phi$}\xspace}

\newcommand{\pp}{\mbox{$p+p$}\xspace}
\newcommand{\dau}{\mbox{$d+\mathrm{Au}$}\xspace}

\newcommand{\pau}{\mbox{$p+\mathrm{Au}$}\xspace}

\newcommand{\pPb}{\mbox{$p+\mathrm{Pb}$}\xspace}

\newcommand{\heau}{\mbox{$^{3}\mathrm{He}+\mathrm{Au}$}\xspace}
\newcommand{\vttsub}{\mbox{$v_{22}^{\rm sub}$}\xspace}
\newcommand{\vtt}{\mbox{$v_{22}$}\xspace}
\newcommand{\vththsub}{\mbox{$v_{33}^{\rm sub}$}\xspace}
\newcommand{\vthth}{\mbox{$v_{33}$}\xspace}

\newcommand{\methodi}{\mbox{\textit{Method 1}}\xspace}
\newcommand{\methodii}{\mbox{\textit{Method 2}}\xspace}
\newcommand{\methodb}{\mbox{\textit{Method 1} and \textit{2}}\xspace}
\newcommand{\pythia}{\mbox{{\sc pythia8}}\xspace}
\newcommand{\ampt}{\mbox{{\sc ampt}}\xspace}
\newcommand{\hijing}{\mbox{{\sc hijing}}\xspace}

\begin{document}



\title{Examination of flow and nonflow factorization methods in small collision systems}

%

\newcommand{\colorado}{University of Colorado, Boulder, Colorado 80309, USA}
\newcommand{\uncg}{University of North Carolina, Greensboro, North Carolina 27413, USA}

\affiliation{\colorado}
\affiliation{\uncg}

\author{S.H.~Lim} \affiliation{\colorado}
\author{Q.~Hu} \affiliation{\colorado}
\author{R. Belmont} \affiliation{\uncg}
\author{K.K.~Hill} \affiliation{\colorado}
\author{J.L.~Nagle} \affiliation{\colorado}
\author{D.V.~Perepelitsa} \affiliation{\colorado}


\date{\today}

\begin{abstract}
Two-particle correlations have been used extensively to study hydrodynamic flow patterns in heavy-ion collisions.
In small collision systems, such as $p$$+$$p$ and $p$$+$$A$, where particle multiplicities are much smaller than in $A$$+$$A$
collisions, nonflow effects from jet correlations, momentum conservation, particle decays, etc. can be significant,
even when imposing a large pseudorapidity gap between the particles.
A number of techniques to subtract the nonflow
contribution in two-particle correlations have been developed by experiments at the Large Hadron Collider (LHC) and
then used to measure particle flow in $p$$+$$p$ and $p$$+$Pb collisions.
Recently, experiments at the BNL Relativistic Heavy
Ion Collider (RHIC) have explored the possibility of adopting these techniques for small collision systems at lower energies.
In this paper, we test these techniques using
Monte Carlo generators \textsc{pythia} and \textsc{hijing}, which do not include any collective flow, and
\textsc{ampt}, which does.
We find that it is crucial to examine the results of such tests both
for correlations integrated over particle transverse momentum $p_T$ and differentially as a function of $p_T$.
Our results indicate reasonable nonflow subtraction for $p$$+$$p$ collisions at the highest LHC energies, while
failing if applied to $p$$+$$p$ collisions at RHIC.
In the case of $p$$+$Au collisions at RHIC, both \textsc{hijing} and \textsc{ampt} results indicate a substantial over-estimate of nonflow for
$p_{T}\gtrsim1~{\rm GeV}/c$ and hence an underestimate of elliptic flow and overestimate of triangular flow.

\end{abstract}





\maketitle



\section{Introduction}
\label{sec:intro}

The standard evolution model of relativistic heavy ion collisions involves the translation of
initial geometric anisotropies into final momentum correlations via nearly perfect hydrodynamic flow followed
by hadronic rescattering~\cite{Romatschke:2017ejr,Heinz:2013th}.   Starting with first indications in
\pp collisions at the CERN Large Hadron Collider (LHC)~\cite{Khachatryan:2010gv}, there is now a wealth of data indicating similar translation of geometry into flow in smaller collisions systems at the BNL Relativistic Heavy Ion Collider (RHIC) and the LHC~\cite{Nagle:2018nvi}.   Here we focus
on \pp collisions at the LHC and \pp and \pau collisions at RHIC where one is pushing the limits of
how small a system and how small a particle multiplicity still results in significant final flow signatures.
It is imperative to consider both collision energies and systems since a simultaneous description within the
perfect fluid paradigm currently gives the best data description~\cite{Weller:2017tsr,PHENIX:2018lia}.

In these small systems, there are many contributions to the final particle correlations and some of these
contributions increase in relative strength with decreasing multiplicity.   There are many sources of correlations among only a subset of the final particles that result in an increased probability of particles to be nearby in momentum space (i.e.,
close in both pseudorapidity $\Delta \eta \approx 0$ and azimuthal angle $\Delta \phi \approx 0$).    Examples
include the decay of heavy resonances, correlations due to quantum statistics (HBT effects)~\cite{Lisa:2005dd}, and the fragmentation
of a high momentum quark or gluon---the last of these resulting in what are referred to as jet correlations.    These correlations are effectively minimized
by requiring the two particles to have a substantial gap in pseudorapidity, typically $|\Delta \eta| > 2$.    There
are other correlations that survive such a cut including dijet correlations.   In a leading-order hard scattering, a
parton (quark or gluon) and partner parton will be nearly back-to-back in azimuth ($\Delta \phi \approx \pi$) and can be widely separated in pseudorapidity, since the incoming partons need not be longitudinally momentum balanced.  As such, one may expect
two-particle correlations with an enhancement near $\Delta \phi \approx \pi$ that extends long range in pseudorapidity.

\begin{figure}[htb]
\includegraphics[width=1.00\linewidth]{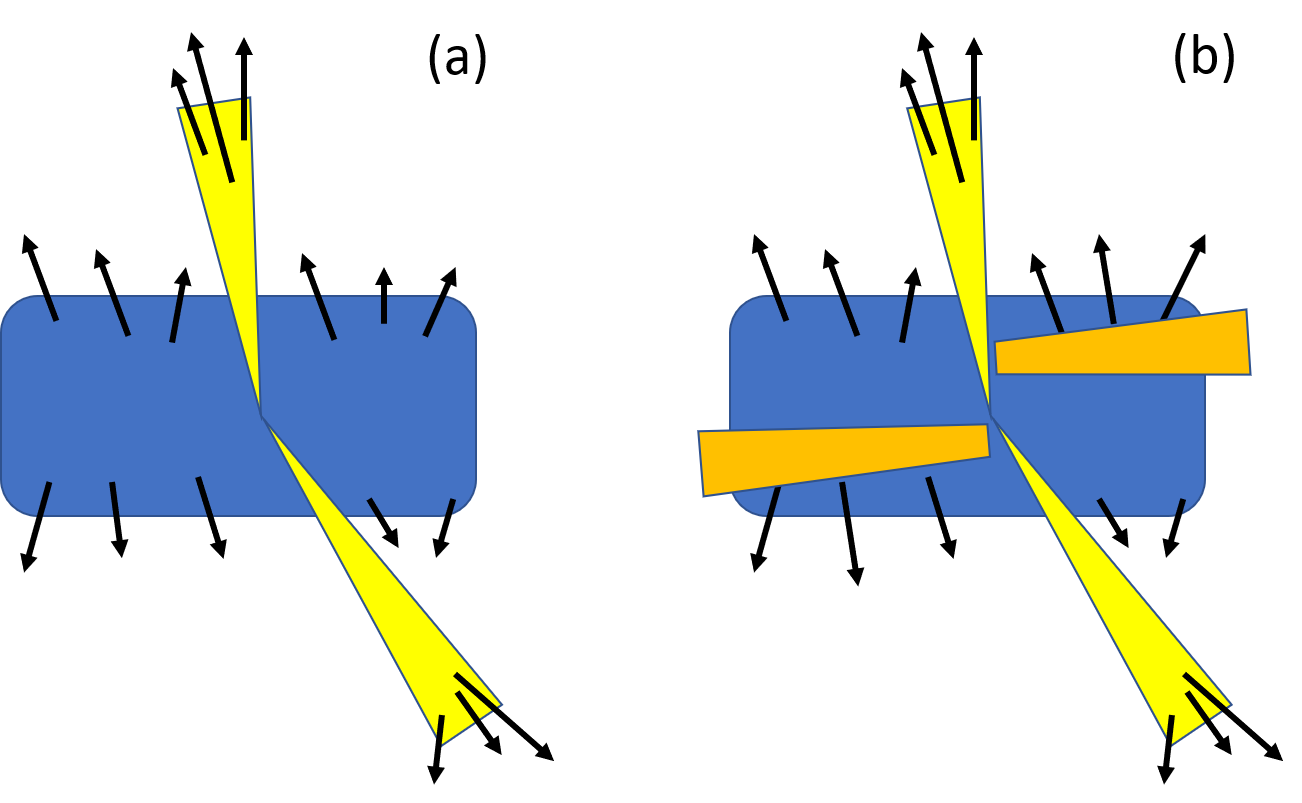}
\caption{\label{fig:cartoon}
(a) A diagram of the simplest scenario with medium particles emitted from the bulk (blue region) and jet particles emitted from two fragmenting partons (yellow cones).   (b) A diagram with the addition of two longitudinally oriented strings such that momentum conservation effects may be present over large rapidity regions.}
\end{figure}

\begin{figure*}[htb]
\includegraphics[width=0.49\linewidth]{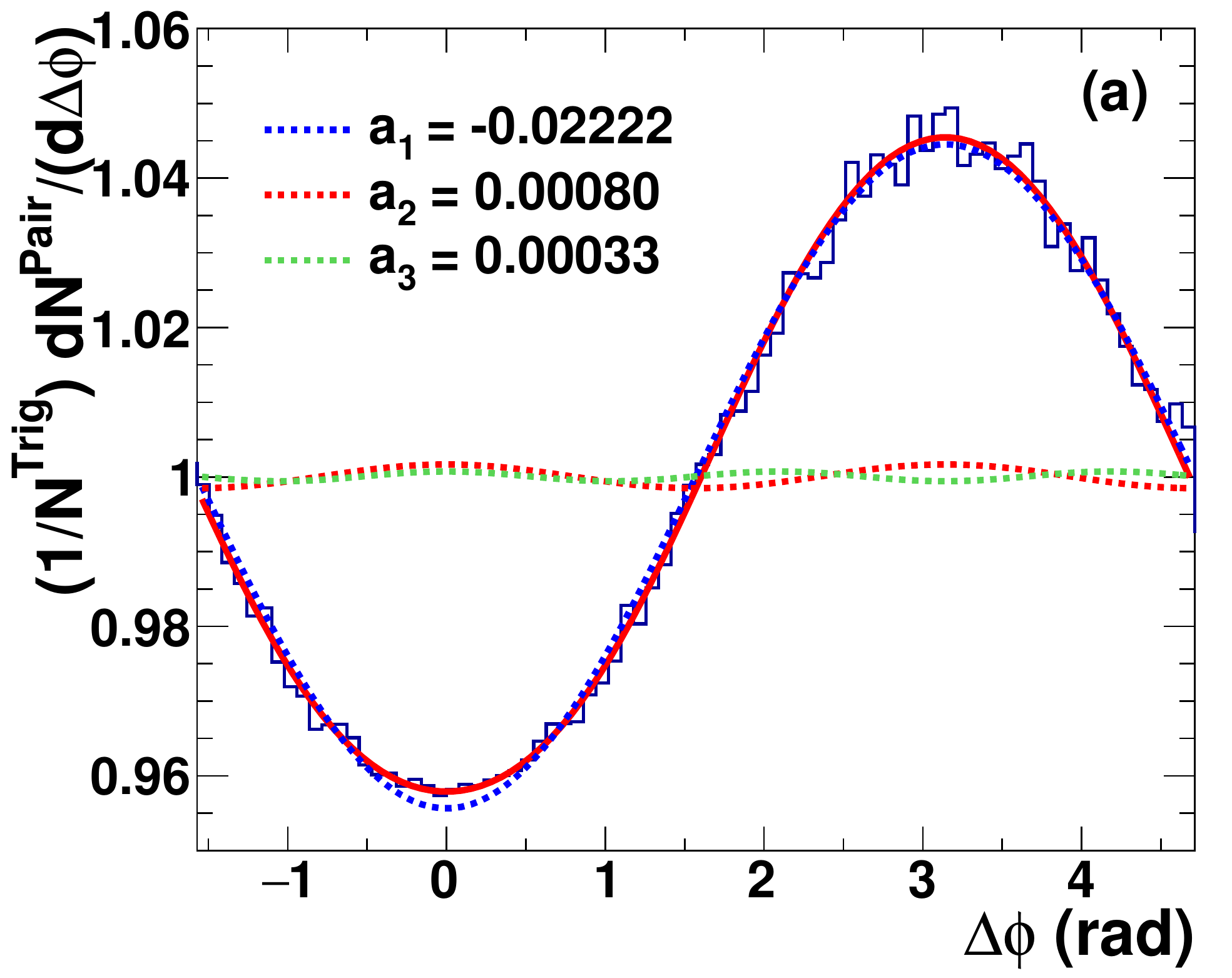}
\includegraphics[width=0.49\linewidth]{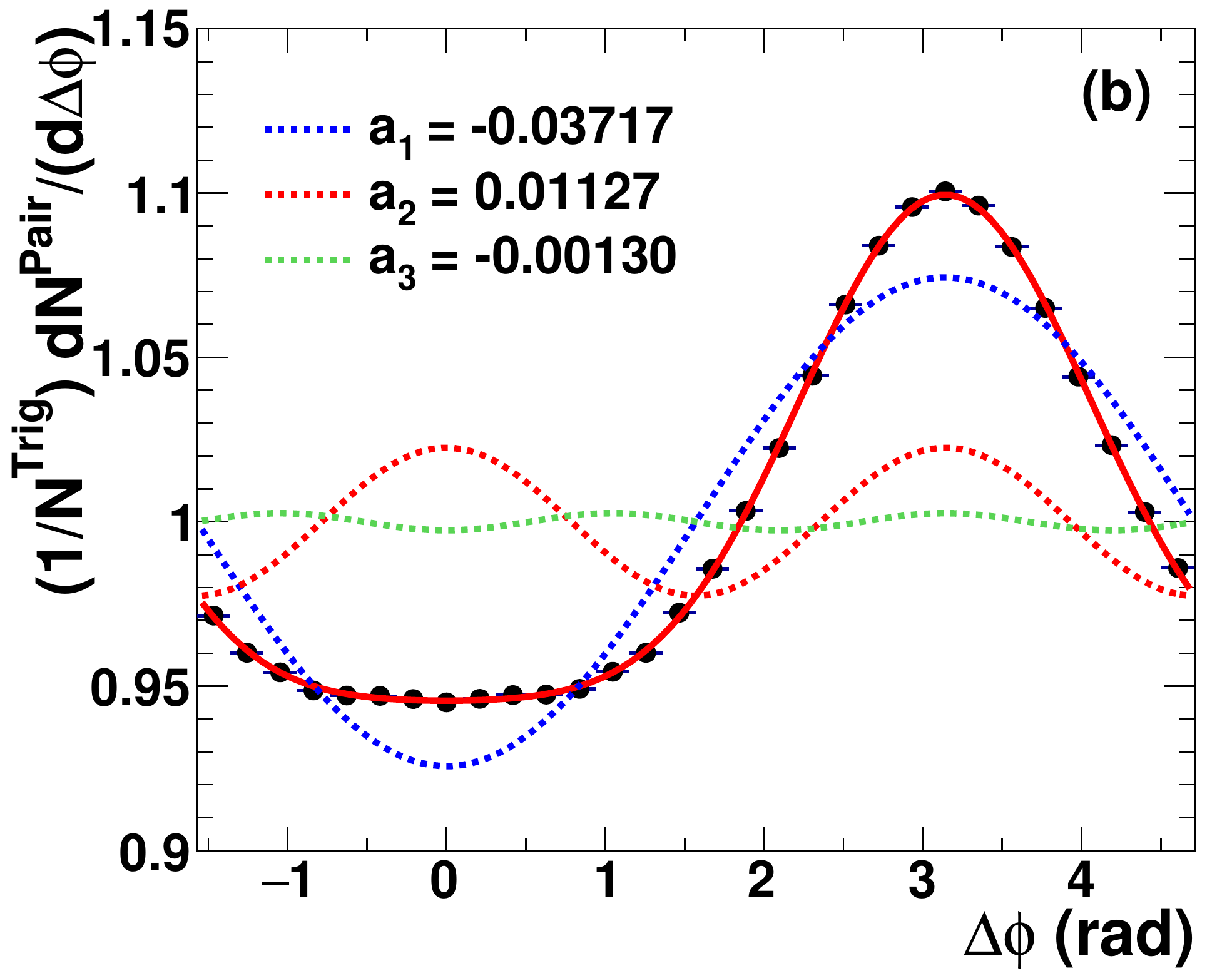}
\caption{\label{fig:phasespace}
(a) Two-particle correlation for pairs generated from pure $N$-body phase space with a pseudorapidity gap $|\Delta\eta|>2$ requirement.
(b) Two-particle correlation for pairs generated in \pythia with a pseudorapidity gap $|\Delta\eta|>2$ requirement.
Fit values for the Fourier decomposition ($a_{1}$, $a_{2}$, and $a_{3}$) are also shown.}
\end{figure*}

As shown in Figure~\ref{fig:cartoon}, one can imagine a toy scenario where a medium indicated by the blue region is created in a collision.   In addition, a hard scattering results in two partons that largely fragment into hadrons outside of the medium.   In a perfectly factorized picture, the medium particles may have angular correlations that
reflect the flowing fluid while the jet particles are correlated with each other but are uncorrelated with the medium geometry.   As one examines collisions with larger final particle multiplicities and larger volume, a larger fraction of the particles come from the medium and thus the influence of the jet correlations is reduced.   In this simplest
scenario the correlations from the jet particles are referred to as ``nonflow'' and those from the medium particles
are referred to as ``flow.''   We highlight that it is impossible for this factorization to be perfectly true.
The jet partons cannot hadronize without some color interaction with the medium or underlying event,
and in some case many of the jet partons scatter with medium partons.
For that reason there is no direct separability of the two, rather they are related through a convolution.

Another key source of nonflow correlations is simply global momentum conservation.  The simplest case would be to imagine $N$ particles emitted isotropically but obeying momentum conservation. The \textsc{root} software package~\cite{Brun:1997pa} has a class \texttt{TGenPhaseSpace} that allows for $N$-body decays simply following phase space filling
and momentum conservation.

Of course, we do not expect that the $N$ particles are distributed simply according to phase space rules.   Within a model such as \pythia~\cite{Sjostrand:2007gs}, one can think of every two-jet event as really being at least a four-jet event.     As shown in
panel (b) of Fig.~\ref{fig:cartoon}, if two incoming partons have a large momentum transfer, the resulting beam remnants extend nearly longitudinal color strings (i.e., the other two jets).   These strings can have a transverse momentum kick such that particles emitted from the upper (orange) string have a slight trend upward and from the lower (orange) string a slight trend downward.    Thus, the partons and resulting hadrons will have a momentum conservation correlation that may be convolved with a flow correlation if these particles undergo additional final state scattering.

The $\Delta \phi$ correlation function with the requirement $|\Delta \eta| > 2$
is shown in Fig.~\ref{fig:phasespace},
where the $N$-body
phase space calculation is shown in the left panel and the
jet-type correlation from \pythia is shown in the right panel.
Most correlation analyses characterize the two-particle correlation in $\Delta \phi$ in terms of a Fourier decomposition~\cite{Voloshin:1994mz}:
\begin{equation}
    f(\Delta \phi) = G \bigg\{1 + 2 \sum_{n=1}^{\infty} a_{n} \cos(n\Delta\phi) \bigg\},
\label{eq:fourier}
\end{equation}
where the coefficients $a_{n}$ are the Fourier coefficient at order $n$,
$G$ is the normalization factor corresponding to the average number of associated particles per trigger particle in the sample, and all sine terms vanish due to symmetry.
The coefficients $a_{n}$ are shown for the two nonflow contributions in Fig.~\ref{fig:phasespace}.  In the case of the $N$-body phase space, the dominant contribution is $c_{1}$ with a depletion near $\Delta \phi \approx 0$ and an enhancement of pairs near $\Delta \phi \approx \pi$.   However, there are nonzero higher order contributions and their relative strengths will of course depend on the other correlations embedded along with momentum conservation.  For the jet-type correlations, the dominant term is also $a_{1}$ due to the strong long-range away-side peak. However, the successive terms contribute significantly
and with alternating signs in order to describe the nearly flat region around $\Delta\phi \approx 0$ and the peak around $\Delta\phi \approx \pi$.

In the following sections we define the nonflow subtraction methods and then detail the resulting
tests of these methods using the Monte Carlo models {\sc pythia}~\cite{Sjostrand:2007gs}, {\sc hijing}~\cite{Gyulassy:1994ew}, and {\sc ampt}~\cite{Lin:2004en} in
different collisions systems and and at different energies.

\section{Definition of Methods}
\label{sec:method}

The most important assumption in all nonflow subtraction methods is the
assumption that the correlation coefficients $a_n$ in Eq.~(\ref{eq:fourier})
can be separated into two linearly additive contributions $a_n = c_n + d_n$,
where $c_{n}$ is the flow coefficient quantifying correlations related to the initial geometry
and $d_{n}$ is the nonflow coefficient of pair correlations.
As discussed in Sec.~\ref{sec:intro}, this assumption may not be realistic, but this
is the starting point for all nonflow subtraction methods, and so we begin our discussion here.
%
%
With this assumption, Eq.~(\ref{eq:fourier}) can be rewritten as
\begin{linenomath}
\begin{equation}
f(\Delta \phi) = G\bigg\{1 + 2 \sum_{n=1}^{\infty} (c_{n} + d_{n}) \cos(n\Delta\phi)\bigg\}.
\label{eq:subtraction}
\end{equation}
\end{linenomath}

In order to extract the true flow, $c_n$, it is necessary to remove the nonflow contributions, $d_{n}$.
This is particularly important in small systems where the nonflow correlations can dominate the flow signal.
%
%
We can define the nonflow contribution $J(\Delta\phi)$ to $f(\Delta\phi)$ as
\begin{linenomath}
\begin{equation}
J(\Delta\phi) = G\bigg\{2 \sum_{n=1}^{\infty} d_{n} \cos(n\Delta\phi)\bigg\},
\label{eq:jetpart}
\end{equation}
\end{linenomath}
so that Eq.~(\ref{eq:subtraction}) becomes
\begin{linenomath}
\begin{equation}
f(\Delta \phi) = J(\Delta\phi) + G\bigg\{1 + 2 \sum_{n=1}^{\infty} c_{n} \cos(n\Delta\phi)\bigg\}.
\label{eq:subJ}
\end{equation}
\end{linenomath}

All techniques attempting to disentangle flow and nonflow operate by comparing
correlations between two data selection samples: one at higher multiplicity (HM), where
flow is expected to have a larger influence; and one at lower multiplicity (LM), where
flow is expected to have a smaller influence.
The flow in LM events is at times assumed to be negligible or non-existent; however,
we need not consider such differences just yet.  The above equations simply need to
be trivially relabeled so that the LM and HM categories are distinct.  Here
we use LM (HM) as a superscript on all relevant quantities to indicate LM (HM) events.

The terms $c_{n}^\mathrm{HM}$ and $c_{n}^\mathrm{LM}$ are the flow correlation coefficients at low and high multiplicity, respectively;
$d_{n}^\mathrm{HM}$ and $d_{n}^\mathrm{LM}$ are the nonflow correlation coefficients at low and high multiplicity, respectively.

Applying these labels and rewriting Eq.~(\ref{eq:jetpart}) we obtain
\begin{linenomath}
\begin{equation}
\frac{J^\mathrm{LM(HM)}(\Delta\phi)}{G^\mathrm{LM(HM)}} = 2 \sum_{n=1}^{\infty} d^\mathrm{LM(HM)}_{n} \cos(n\Delta\phi),
\label{eq:jetpartratio}
\end{equation}
\end{linenomath}
meaning the $J^\mathrm{LM}(\Delta\phi)$ and $J^\mathrm{HM}(\Delta\phi)$ are related to each other as
\begin{linenomath}
\begin{equation}
\frac{J^\mathrm{HM}(\Delta\phi)}{G^\mathrm{HM}} = \frac{\sum_{n=1}^{\infty} j_nd^\mathrm{LM}_{n} \cos(n\Delta\phi)}{\sum_{n=1}^{\infty} d^\mathrm{LM}_{n} \cos(n\Delta\phi)}\frac{J^\mathrm{LM}(\Delta\phi)}{G^\mathrm{LM}},
\label{eq:jetpartratiorelation}
\end{equation}
\end{linenomath}
where the relational coefficients are $j_{n} = d_{n}^\mathrm{HM} / d_{n}^\mathrm{LM}$.
By estimating the $j_n$ and the flow coefficients at low multiplicity, $c_{n}^\mathrm{LM}$,
the nonflow contribution at high multiplicity can be determined and subtracted.
There are a few different nonflow subtraction methods available on the market based on different assumptions regarding the $c_{n}^\mathrm{LM}$ and $j_{n}$ coefficients.


\subsection{\methodi}
The first method we detail has been developed by the ATLAS Collaboration, as originally
applied in Ref.~\cite{Aad:2015gqa}.   Here we detail the key assumptions made in using this
method.
The first assumption is that the nonflow correlation coefficient
$j_n=d_{n}^\mathrm{HM}/d_{n}^\mathrm{LM}$ is independent of the harmonic number $n$, so that $j_n = j^\mathrm{ATLAS}$ for all $n$.
This allows Eq.~(\ref{eq:jetpartratiorelation}) to be rewritten as:
\begin{linenomath}
\begin{equation}
\frac{J^\mathrm{HM}(\Delta\phi)}{G^\mathrm{HM}} = j^\mathrm{ATLAS}\frac{J^\mathrm{LM}(\Delta\phi)}{G^\mathrm{LM}}.
\label{eq:jetpartratiorelationATLAS}
\end{equation}
\end{linenomath}
This assumption follows from the idea that at the LHC the nonflow correlation is dominated by the jet
contribution.   The additional requirement is then that the shape of the jet correlation does not change
with multiplicity event class.

A second assumption made is that the flow contribution at the lowest harmonic $c_1$ is negligible compared to the nonflow $d_1$,
i.e., $c_1 \ll d_1$.  Hence $c_1$ is completely ignored; alternately one can think of it as being absorbed via redefinition of $d_1$.
Following these assumptions, we can rewrite the angular correlation distributions in Eq.~(\ref{eq:subJ}) for HM events in relation to
LM events,
\begin{linenomath}
\begin{align}
f^\mathrm{LM}(\Delta\phi) &= J^\mathrm{LM}(\Delta\phi) + G^\mathrm{LM} \Big\{ 1+2\sum_{n=2}^{\infty} c_n^\mathrm{LM}\cos(n\Delta\phi) \Big\}, \\
f^\mathrm{HM}(\Delta\phi) &= J^\mathrm{HM}(\Delta\phi) + G^\mathrm{HM} \Big\{ 1+2\sum_{n=2}^{\infty} c_n^\mathrm{HM}\cos(n\Delta\phi) \Big\},
\end{align}
\end{linenomath}
and then combining the two using Eq.~\ref{eq:jetpartratiorelationATLAS} to obtain
\begin{widetext}
\begin{align}
f^\mathrm{HM}(\Delta\phi)
                          &= \frac{G^\mathrm{HM}j^\mathrm{ATLAS}}{G^\mathrm{LM}}f^\mathrm{LM}(\Delta\phi) + G^\mathrm{HM} ( 1- j^\mathrm{ATLAS})
                             \Big\{1+2\sum_{n=2}^{\infty} \Big(\frac{c_{n}^\mathrm{HM} - j^\mathrm{ATLAS} c_{n}^\mathrm{LM}}{1-j^\mathrm{ATLAS}}\Big)\cos(n\Delta\phi) \Big\} \\
                          &= F^\mathrm{temp}f^\mathrm{LM}(\Delta\phi) + G^\mathrm{temp}\bigg\{1+2\sum_{n=2}^{\infty}{c_{n}^\mathrm{temp}}\cos(n\Delta\phi)\bigg\},
\end{align}
\end{widetext}
where $F^\mathrm{temp}$, $G^\mathrm{temp}$, and $c_{n}^\mathrm{temp}$ are parameters
defined by

\begin{align}
F^\mathrm{temp} &= \frac{G^\mathrm{HM}j^\mathrm{ATLAS}}{G^\mathrm{LM}}, \label{eq:Ftemp} \\
G^\mathrm{temp} &= G^\mathrm{HM}(1-j^\mathrm{ATLAS}) \label{eq:Gtemp}, \\
c_n^\mathrm{temp} &= \frac{c_{n}^\mathrm{HM} - j^\mathrm{ATLAS} c_{n}^\mathrm{LM}}{1-j^\mathrm{ATLAS}} \label{eq:cntemp}.
\end{align}
There parameters are obtained from fitting $f^\mathrm{HM}(\Delta\phi)$ with $f^\mathrm{LM}(\Delta\phi)$.
Thus, this approach is often referred to as a ``template-fitting'' method because the LM correlation function serves
as a template for the HM correlation function.

In the special case where the flow coefficients are identical at all orders between the
low multiplicity and high multiplicity classes, i.e., $c_{n}^{\rm HM} = c_{n}^{\rm LM}$ for all $n$, then
the extracted $c_{n}^\mathrm{temp}$ is exactly equal to $c_{n}^\mathrm{HM}$.  This is a simplifying
assumption that has no concise physics motivation.   This special case was assumed in the
ATLAS publications~\cite{Aad:2015gqa,Aaboud:2016yar}.

Relaxing this special case, an additional correction should be applied to the fitted value to obtain $c_{n}^\mathrm{HM}$:

\begin{equation}
	c_{n}^\mathrm{HM} = c_{n}^\mathrm{temp} - j^\mathrm{ATLAS} (c_{n}^\mathrm{temp} - c_{n}^\mathrm{LM}),
	\label{eq:atlas_corr}
\end{equation}
where $j_n$ can be obtained using fit parameters $j^\mathrm{ATLAS} = G^\mathrm{LM}F^\mathrm{temp}/G^\mathrm{HM}$ and $c_{n}^\mathrm{LM}$ is usually estimated using the flow coefficient measured in the second lowest multiplicity sample.   The ATLAS Collaboration applies
this additional correction in Ref.~\cite{Aaboud:2018syf}.


\subsection{\methodii}
The second method we detail has been developed by the CMS Collaboration, as originally
applied in Ref.~\cite{Khachatryan:2016txc}.  Similar approaches have also been explored by the ATLAS and ALICE collaborations~\cite{Aad:2012gla,Abelev:2012ola}.    Here we detail the key assumptions made in using this
method.

Like the ATLAS method, this method also assumes that the nonflow correlation shape does not change with multiplicity class,
hence
$j_n=d_{n}^\mathrm{HM}/d_{n}^\mathrm{LM}$ is independent of the harmonic number $n$, so that $j_n = j^\mathrm{CMS}$ for all $n$.
This allows Eq.~(\ref{eq:jetpartratiorelation}) to be rewritten as
\begin{linenomath}
\begin{equation}
\frac{J^\mathrm{HM}(\Delta\phi)}{G^\mathrm{HM}} = j^\mathrm{CMS}\frac{J^\mathrm{LM}(\Delta\phi)}{G^\mathrm{LM}},
\label{eq:jetpartratiorelationCMS}
\end{equation}
\end{linenomath}
in exact analogy to the ATLAS case.  However, the method by which $j^\mathrm{CMS}$ is determined is different.
They assume that the nonflow correlation coefficients, at least for $n>1$, are dominated by jet contributions.  Since
they assume the jet shape is independent of multiplicity class, they only need to determine the
relative jet yields in the different multiplicity classes.   They do this by measuring the near-side jet
yield in the different multiplicity classes via the short range (SR) correlation, i.e., with
$\Delta \phi \approx 0$ and $\Delta \eta \approx 0$.    They isolate the near-side jet peak by taking the difference
between the SR and long-range (LR) correlations as follows:
\begin{linenomath}
\begin{equation}
j^\mathrm{CMS} = \frac{G^\mathrm{LM}\int_{-1.2}^{1.2} (f^\mathrm{HM}_\mathrm{SR}(\Delta\phi) - f^\mathrm{HM}_\mathrm{LR}(\Delta\phi))\mathrm{d}\Delta\phi}{G^\mathrm{HM}\int_{-1.2}^{1.2} (f^\mathrm{LM}_\mathrm{SR}(\Delta\phi) - f^\mathrm{LM}_\mathrm{LR}(\Delta\phi))\mathrm{d}\Delta\phi},
\end{equation}
\end{linenomath}
where $f^\mathrm{HM (LM)}_\mathrm{SR}(\Delta\phi)$ is the short-range correlation with $|\Delta\eta| < 1$
and $f^\mathrm{HM (LM)}_\mathrm{LR}(\Delta\phi)$ is the long-range correlation with $2 < |\Delta\eta| < 5$.
The short-range and long-range correlations are usually normalized by number of trigger particles such that the integral of their difference over $-1.2 < \Delta\phi < 1.2$ is close to the number of particles produced per jet.

Next, direct Fourier fits are applied to the long-range correlations at low and high multiplicities, and the Fourier coefficient $a_n^\mathrm{HM (LM)}$ are extracted.
This approach assumes that there is no flow contribution at all in the low multiplicity event
selection, i.e., $c_{n}^\mathrm{LM} = 0$.   This is an extreme assumption that forces any extraction
of flow coefficients as a function of multiplicity class to approach zero at low multiplicity, and
is quite different from the ATLAS approach.
Then at each order $n$ the flow coefficient at high multiplicity is given by
\begin{linenomath}
\begin{equation}
c_{n}^\mathrm{HM} = a_{n}^\mathrm{HM} - j^\mathrm{CMS}a_{n}^\mathrm{LM}.
\label{eq:cms_raw}
\end{equation}
\end{linenomath}
As noted above, as measurements move towards low multiplicity, such that $c_n^\mathrm{HM} \approx c_n^\mathrm{LM}$, $c_n^\mathrm{HM}$ will converge to zero by construction.  The CMS results using this method thus always trend to zero.
However, if a nonzero flow coefficient in the low multiplicity sample is allowed, Eq.~(\ref{eq:cms_raw}) could be rewritten as
\begin{linenomath}
\begin{equation}
c_{n}^\mathrm{HM} =a_{n}^\mathrm{HM} - j^\mathrm{CMS}(a_{n}^\mathrm{LM} - c_{n}^\mathrm{LM}),
\label{eq:cms_corr}
\end{equation}
\end{linenomath}
where $c_{n}^\mathrm{LM}$ can be estimated in the same way as in the ATLAS method.
Once the generalized results as shown in Eq.~(\ref{eq:atlas_corr}) and Eq.~(\ref{eq:cms_raw}) are used, one should be able to obtain similar results with the two methods.

One remaining difference between the methods would arise from the different ways of estimating the jet variable $j_n$.    The nonflow correlation can
have contributions from jets as well as overall momentum conservation.   The momentum conservation contribution is predominantly in the $d_1$ coefficient.
Thus, in the CMS method where they subtract the nonflow order by order, even though the $j^\mathrm{CMS}$ is determined from jets alone, it does not matter since
they are not extracting a first-order $c_1$ at high multiplicity.   In contrast, in the ATLAS template-fitting method, it assumes the nonflow shape
(combining both jets and momentum conservation) scale in the same way.   This seems unlikely given the findings by CMS via the short-range correlations.
Since neither experiment extracts a $c_1$ at high multiplicity where they would get very different results, there is only the potential for a residual effect on the higher order $c_{n}$ via the ATLAS fitting procedure.


\begin{figure}[htb]
\includegraphics[width=1.0\linewidth]{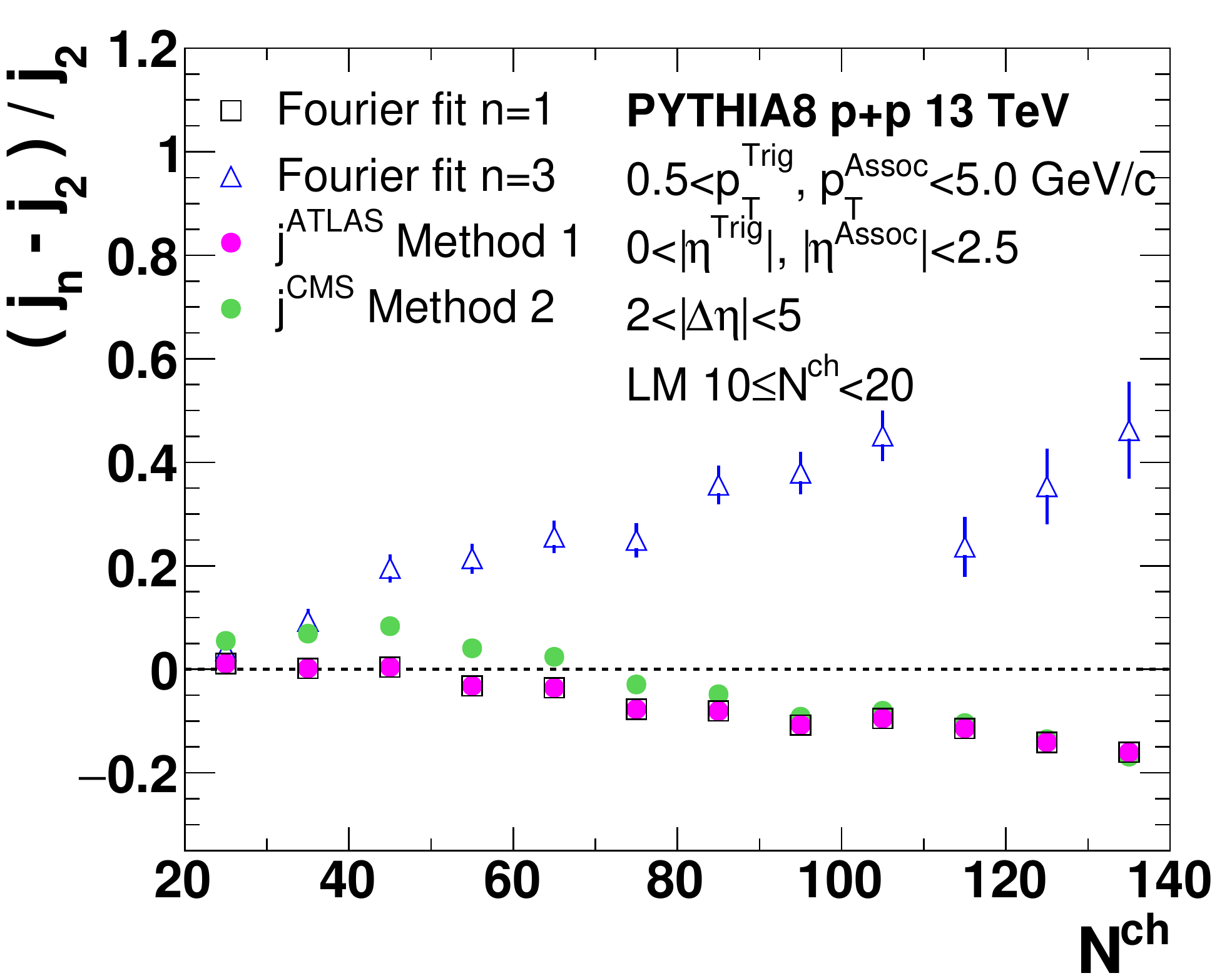}
\caption{Relative difference between the ratio of nonflow correlation coefficient at high multiplicity to that at low multiplicity, $j_n = d_{n}^\mathrm{HM} / d_{n}^\mathrm{LM}$, from different methods to that from direct Fourier fit of the second order, $j_2$, as a function of multiplicity. A pseudorapidity gap of $2 < |\Delta\eta| < 5$ is required.}
\label{fig:pp13TeV_dn_ratio}
\end{figure}

\begin{figure}[htb]
\includegraphics[width=1.0\linewidth]{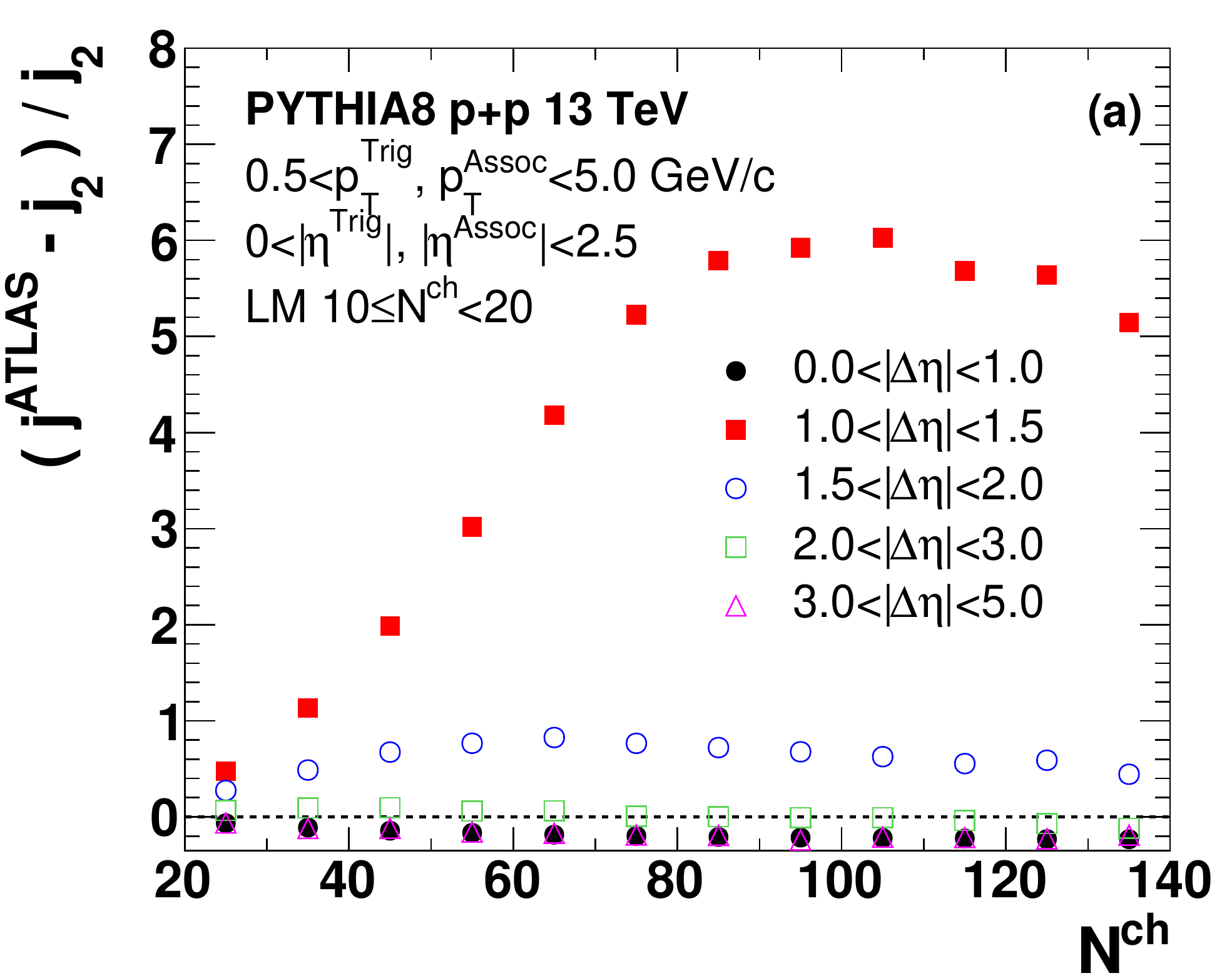}
\includegraphics[width=1.0\linewidth]{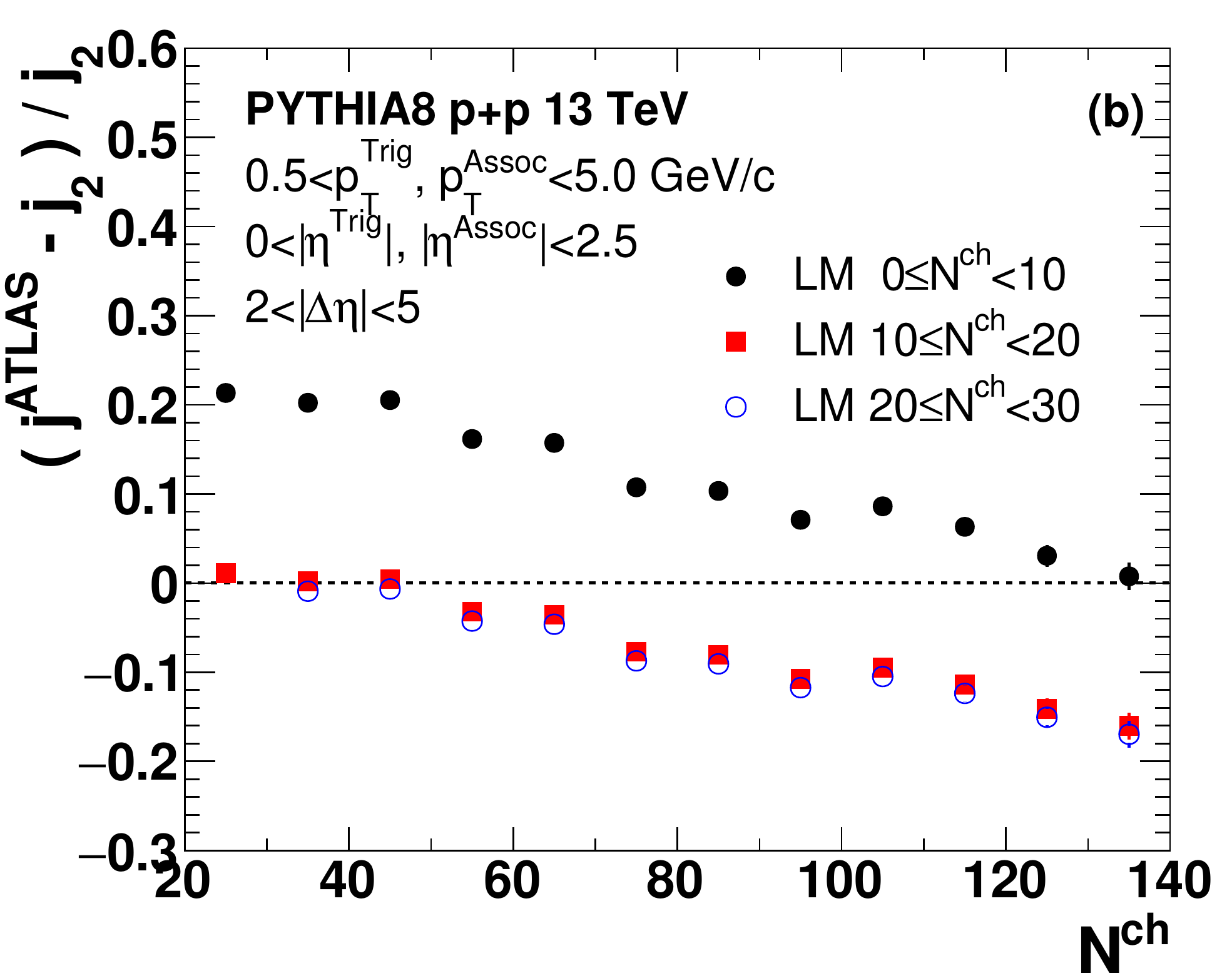}
\caption{The relative difference between $j^\mathrm{ATLAS}$ and $j_2$ from direct Fourier fit (a) as a function of multiplicity for different gaps and (b) different choices of low multiplicity sample.}
\label{fig:pp13TeV_dn_dependence}
\end{figure}

\subsection{Jet Shape Assumption}

\methodb are derived from the same expression and heavily rely on the assumption that the jet shape is the same at low and high multiplicities.
This assumption can be tested in \pythia, in which no flow contribution is expected, i.e., $c_n = 0$.  We perform a direct Fourier coefficient
extraction in the low and high multiplicity event classes and then compute $j_n = d_{n}^\mathrm{HM} / d_{n}^\mathrm{LM}$.   To examine if all $j_n$
are the same, we plot $(j_n - j_2) / j_2$ as shown in Fig.~\ref{fig:pp13TeV_dn_ratio}.    The points for $n=1$ and $n=3$ indicate that the
assumption of a common $j_n$ is violated in \pythia and that violation increases with higher multiplicity, reaching a level for $n=1$ ($n=3$)
relative to $n=2$ of $-20\%$ (+40\%).   We note that the $j_2$ is decreasing with multiplicity, so the impact of this violation might be
actually smaller at higher multiplicity.    We also show the $j^\mathrm{ATLAS}$ and $j^{\mathrm{CMS}}$ extracted values.   One can see that
they closely agree with the $n=1$ case and deviate significantly from the $n=3$ case.  Note that the $j^\mathrm{ATLAS}$ is actually identical
to the $n=1$. This is because there is no first-order flow coefficient ($c_{1}=0$) by construction. The fit procedure results in $j^\mathrm{ATLAS}$ being determined completely by the first-order nonflow correlation coefficient ($d_1$), while the contribution from higher order nonflow correlation coefficients is absorbed into the higher order flow correlation coefficients.

We note that the $j_n$ assumption can also be sensitive to the particular pseudorapidity gap chosen and the low multiplicity reference selection.
The sensitivity of the $j_n$ estimated by \methodi to the choice of pseudorapidity gap and low multiplicity selection is shown in Fig.~\ref{fig:pp13TeV_dn_dependence}.  The method has particular problems when the pseudorapidity gap is $1 < |\Delta \eta| < 1.5$, because
there are contributions from both the near- and away-side jet.    The method also is most sensitive when the low multiplicity selection is
at its lowest.

All of these violations of the assumptions in \methodb can only be gauged in terms of consequences on the extracted flow coefficients $c_n$
by testing the methods on various Monte Carlo physics models.  In the following sections, we examine the results of such tests.


\section{Closure Tests with \pythia and \hijing}

We now test these procedures on Monte Carlo generators.
In Monte Carlo generators such as \pythia~\cite{Sjostrand:2007gs} and \hijing~\cite{Gyulassy:1994ew},
there is no collective flow in a sense that there are no final state interactions
to translate a spatial geometry into momentum anisotropies. Thus, one expects that
the application of a successful nonflow subtraction method should result in flow
anisotropy coefficients of exactly zero in these cases.
These are thus referred to as closure tests.  By measuring the residual of
these coefficients $c_n$, the level of closure can be quantified. We have applied the methods
described in Sec.~\ref{sec:method} to determine the level of closure in various Monte Carlo
generators.

Before proceeding we want to define the nomenclature used in the following sections.  In our studies we focus on the
extraction of elliptic flow $n=2$, and show all results in terms of \vtt and \vttsub extracted
from the two-particle correlations.
First we define $\vtt = a_2(p_{T,1},p_{T,2})$,
where the trigger and associated particles are in momentum selections around $p_{T,1}$ and $p_{T,2}$, respectively.
Once the nonflow subtraction technique is applied and the $c_n$ coefficients are estimated, we define
$\vttsub = c_2(p_{T,1},p_{T,2})$.
These quantities are related but not equal to $v_2^2$.  The standard differential $v_2$ as a function of \pt
would then be extracted as
$v_2(p_{T,1}) = \vtt/\sqrt{a_2(p_{T,2},p_{T,2})}$
and
$v_2^{\rm sub}(p_{T,1}) = \vttsub/\sqrt{c_2(p_{T,2},p_{T,2})}$.
In the case where both $p_{T,1}$ and $p_{T,2}$ represent the same broad range in \pt,
e.g., $0.5<\pt<5~{\rm GeV}/c$,
these expressions reduce to
$v_2 = \sqrt{\vtt} = \sqrt{a_2}$
and
$v_2^{\rm sub} = \sqrt{\vttsub} = \sqrt{c_2}$,
representing integral $v_2$ as a function of some event-level variable (multiplicity, centrality, etc.).

We also highlight that experiments have different techniques for selecting lower and higher multiplicity events.   If the
selection is based on charged particle multiplicity centered around midrapidity, i.e., in the same range as the two particles
used for the correlations, we label the event categories in terms of \nch.   In contrast, other measurements utilize
multiplicity or energy in a forward or backward rapidity range, for example in the Pb-going direction in \pPb collisions, and
thus outside the range of the two particles used for the correlations.  In this second case, we refer to the event selections by
``centrality.''

\subsection{LHC \pp Case}
\label{sec:lhc_pp}

\begin{figure*}[htb]
\includegraphics[width=0.49\linewidth]{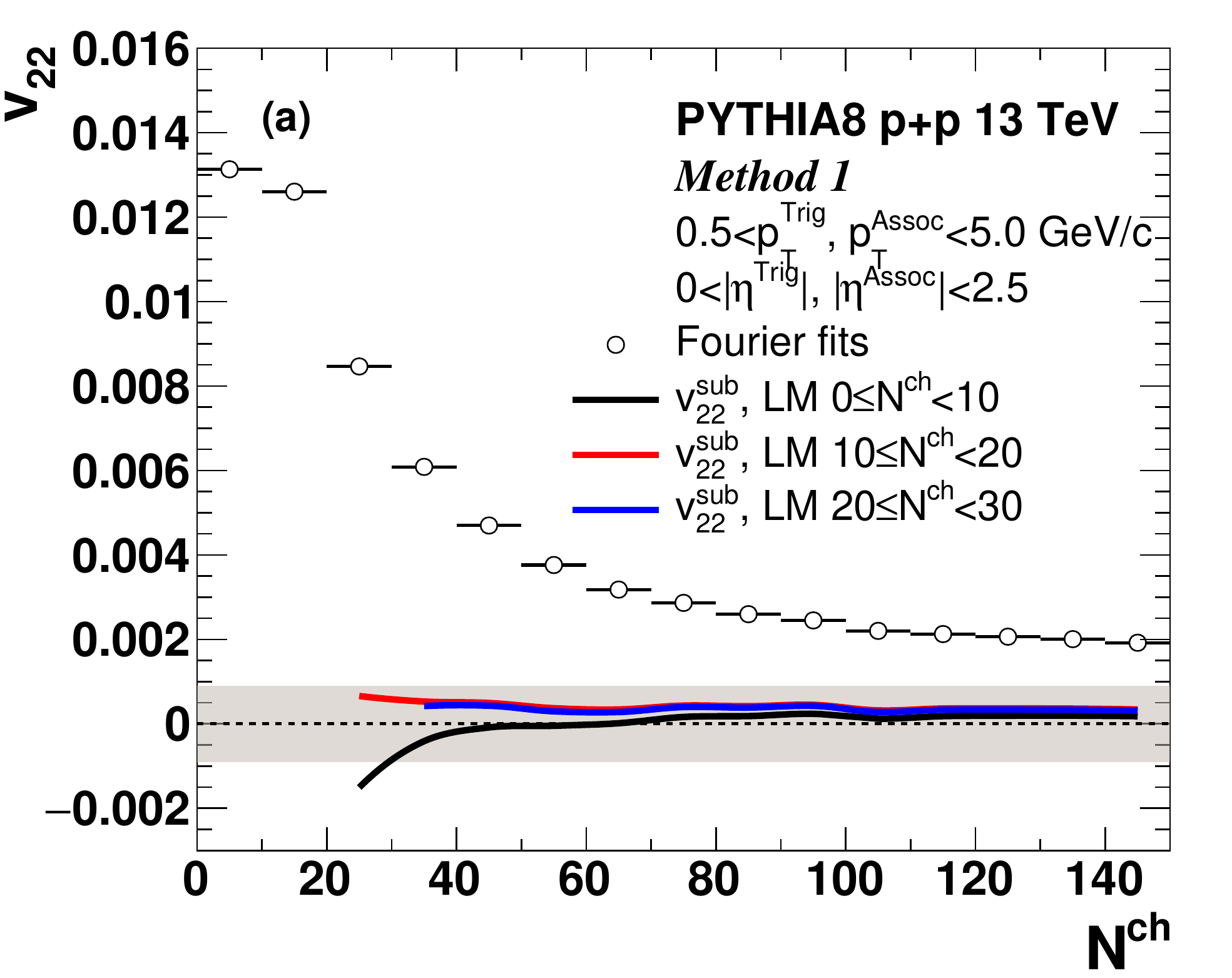}
\includegraphics[width=0.49\linewidth]{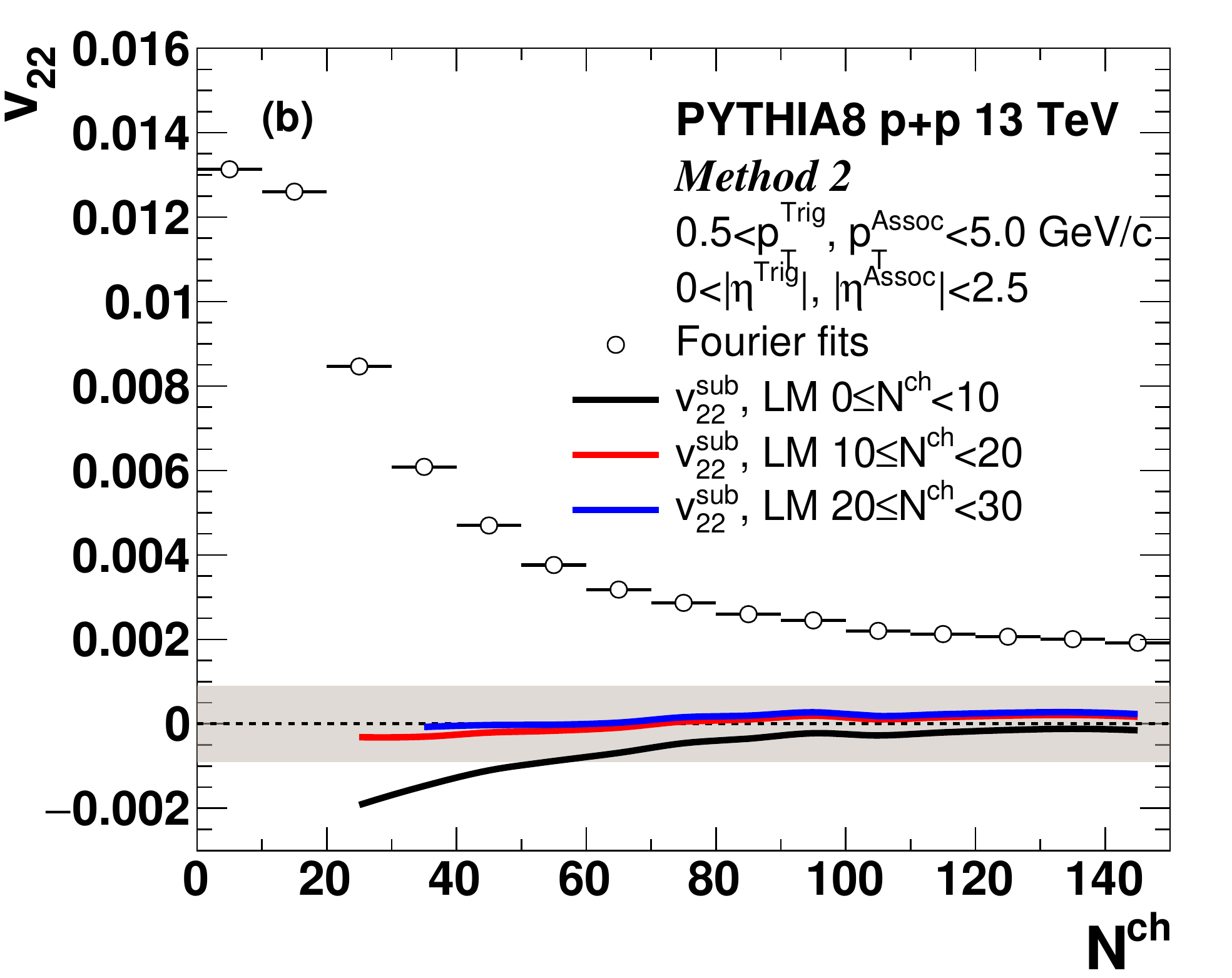}
\caption{\label{fig:pp13TeV_v22_mult}
The second-order Fourier coefficient \vtt of long-range ($2<|\deta|<5$) two-particle correlation as a function of charged hadron multiplicity in \pp collisions at \sqs~=~13~TeV from \pythia before and after nonflow subtraction. Multiplicity is defined as the number of charged hadrons in $\pt>0.4~{\rm GeV}/c$ and $|\eta|<2.5$. Gray bands correspond to a 3\% $|v_2|$ window.}
\end{figure*}

\begin{figure*}[htb]
\includegraphics[width=0.49\linewidth]{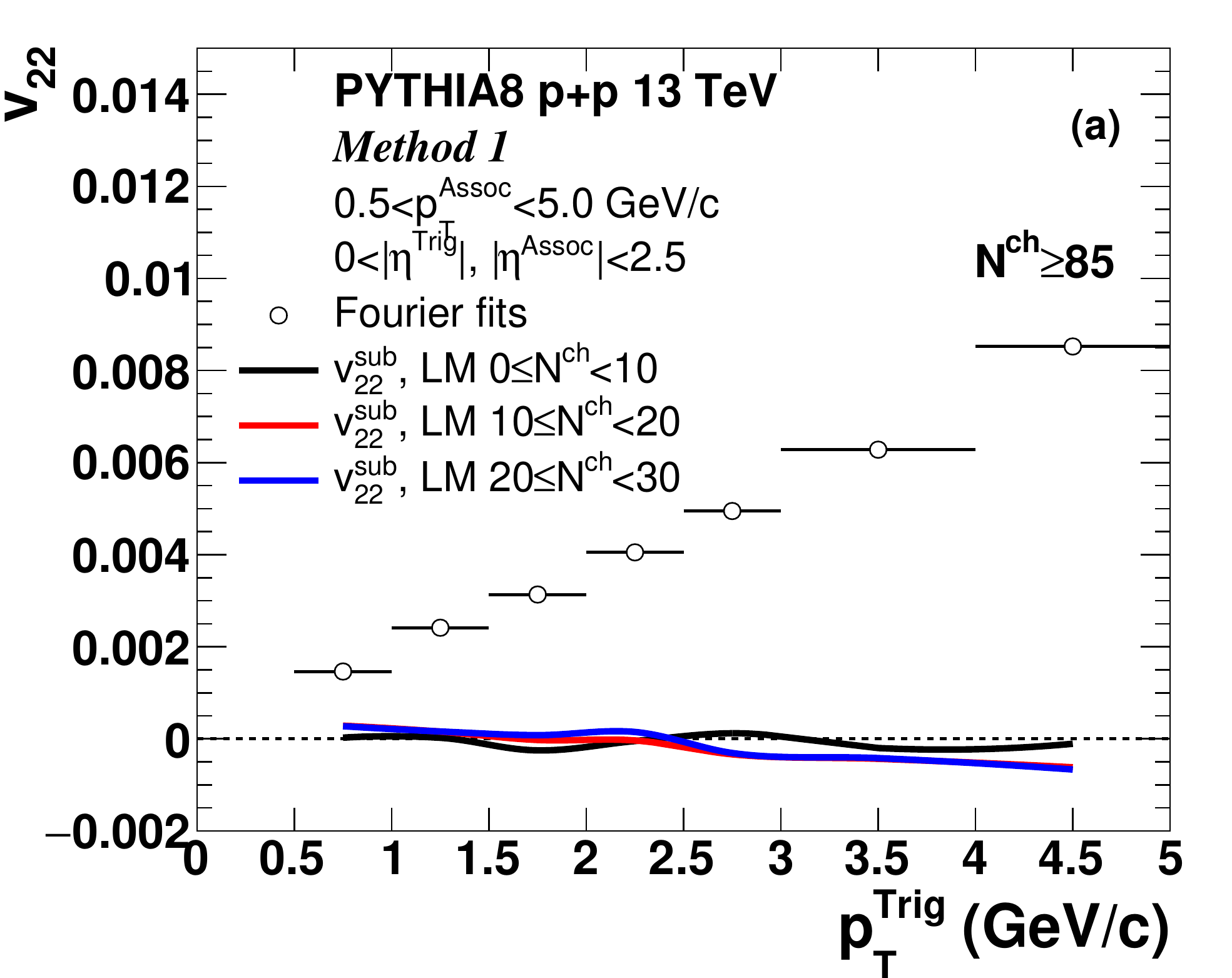}
\includegraphics[width=0.49\linewidth]{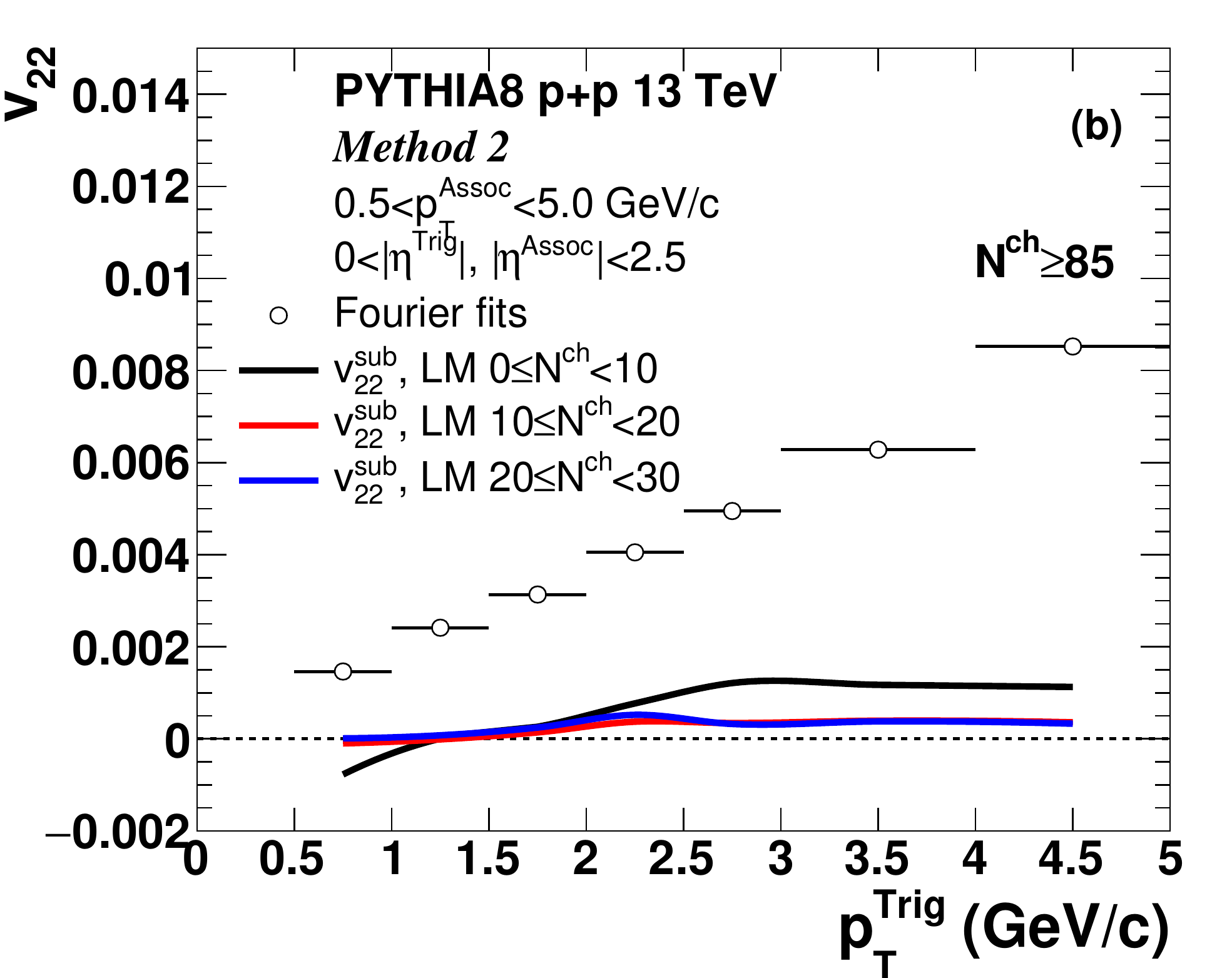}
\caption{\label{fig:pp13TeV_v22_pT}
The second-order Fourier coefficient \vtt of long-range ($2<|\deta|<5$) two-particle correlation as a function of \pt in \pp collisions of $\nch\geq85$ at \sqs~=~13~TeV from \pythia before and after nonflow subtraction. Multiplicity is defined as the number of charged hadrons in $\pt>0.4~{\rm GeV}/c$ and $|\eta|<2.5$.}
\end{figure*}

These methods were developed particularly by the ATLAS and CMS Collaborations for use
in the highest energy \pp collisions, and we examine that collision system first.   We note that
the CMS Collaboration has presented the results of a \pythia closure test for particles
integrated over \pt in their paper (see Fig. 4 of Ref.~\cite{Khachatryan:2016txc}),
but such a study has not been published by ATLAS.   We consider acceptance selection cuts similar to those in the experiments, though use one set so that all methods are compared
apples-to-apples. Correlations are determined from all charged hadrons within $|\eta|<2.5$
from $10^{8}$ \pythia \pp events (\texttt{SoftQCD:nonDiffractive = on}) at \sqs~=~13~TeV.
Charged hadrons with $\pt>0.4~{\rm GeV}/c$ ($\pt>0.5~{\rm GeV}/c$) are used for event
multiplicity categorization (particle correlations). We note that a slightly different $\pt$ selection
($\pt>0.3~{\rm GeV}/c$) has been used for data analysis by the CMS collaboration.

Figure~\ref{fig:pp13TeV_per_trig_yield} (shown in the Appendix for clarity) shows one
dimensional \dphi two-particle correlation functions for short-range ($|\deta|<1$) and
long-range ($2<|\deta|<5$) regions.  Charged hadrons are required to satisfy $\pt>0.5~{\rm GeV}/c$ and $|\eta|<2.5$.
Each panel presents a different range of
charged hadron multiplicity event selection \nch. In this case,
there is no visually obvious shape variation of the \dphi correlation
functions both in short and long ranges throughout the entire multiplicity range.   This implies that the $j_n$ values
are going to be approximately independent of $n$, though not exactly as demonstrated in the previous discussion.
The dashed lines are fits to the distributions of long-range correlations to extract Fourier coefficients described in
Eq.~(\ref{eq:fourier}).

Figure~\ref{fig:pp13TeV_v22_mult} shows the second-order Fourier
coefficients \vtt extracted directly from the correlation functions and after the nonflow
subtraction technique is applied.   Results are shown for charged hadrons with $0.5<\pt<5~{\rm GeV}/c$ as a function of event
multiplicity \nch.   Figure~\ref{fig:pp13TeV_v22_pT} show the results for the highest multiplicity selection and for trigger particles
as a function of \pt.  The left and right panels of both figures apply the \methodi (ATLAS) and \methodii (CMS) methods, respectively.

The nonzero value of the coefficients after subtraction indicate that the closure test is not perfectly satisfied.   This is
mainly due to remaining jet correlations from the near-side ($\dphi\approx 0$), even with the
large \deta gap, and/or a small shape change from the away-side ($\dphi\approx \pi$).
The nonflow effect on \vtt is larger at low multiplicity events, and
it becomes smaller as the event multiplicity increases. The two nonflow subtraction methods are
applied with three different ranges of low multiplicity selection, and the results of
the subtracted \vttsub from each method are shown in each panel.

In Fig.~\ref{fig:pp13TeV_v22_mult} the gray bands correspond
to $|v_{2}|=0.03$ (thus equivalent to $\vtt = 0.03 \times 0.03$).    There is some sensitivity to the
low multiplicity selection.  The \vttsub with low multiplicity selection $10\leq\nch<20$ and
$20\leq\nch<30$ are within the level of $|v_{2}|<0.03$ over the entire multiplicity range.
Using events in $0\leq\nch<10$ as a low multiplicity bin results in \vttsub with a
slightly larger deviation from zero in the lower multiplicity range, but it converges
with the \vttsub from the other cases at higher multiplicities.

The same test is done as a function of the \pt of the trigger particle (\pttrig) for
the highest 5\% multiplicity events ($\nch\geq85$). Figure~\ref{fig:pp13TeV_per_trig_yield_pTbin}
in the Appendix shows the \dphi correlation functions for low
($0\leq\nch<10$) and high ($\nch\geq85$) multiplicity events. Each panel shows a different
\pttrig range, and the \pt range of associated particles is $0.5<\ptassoc<5~{\rm GeV}/c$.
The shape of the near-side peak in the short-range correlation function
becomes narrower as \pttrig increases, and there is no significant shape difference
between low and high multiplicity events at the same \pttrig.

Figure~\ref{fig:pp13TeV_v22_pT}
shows the \vtt of long-range \dphi correlations as a function of \pttrig. Here,
the \vtt increases with \pttrig indicating a stronger nonflow effect at higher \pttrig.
The lines represent \vttsub using events from three different multiplicity ranges for the
nonflow subtraction. As with the nonflow results as a function of multiplicity, the
\vttsub with reference events in $10\leq\nch<20$ and $20\leq\nch<30$ are quite consistent,
and the \vttsub values are smaller than 0.001 in $0.5<\pttrig<5~{\rm GeV}/c$.
The \vttsub with events of the lowest multiplicity range ($0\leq\nch<10$) is slightly
different from the other two cases and shows a greater degree of nonclosure at higher \pt using \methodii.

To summarize the closure test in \pp collisions at \sqs~=~13~TeV from \pythia:

\begin{enumerate}
    \item The \dphi two-particle correlation functions in \pythia exhibit minor violations of the assumptions in the nonflow
    subtraction methods examined.
    \item The \pythia resulting \vttsub using \methodb pass the closure test much better than $|v_{2}| < 0.03$ as long as one
    avoids the lowest multiplicity range for the reference selection.
    \item The modest degree of nonclosure with \pythia may be considered as a systematic uncertainty on the final extracted $v_{2}$ results.
\end{enumerate}

\subsection{RHIC \pp Case}

\begin{figure*}[htb]
\includegraphics[width=0.49\linewidth]{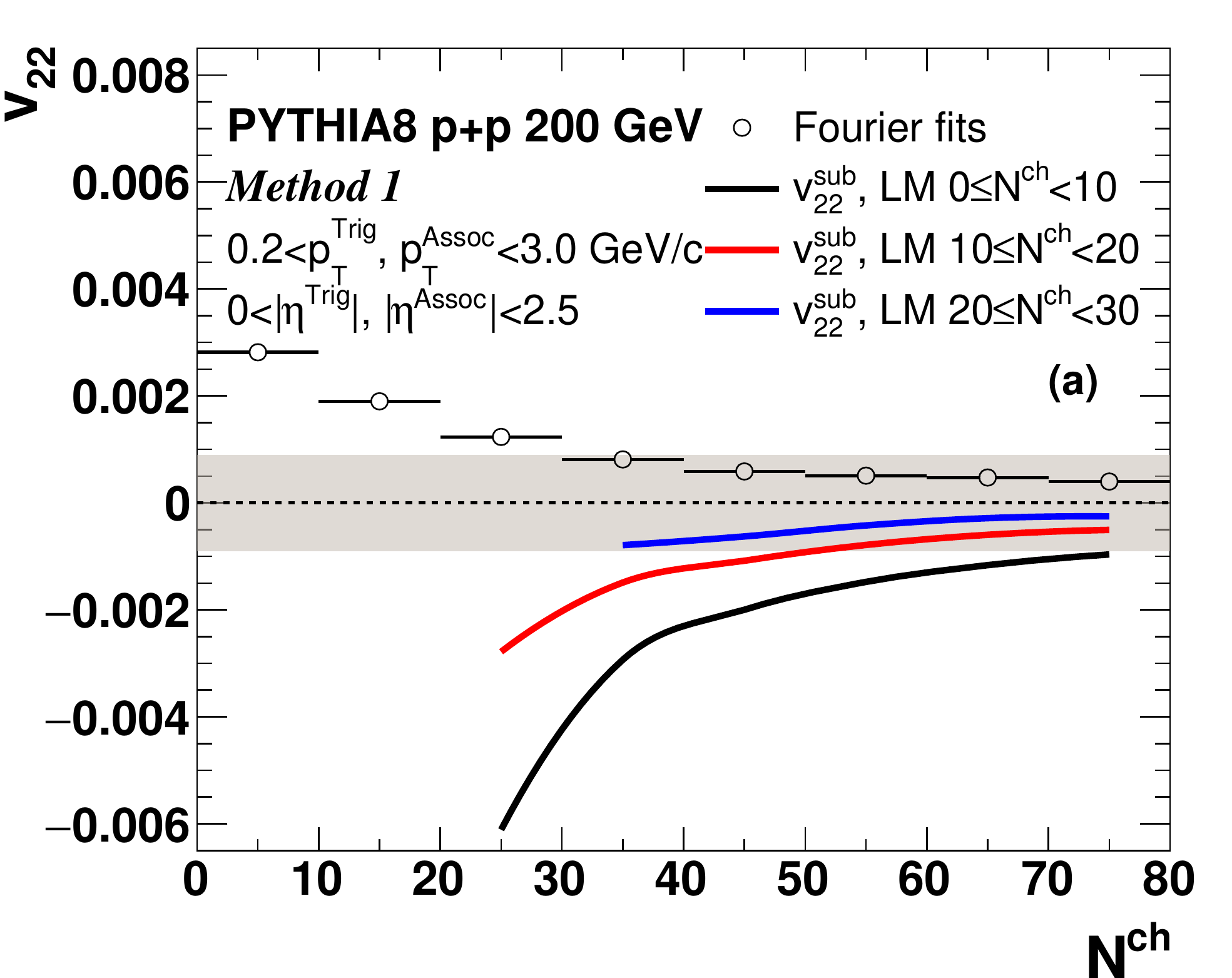}
\includegraphics[width=0.49\linewidth]{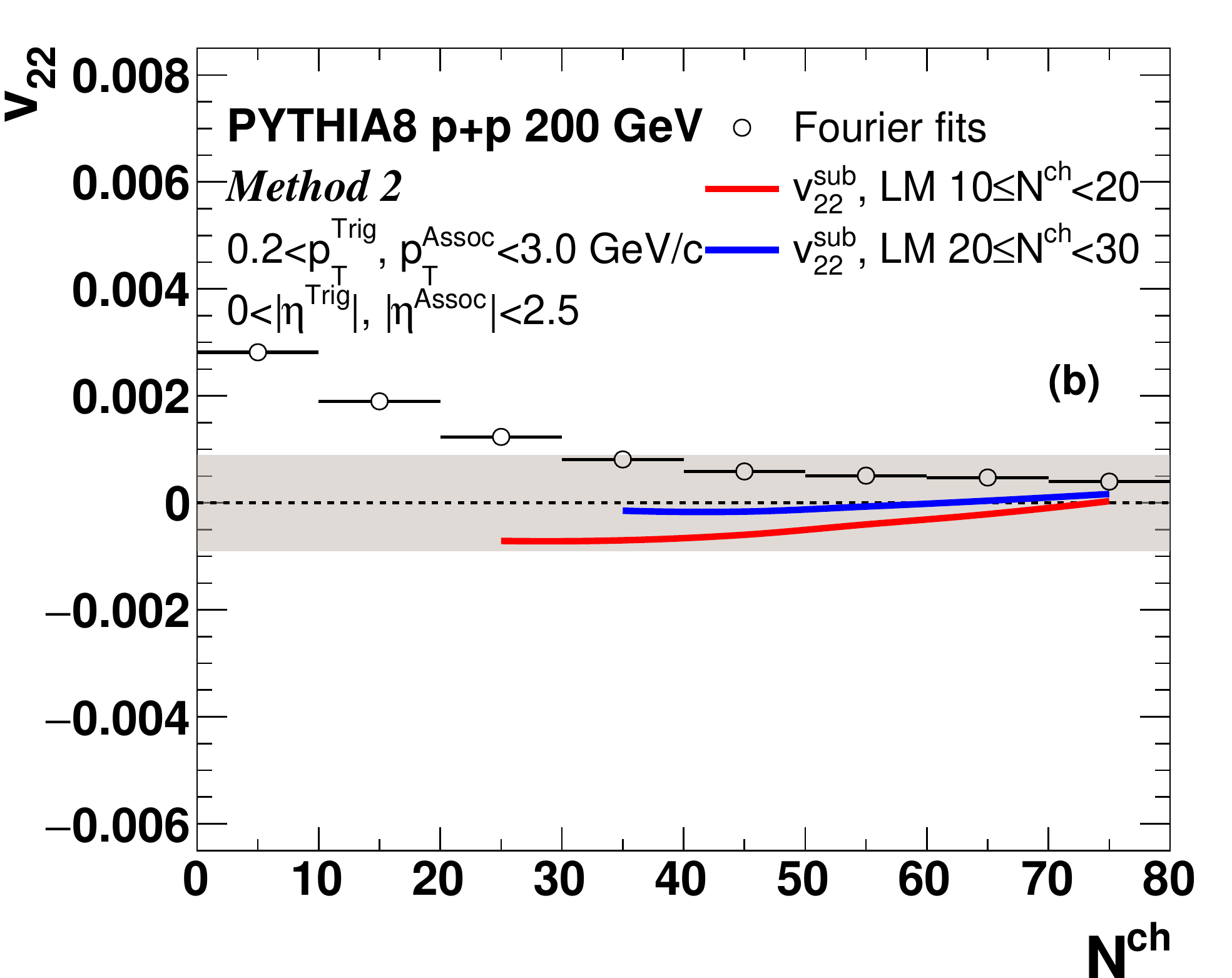}
\caption{\label{fig:pp200GeV_v22_mult}
The second order Fourier coefficient \vtt of long-range ($2<|\deta|<5$) two-particle correlation as a function of charged hadron multiplicity in \pp collisions at \sqs~=~200~GeV from \pythia before and after nonflow subtraction. Multiplicity is defined as the number of charged hadrons in $\pt>0.2~{\rm GeV}/c$ and $|\eta|<2.5$. Gray bands are corresponding to a 3\% $|v_2|$ window.}
\end{figure*}

\begin{figure*}[htb]
\includegraphics[width=0.49\linewidth]{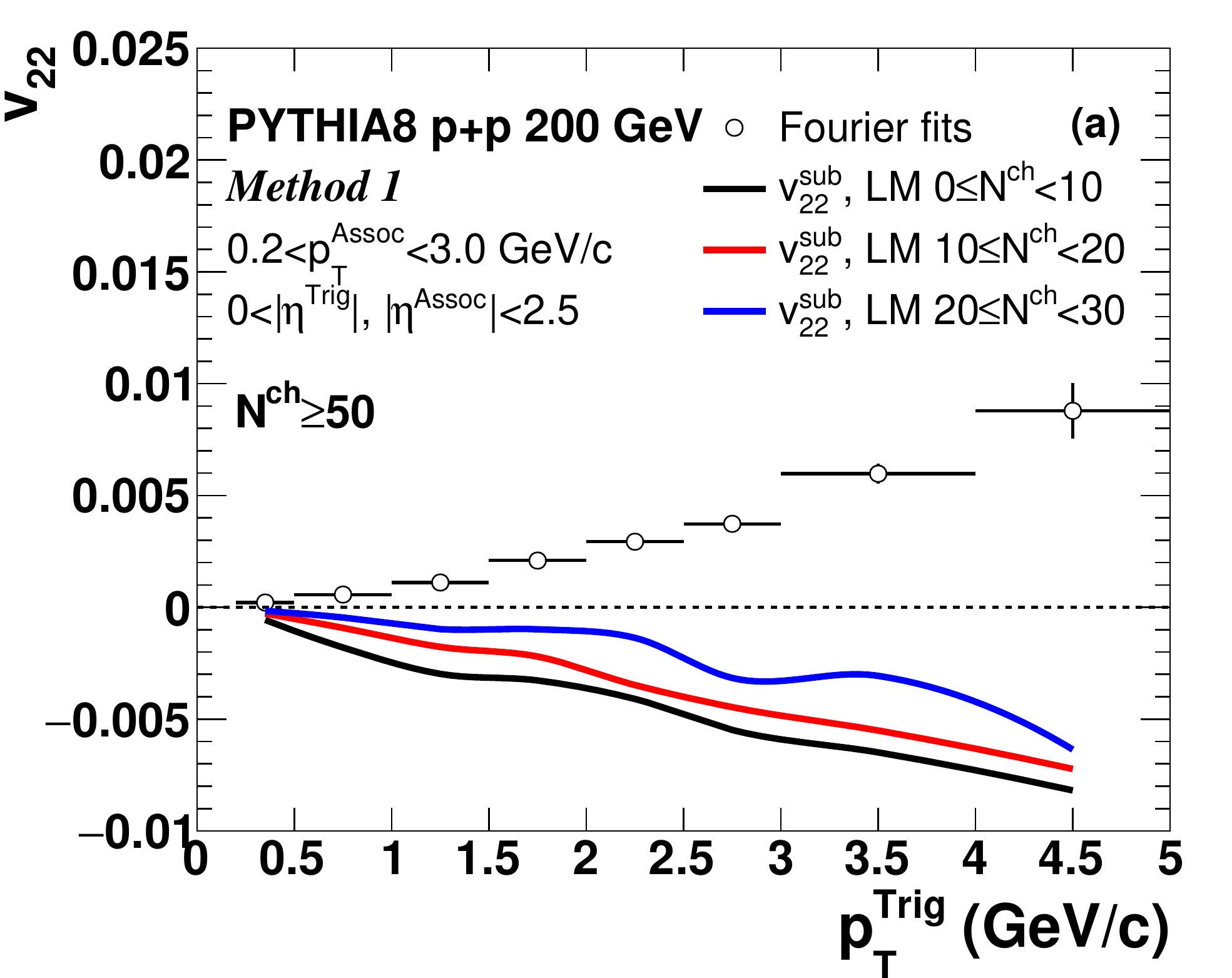}
\includegraphics[width=0.49\linewidth]{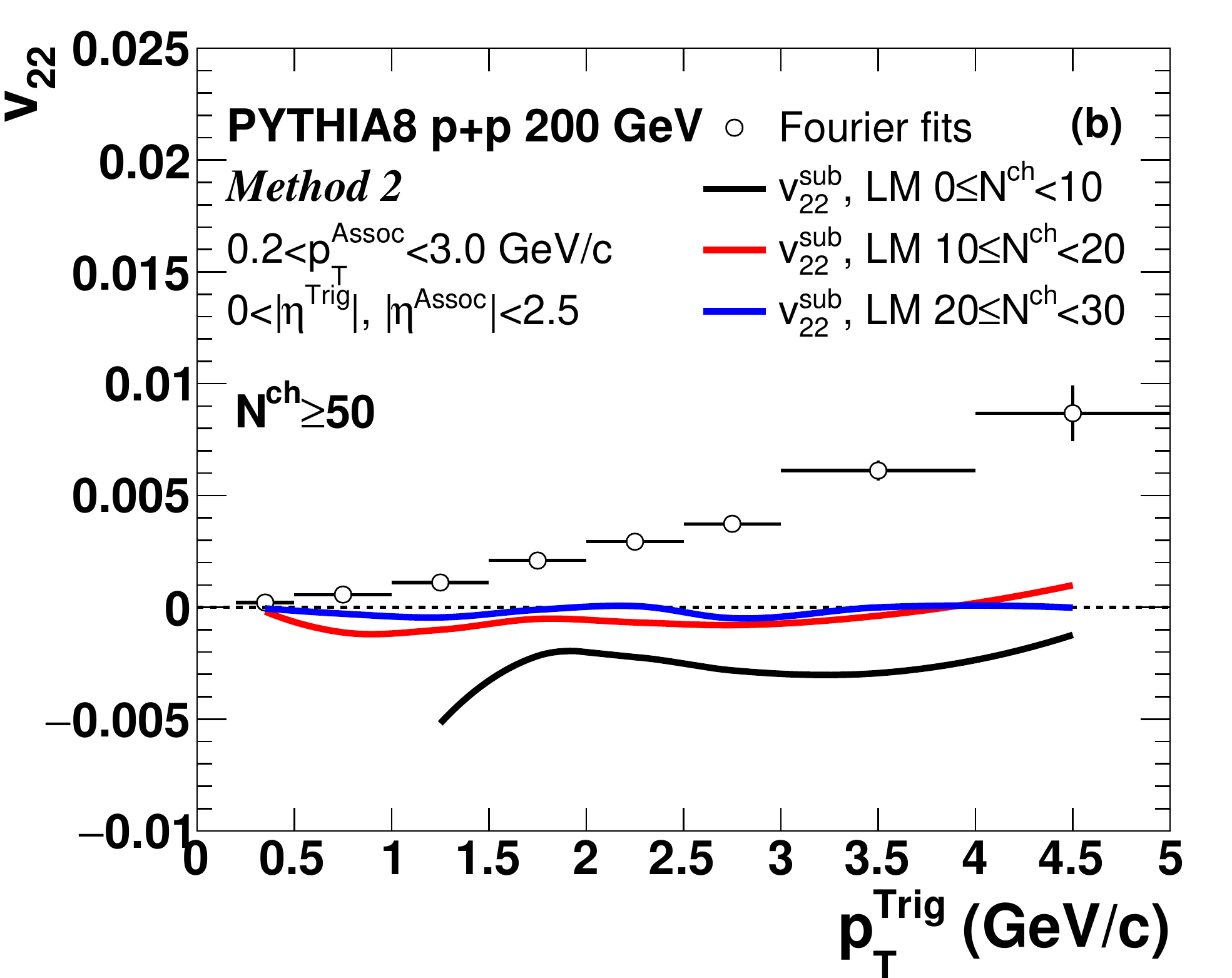}
\caption{\label{fig:pp200GeV_v22_pT}
The second-order Fourier coefficient \vtt of long-range ($2<|\deta|<5$) two-particle correlation as a \pt in \pp collisions of $\nch\geq50$ at \sqs~=~200~GeV from \pythia before and after nonflow subtraction. Multiplicity is defined as the number of charged hadrons in $\pt>0.2~{\rm GeV}/c$ and $|\eta|<2.5$.}
\end{figure*}

There are currently no measurements of extracted flow coefficients in \pp collisions at RHIC.
These are challenging measurements due to multiple collision pileup, much lower
multiplicities compared with LHC collisions, and the more limited phase space acceptance of
the RHIC experiments.    However, we include this case for completeness and to
inform future studies. For this test, we use all charged hadrons within $|\eta|<2.5$
from $10^{9}$ \pythia \pp events (\texttt{SoftQCD:nonDiffractive = on}) at \sqs~=~200~GeV, and charged
hadrons in $\pt>0.2~{\rm GeV}/c$ are used for event multiplicity categorization.
For these studies, we have modeled a very large acceptance similar
to that of the LHC experiments.

Two-particle \dphi and \deta correlation functions are made with the same definition
introduced in Sec.~\ref{sec:lhc_pp}. One-dimensional \dphi correlation functions in
short ($|\deta|<1$) and long ($2<|\deta|<5$) ranges in various multiplicity selections are
presented in Fig.~\ref{fig:pp200GeV_per_trig_yield} in Appendix~\ref{sec:app:cf}.
Charged hadrons in $0.2<\pt<3~{\rm GeV}/c$ and $|\eta|<2.5$ are used for the correlation function.

One obvious difference from the \dphi correlation functions of the LHC case
(Figure~\ref{fig:pp13TeV_per_trig_yield}) is that the shape of short-range \dphi
correlation function significantly changes in the multiplicity range of $0\leq\nch<40$ at
\sqs~=~200~GeV. In the case of the lowest multiplicity bin, ($0\leq\nch<10$), the
per-trigger-yield in the short-range correlation is smaller than that in the long-range
correlation, so \methodii is simply not applicable within this multiplicity bin because it
results in a negative jet yield.

Figure~\ref{fig:pp200GeV_v22_mult} shows the \vtt from Fourier fits to the long-range \dphi
correlation functions plotted versus multiplicity \nch. Additionally plotted as solid lines are
the \vttsub values using three different low multiplicity reference ranges. The gray bands
correspond to a window of $|v_{2}|=0.03$. In results from \methodi shown in the left panel.
The \vttsub clearly depends on the selection of low multiplicity reference, and the
\vttsub greatly deviate from zero in lower multiplicity ranges. This is because
of the dramatic shape variation of the \dphi correlations in events of $0\leq\nch<40$. When using
events in the $20\leq\nch<30$ multiplicity range as reference, the \vttsub is within the
window of $|v_{2}|=0.03$.

In the right panel of Fig.~\ref{fig:pp200GeV_v22_mult}, the nonflow subtraction
results using \methodii
are presented; however, the lowest multiplicity bin ($0\leq\nch<10$) is not included due
to the negative jet yield as discussed earlier. The \vttsub from \methodii using the
remaining reference multiplicity ranges are within the window of $|v_{2}|=0.03$. One possible
reason for these results differing from \methodi is an interplay of the shape variation of
\dphi correlation functions at both short and long range.
In \methodi, the particular multiplicity dependence of the long-range correlation function results in a large deviation of the \vttsub from zero. However in \methodii, which also uses information about the scaling of the jet yield at short range, this effect is partially compensated and the deviation is significantly smaller.
As shown in Fig.~\ref{fig:pp200GeV_per_trig_yield}, the difference of minimum per-trigger yields in short-range and long-range correlations strongly changes with multiplicity, and the additional jet yield in low multiplicity events possibly results in a smaller scaling for \vtt of the low multiplicity events.

The outcome of the nonflow subtraction as a function of \pt in \pp collisions at \sqs~=~200 GeV has been studied
as well. \dphi correlation functions with different \pttrig ranges at short
and long ranges are presented in Fig.~\ref{fig:pp200GeV_per_trig_yield_pTbin}
in the Appendix.
Charged hadrons in $|\eta|<2.5$ are used for the two-particle correlation, and the \pt
range of associated particles is $0.2<\ptassoc<3~{\rm GeV}/c$. Events of $\nch\geq50$
corresponding to the highest 5\% multiplicity are selected as high multiplicity events.
The shape of short-range \dphi correlation functions in the lowest multiplicity range
($0\leq\nch<10$) show a large \pt dependence in $0.2<\pttrig<2~{\rm GeV}/c$, and the
short-range \dphi correlation function of $0.2<\pttrig<0.5~{\rm GeV}/c$ in the high
multiplicity bin also show a quite different shape compared to the \dphi correlation
functions of other \pttrig bins.

The \vtt from Fourier fits as well as \vttsub are presented in Fig.~\ref{fig:pp200GeV_v22_pT},
and the left (right) panel shows the \vttsub with \methodi (\methodii).
In the nonflow subtraction results with \methodi using only long-range \dphi
correlation functions, the \vttsub strongly depends on \pttrig, and the deviation from
zero becomes larger as \pttrig increases. The difference is largest when using events in
the lowest multiplicity range ($0\leq\nch<10$) as the reference. These \vttsub results indicate
that the shape of the long-range \dphi correlation function changes with
multiplicity so that the scaled low multiplicity \dphi correlation does not perfectly
describe those in high multiplicity events. The \vttsub from \methodii in the right
panel shows a smaller deviation from zero than that of \methodi, and the nonflow
subtraction result using the lowest multiplicity reference ($0\leq\nch<10$) is worse
than the two other cases.
It is notable that there is a much smaller shape variation when going from low to high multiplicity events when categorizing event activity at forward rapidity.
This can be considered for future analyses with \pp data at RHIC.

To summarize the closure test in \pp collisions at \sqs~=~200~GeV from \pythia:
\begin{enumerate}
    \item The pseudorapidity coverage $|\eta|<2.5$ was used to compare with the closure test results of the LHC (Sec. \ref{sec:lhc_pp}).
    \item The \dphi two-particle correlation functions exhibit a clear multiplicity and \pttrig dependent shape variation unlike the case in \pp collisions at \sqs~=~13~TeV from \pythia.
    \item This shape variation results in a significant dependence on the low multiplicity reference selection for both methods.
    \item \vttsub from \methodi shows a much larger deviation from zero compared to the results in the LHC case.
    \item \vttsub from \methodii gives a smaller deviation than that of \methodi, but \methodii cannot be applied using the some low multiplicity references and \pttrig ranges due to a negative jet yield.
\end{enumerate}

\subsection{RHIC \pau Case}

\begin{figure*}[htb]
\includegraphics[width=0.49\linewidth]{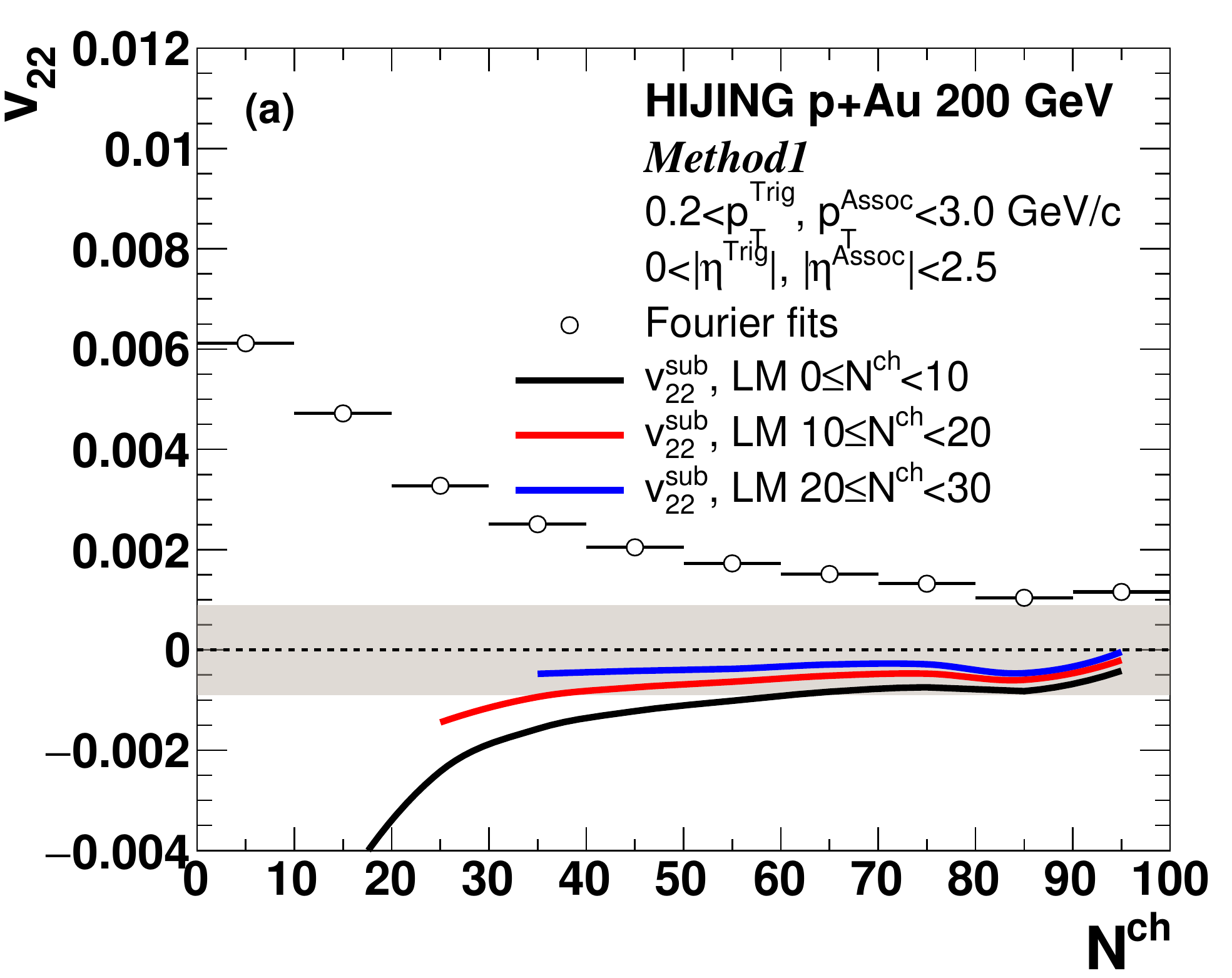}
\includegraphics[width=0.49\linewidth]{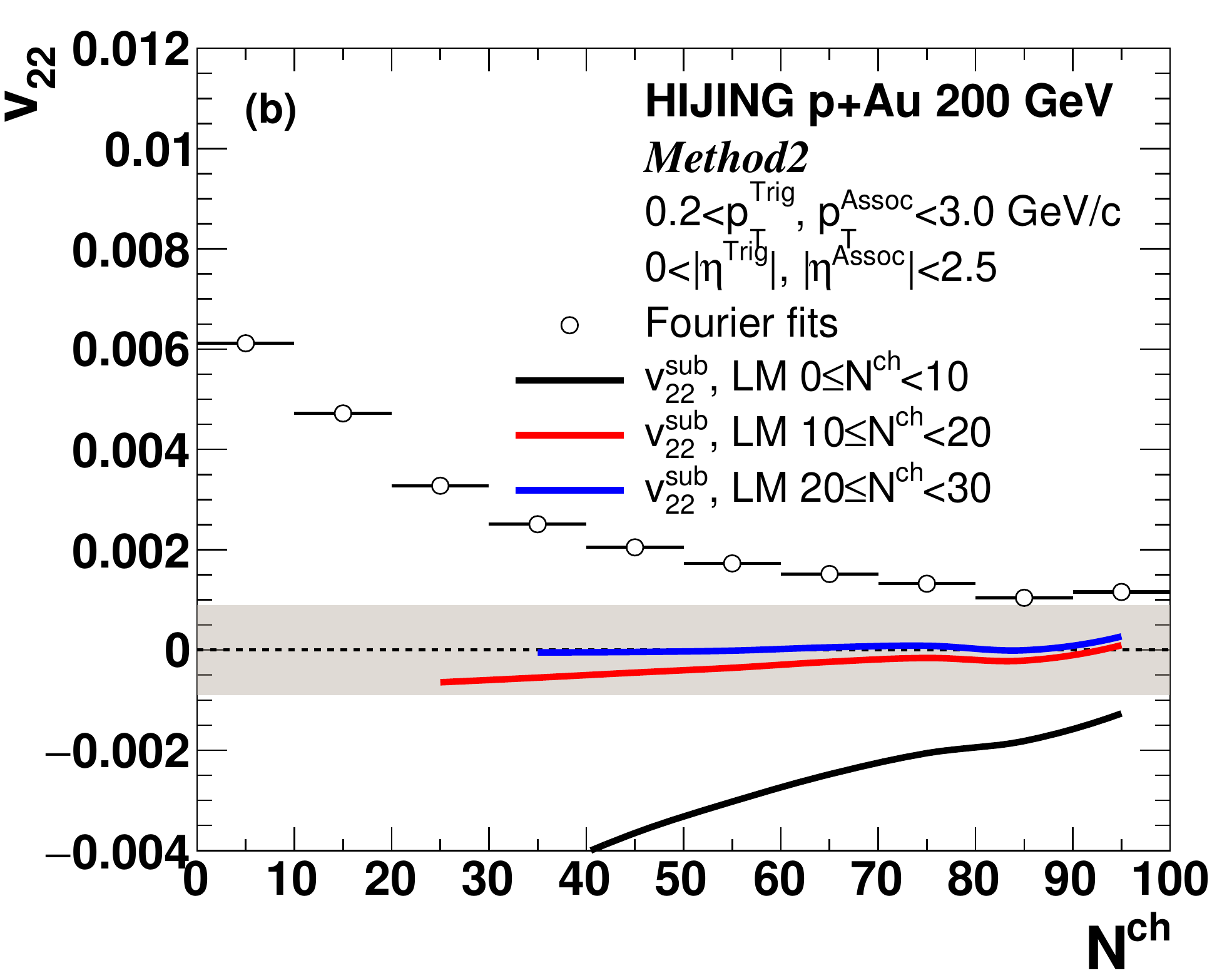}
\caption{\label{fig:pAu200GeV_v22_mult}
The second-order Fourier coefficient \vtt of long-range ($2<|\deta|<5$) two-particle correlation as a function of charged hadron multiplicity in \pau collisions at \sqsn~=~200~GeV from \hijing before and after nonflow subtraction. Multiplicity is defined as the number of charged hadrons in $\pt>0.2~{\rm GeV}/c$ and $|\eta|<2.5$. Gray bands correspond to a 3\% $|v_2|$ window.}
\end{figure*}

\begin{figure*}[htb]
\includegraphics[width=0.49\linewidth]{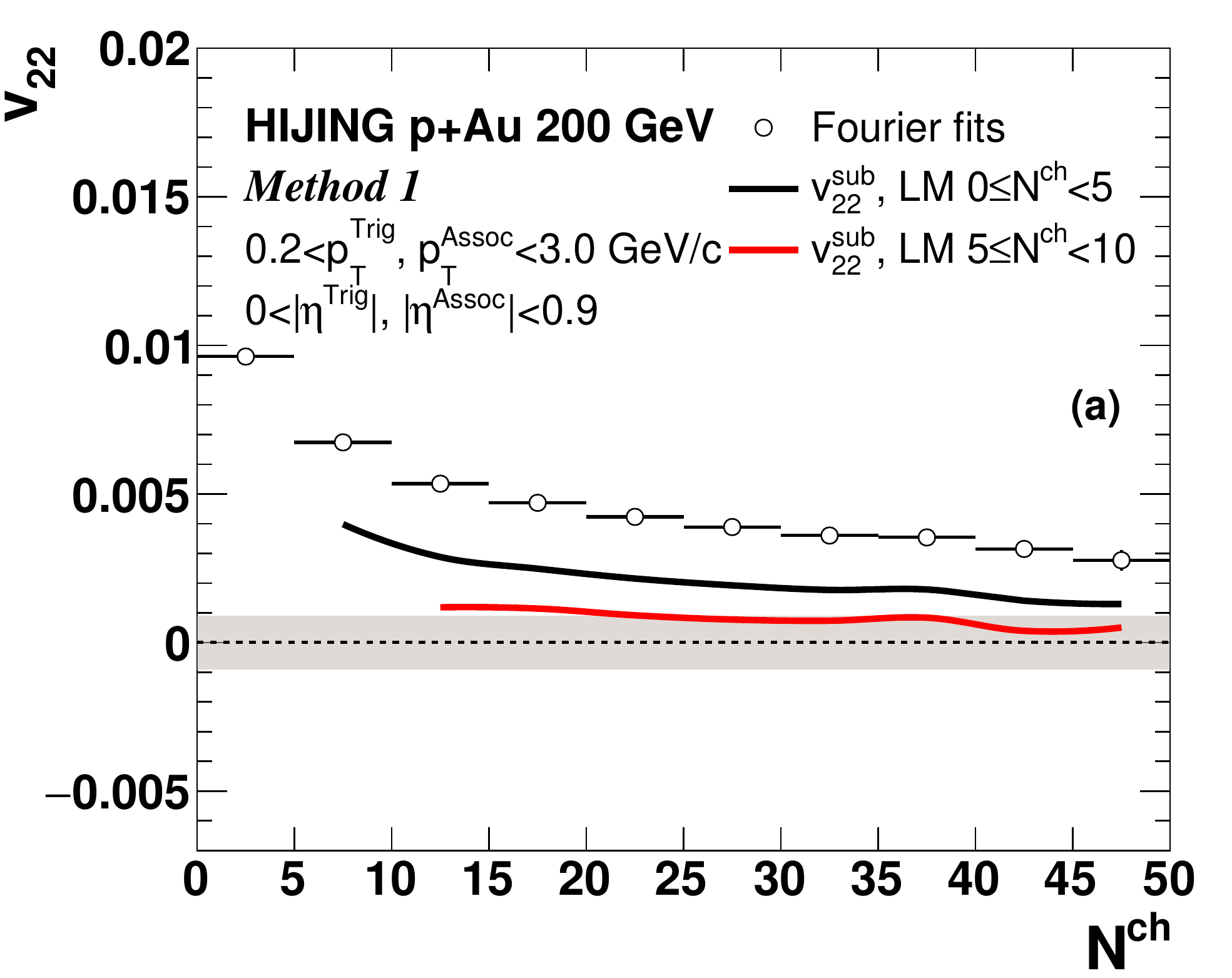}
\includegraphics[width=0.49\linewidth]{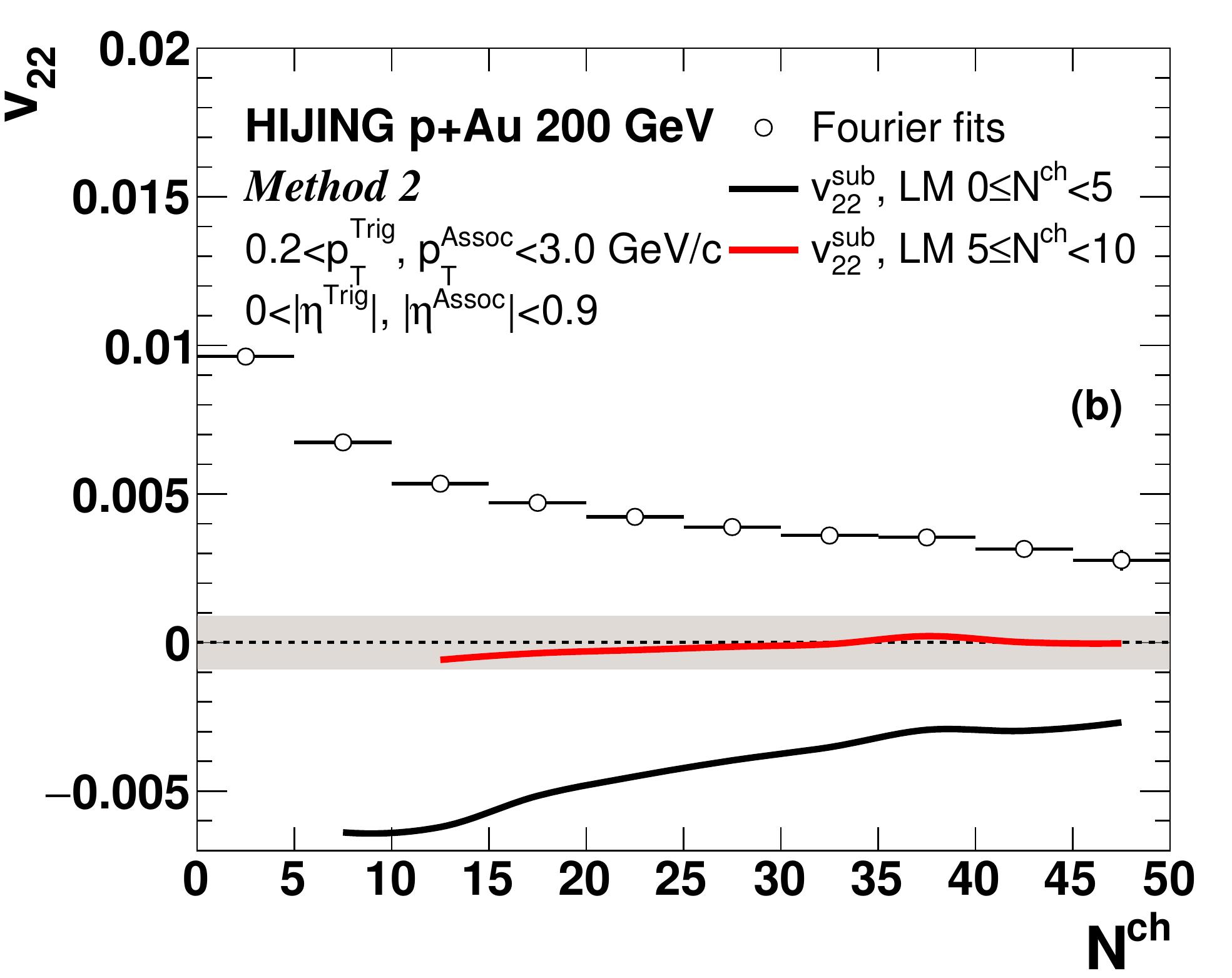}
\caption{\label{fig:pAu200GeV_v22_mult_eta09}
The second-order Fourier coefficient \vtt of long-range ($1<|\deta|<1.8$) two-particle correlation as a function of charged hadron multiplicity in \pau collisions at \sqsn~=~200~GeV from \hijing before and after nonflow subtraction. Multiplicity is defined as the number of charged hadrons in $\pt>0.2~{\rm GeV}/c$ and $|\eta|<0.9$. Gray bands are corresponding to a 3\% $|v_2|$ window.}
\end{figure*}

\begin{figure*}[htb]
\includegraphics[width=0.49\linewidth]{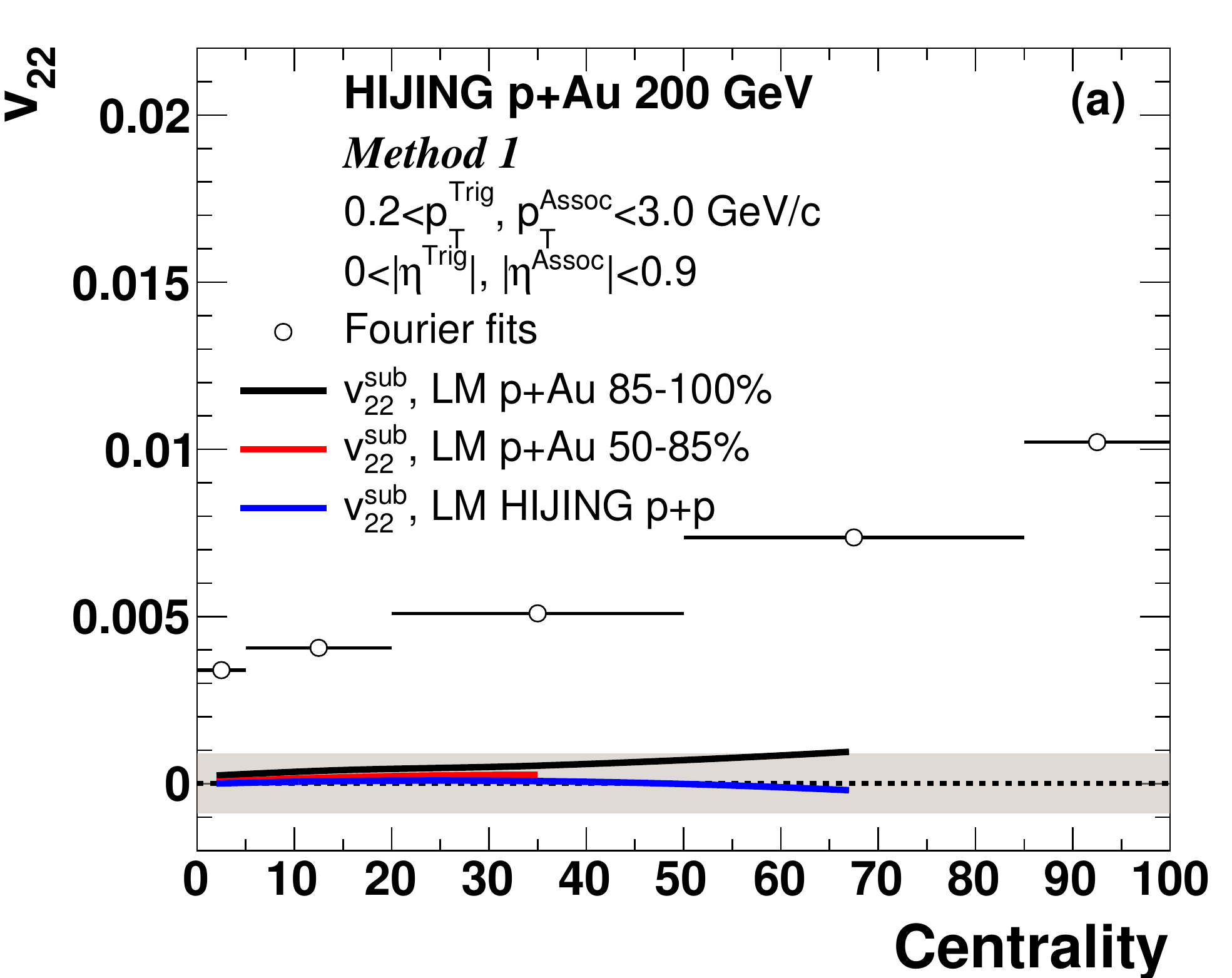}
\includegraphics[width=0.49\linewidth]{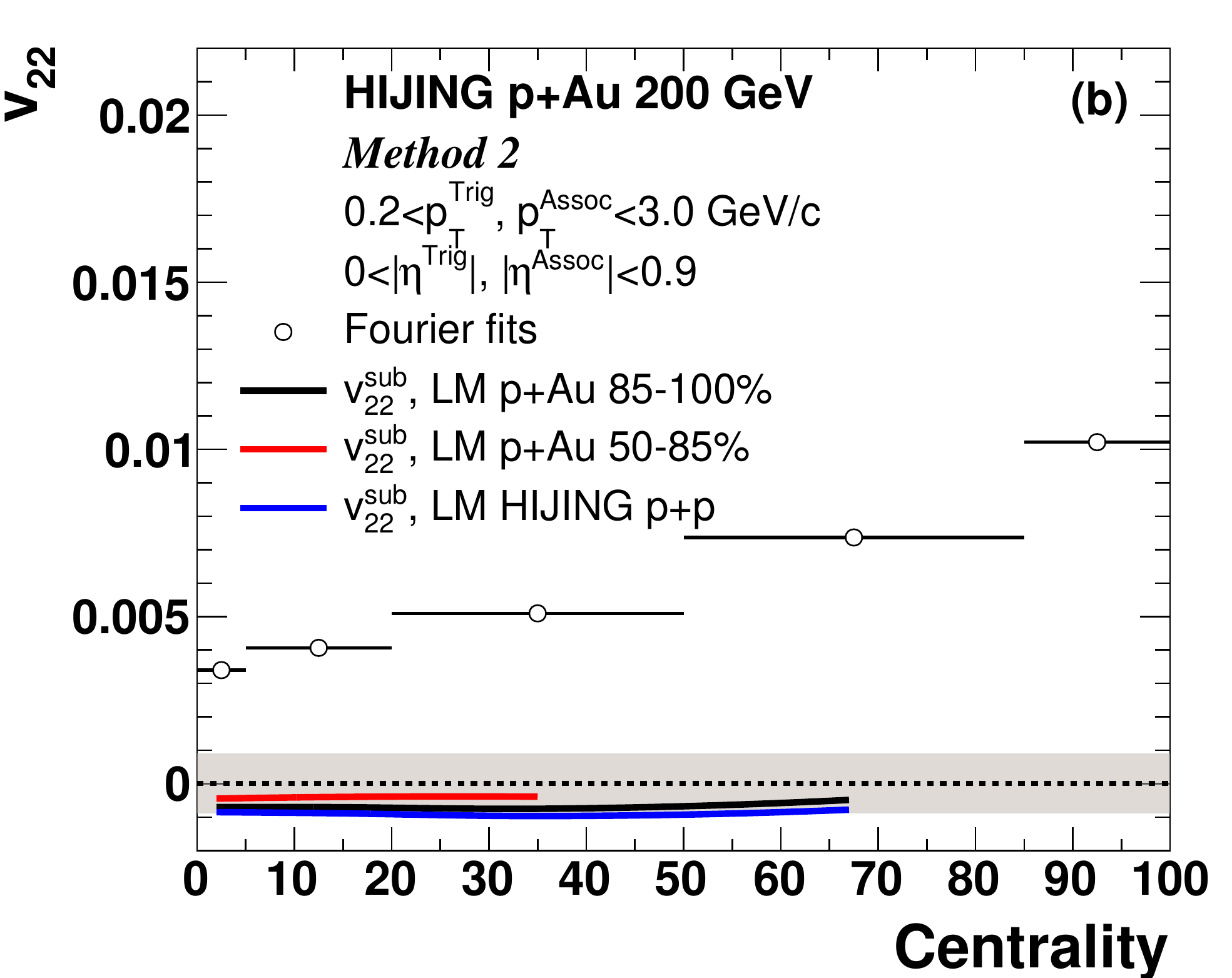}
\caption{\label{fig:pAu200GeV_v22_cent_eta09}
The second-order Fourier coefficient \vtt of long-range ($1<|\deta|<1.8$) two-particle correlation as a function of centrality in \pau collisions at \sqsn~=~200~GeV from \hijing before and after nonflow subtraction. Centrality is defined as the number of charged particles in $-5.0<\eta<-3.3$ (Au-going direction). Gray bands correspond to a 3\% $|v_2|$ window.}
\end{figure*}

\begin{figure*}[htb]
\includegraphics[width=0.49\linewidth]{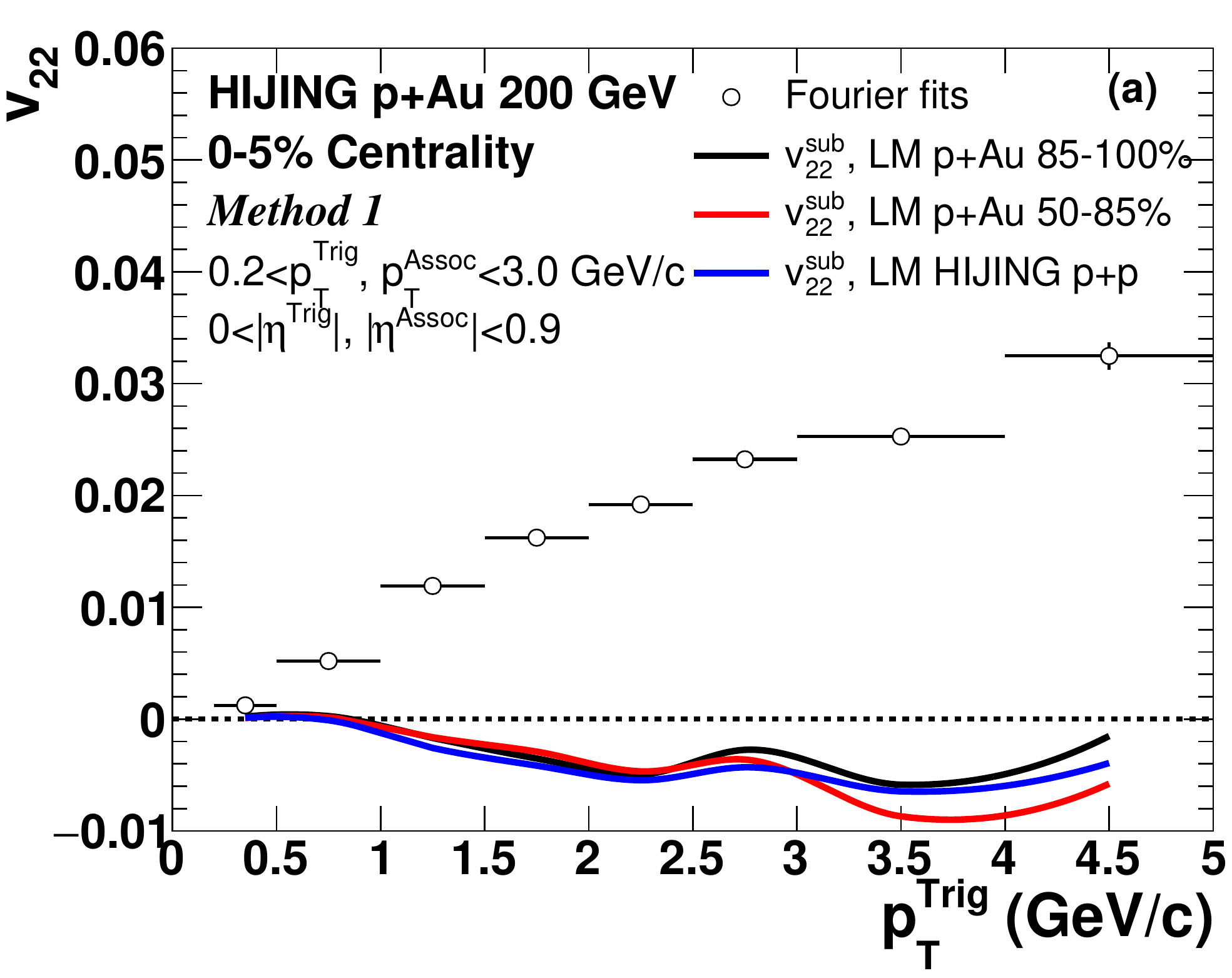}
\includegraphics[width=0.49\linewidth]{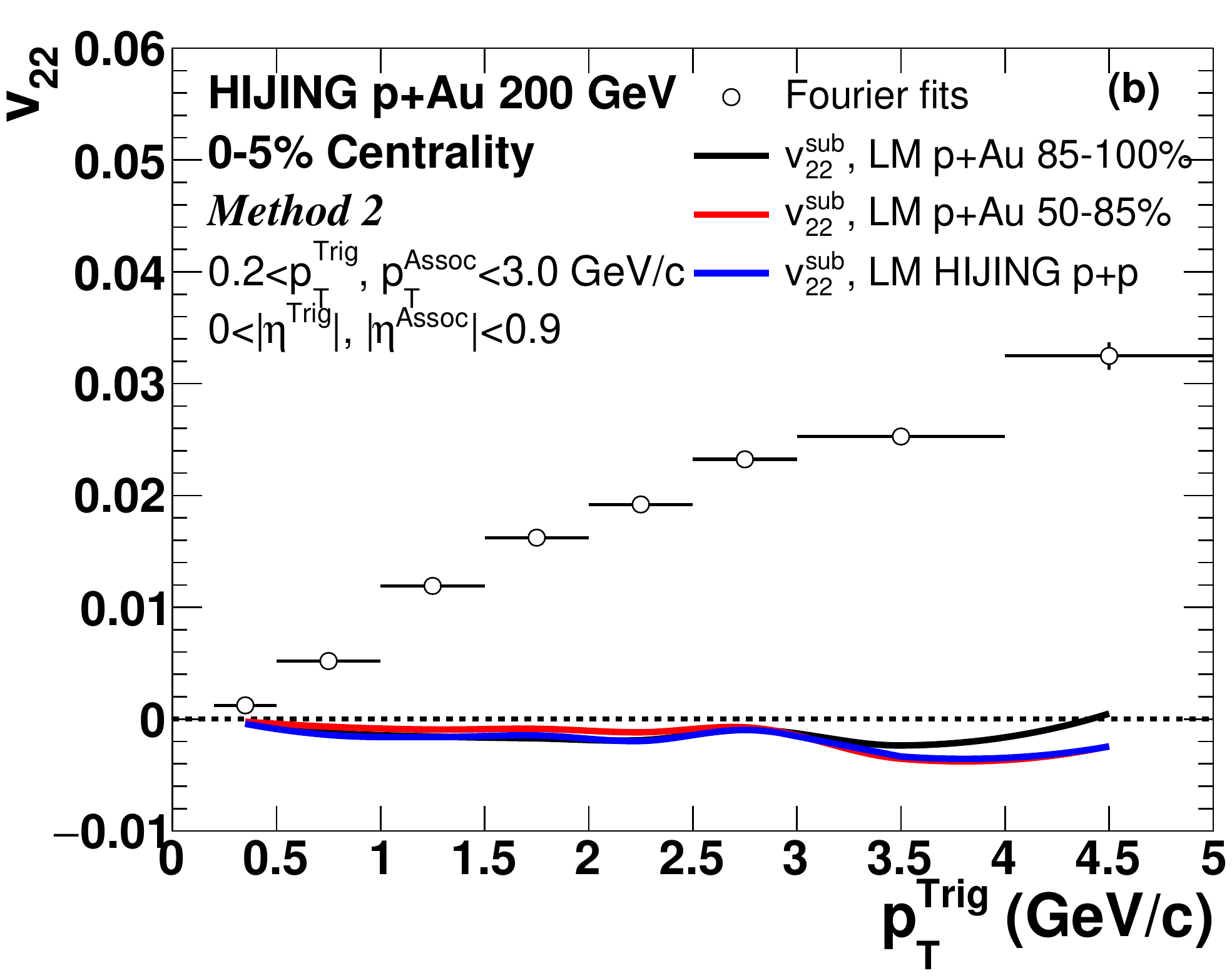}
\caption{\label{fig:pAu200GeV_v22_pt}
The second-order Fourier coefficient \vtt of long-range ($1<|\deta|<1.8$) two-particle correlation as a function of \pt in 0--5\% of \pau collisions at \sqsn~=~200~GeV from \hijing before and after nonflow subtraction. Centrality is defined as the number of charged particles in $-5.0<\eta<-3.3$ (Au-going direction).}
\end{figure*}

\begin{figure*}[htb]
\includegraphics[width=0.49\linewidth]{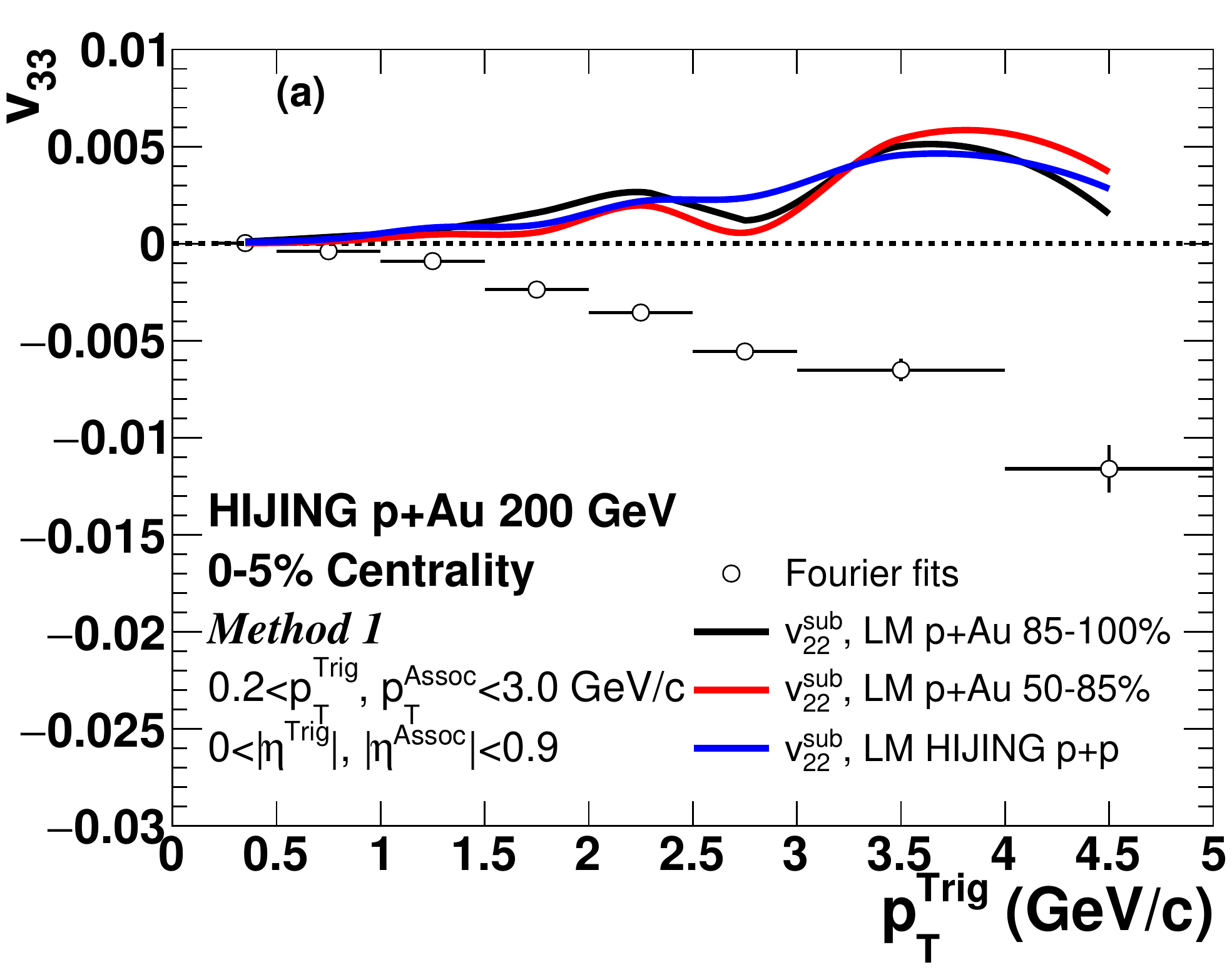}
\includegraphics[width=0.49\linewidth]{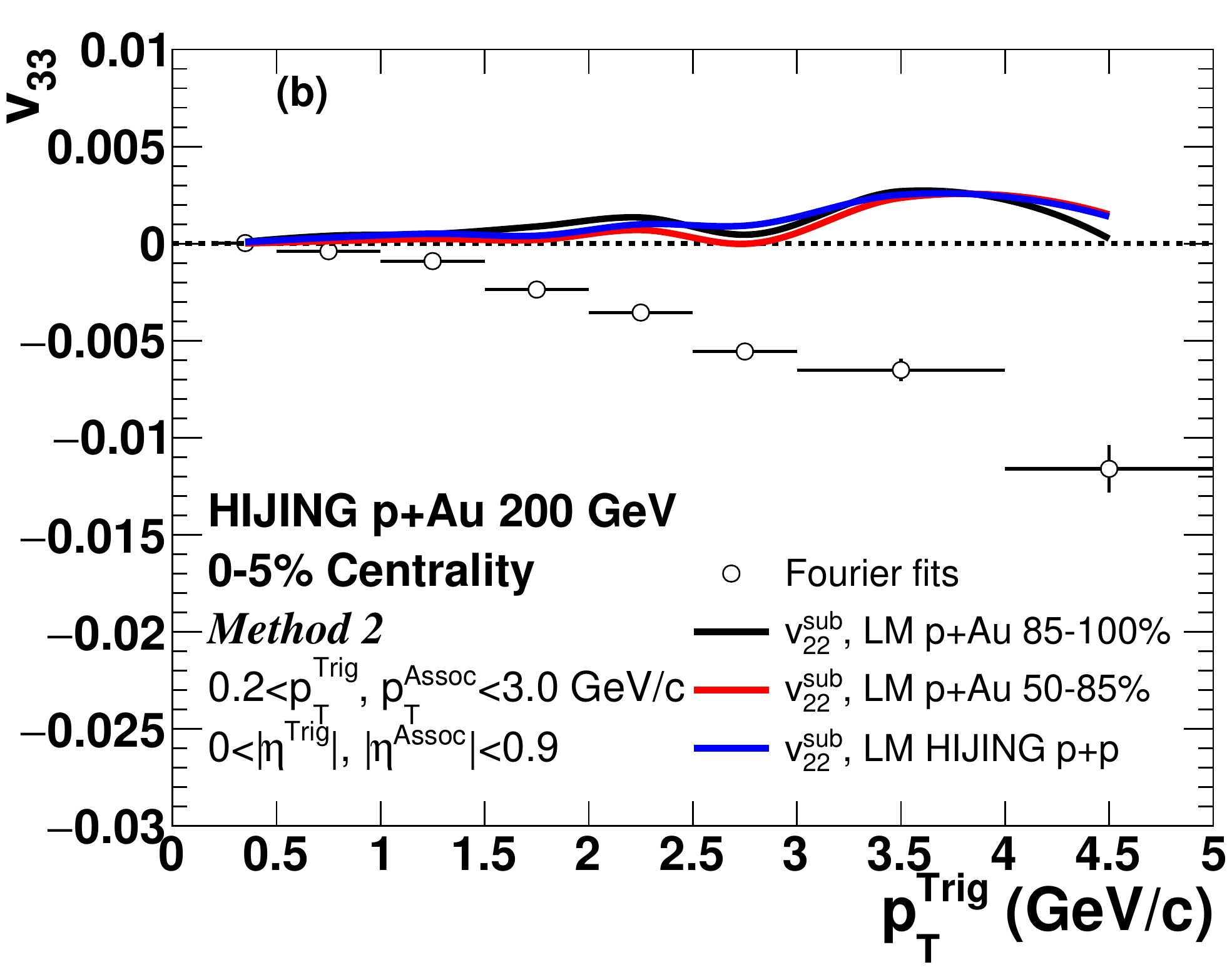}
\caption{\label{fig:pAu200GeV_v33_pt}
The third-order Fourier coefficient \vthth of long-range ($1<|\deta|<1.8$) two-particle correlation as a function of \pt in 0--5\% of \pau collisions at \sqsn~=~200~GeV from \hijing before and after nonflow subtraction. Centrality is defined as the number of charged particles in $-5.0<\eta<-3.3$ (Au-going direction).}
\end{figure*}

There have been numerous extractions of flow coefficients at RHIC in \pau, \dau, and \heau
collisions -- highlighted by the PHENIX publications of elliptic $v_{2}$ and triangular
$v_{3}$ flow in all three systems~\cite{PHENIX:2018lia}. The PHENIX results are shown with no
subtraction of the nonflow and instead with asymmetric systematic uncertainties to
estimate the possible contributions. These correlations have been checked with a
pseudorapidity gap as large as $|\Delta \eta|>2.75$ using the PHENIX central arm tracks
($|\eta|<0.35$) and the Au-going Beam-Beam Counter ($-3.9<\eta<-3.1$).

Recently the STAR experiment has shown preliminary results using tracks in their Time Projection Chamber only, with pseudorapidity $|\eta|<0.9$ and $|\Delta\eta|>1.0$~\cite{Huang:2019rsn}.
The much smaller pseudorapidity gap yields a much larger nonflow contribution with influences on both the near-side $\dphi \approx 0$ and the away-side $\dphi \approx \pi$.
They have employed multiple of the above outlined subtraction techniques to extract preliminary flow coefficients and find smaller $v_{2}$ than the PHENIX results particularly for $\pt>1.5~{\rm GeV}/c$ in high multiplicity \pau and \dau events~\cite{Huang:2019rsn}.

Here we examine the nonflow subtraction in various kinematic ranges and multiplicity
classifications in \pau collisions. $1\times 10^{9}$ \pau and $2\times 10^{9}$ \pp events were generated with
\hijing~\cite{Gyulassy:1994ew}, and charged hadrons were selected for two-particle correlations.
We explore a nonflow subtraction with \pp events from \hijing in
addition to using low multiplicity or peripheral \pau events. First, the case of a wide pseudorapidity coverage ($|\eta|<2.5$) similar to the LHC experiments has been studied to check any
difference with the same kinematic range but in lower collision energy (as was done with
the study in \pythia \pp collisions at \sqs~=~200~GeV).    Then we detail a study modeling the more
limited STAR acceptance.

Figure~\ref{fig:pAu200GeV_per_trig_yield} in the Appendix shows two-particle
\dphi correlation function in short ($|\deta|<1$) and long ($2<|\deta|<5$) range in various
multiplicity bins in \pau collisions at \sqsn~=~200~GeV from \hijing. Charged hadrons in
$0.2<\pt<3~{\rm GeV}/c$ and $|\eta|<2.5$ are used for the correlation functions, and the
multiplicity is defined as the number of charged hadrons in $\pt>0.2~{\rm GeV}/c$ and
$|\eta|<2.5$. Similar to the case of \pythia \pp in \sqs~=~200~GeV, the shape of the two-particle \dphi
correlation function in the lowest multiplicity bin ($0\leq\nch<10$) is quite different
from that in the higher multiplicity ranges.

The \vtt from direct Fourier fits to the two
particle \dphi correlation in long range are presented in
Fig.~\ref{fig:pAu200GeV_v22_mult}, and the lines represent the \vttsub using low
multiplicity events in three different ranges for nonflow subtraction. The \vttsub with
events in the lowest multiplicity range ($0\leq\nch<10$) show the largest deviation from
zero in the results with both methods. The worse closure with \methodii is probably
related to the shape of near-side short-range correlation in low multiplicity events
($0\leq\nch<10$). The \vttsub with events in the two other low multiplicity bins,
$10\leq\nch<20$ and $20\leq\nch<30$, are within the window of $v_{2}=0.03$.

We further explore this closure test using kinematic ranges similar to the STAR
experiment~\cite{Huang:2019rsn}.
Figure~\ref{fig:pAu200GeV_per_trig_yield_eta0_9} in
the Appendix show the two-particle \dphi correlation function in short
($0<|\deta|<0.5$) and long ($1<|\deta|<1.8$) range in various multiplicity bins in \pau
collisions at \sqsn~=~200~GeV from \hijing. Charged hadrons in $0.2<\pt<3~{\rm GeV}/c$ and
$|\eta|<0.9$ are used for the two-particle correlations, and multiplicity is defined as
the number of charged hadrons in $\pt>0.2~{\rm GeV}/c$ and $|\eta|<0.9$. In this narrower
pseudorapidity acceptance, the shape of two-particle \dphi correlation function in the
lowest multiplicity bin ($0\leq\nch<5$) is different from that in higher multiplicity ranges,
as was the case with a wider pseudorapidity acceptance ($|\eta|<2.5$). Another important
thing to point out is that there is a clear peak shape at the near-side ($\dphi\approx 0$) in
long-range from jet correlations which is invisible in \dphi correlation
functions with a larger \deta gap ($2<|\deta|<5$) shown in
Figure~\ref{fig:pAu200GeV_per_trig_yield}.

Figure~\ref{fig:pAu200GeV_v22_mult_eta09} shows
the \vtt from Fourier fits to the two-particle \dphi correlation in long-range
($1<|\deta|<1.8$), and the values are larger than those from the wider \deta gap indicating
a stronger nonflow effect with a smaller \deta gap. The solid lines give \vttsub using two
different low multiplicity bins, $0\leq\nch<5$ and $5\leq\nch<10$. It is interesting that
the \vttsub using the lowest multiplicity bin is significantly different between the two
methods. The positive \vttsub with the \methodi is due to the remaining jet correlation
with a smaller \deta gap in \dphi correlation functions of higher multiplicity bins which
is barely seen in the \dphi correlations of the lowest multiplicity. Therefore, the scaled
correlation function of the low multiplicity bin cannot describe the peak structure on
the near-side, resulting in the positive \vttsub. The results from \methodii show large
negative values, and this is related to the
different shape of short-range \dphi correlation function in the lowest multiplicity bin
introducing a large scaling with jet yields. The \vttsub with the next low multiplicity
bin ($5\leq\nch<10$) show a better closure within the level of $|v_{2}|=0.03$.

Another way to categorize event activity in \pau collisions is to use centrality defined
with charged particle multiplicity in the Au-ion-going (backward) rapidity ($-5.0<\eta<-3.3$
in STAR and $-3.9<\eta<-3.1$ in PHENIX).   Results from the PHENIX and STAR experiments
in small systems are categorized in this manner.  In \pau collisions at \sqsn~=~200~GeV, the
multiplicity correlation between mid and backward rapidity is weak, so the
shape difference of the \dphi correlation functions seen in low multiplicity events
defined at midrapidity may not appear in peripheral events defined at backward rapidity.
Figure~\ref{fig:pAu200GeV_per_trig_yield_eta0_9_centrality} in the Appendix shows two-particle \dphi correlation in short ($|\deta|<0.5$) and long ($1<|\deta|<1.8$)
ranges in \pp and various centrality bins of \pau collisions from \hijing.
Charged hadrons in $0.2<\pt<3~{\rm GeV}/c$ and $|\eta|<0.9$ are used for the correlation
functions, and centrality is defined with charged hadrons in $-5.0<\eta<-3.3$.
As discussed earlier, one thing notably different from the case of multiplicity
categorized at midrapidity shown in Figure~\ref{fig:pAu200GeV_per_trig_yield_eta0_9} is
that the shape of two-particle \dphi correlation function is similar in \pp and all
centrality bins of \pau collisions.
At RHIC, selecting events based on multiplicity in the same kinematic range as the particles selected for correlations introduces a significant undesirable shape variation of correlation functions.

Figure~\ref{fig:pAu200GeV_v22_cent_eta09} shows the \vtt from Fourier fits to the two
particle \dphi correlation at long-range. Here, the nonflow effects
become larger from peripheral events to central events. The lines are \vttsub using events
from \pp and two different peripheral selections (50--85\% and 85--100\%) of \pau events.
Note that it is usually difficult to collect events of 85--100\% centrality of \pau collisions
in experiments due to the difficulty of triggering on events with small multiplicity at forward and backward rapidity,
so the next peripheral bin (50--85\%) is also used for the nonflow subtraction. The \vttsub
from both methods with all three selections of low multiplicity events are within the level
of $|v_{2}|=0.03$ shown as a gray band. This smaller \vttsub with centrality compared to the results using multiplicity at midrapidity is mainly coming
from the similar shape of the \dphi correlation functions from peripheral to central \pau
collisions.

As an additional test in \pau collisions from \hijing, we have checked the \pttrig
dependence in 0--5\% central collisions. Figure~\ref{fig:pAu200GeV_per_trig_yield_eta0_9_pt}
in the Appendix shows \dphi correlations at short ($|\deta|<0.5$)
and long ($1<|\deta|<1.8$) ranges for different \pttrig bins. Charged hadrons in
$|\eta|<0.9$ are used for the two-particle correlation, and the \pt range of associated
particles is $0.2<\ptassoc<3~{\rm GeV}/c$. Centrality is defined with charged hadrons in
$-5.0<\eta<-3.3$ (Au-going direction). The shape of the correlation function in
$0.2<\pttrig<0.5~{\rm GeV}/c$ is different from other \pttrig bins both in \pp and 0--5\% \pau
collisions, but the shapes in \pp and 0--5\% \pau events at the same \pttrig bin look
comparable.

Figure~\ref{fig:pAu200GeV_v22_pt} shows the \vtt from Fourier fit to the long-range \dphi correlations as a function of \pttrig in 0--5\% central \pau
collisions, and the lines represent \vttsub with events in
\pp and two peripheral centrality bins of \pau collisions. Note that the $y$-axis range
is much larger than the previous plots where the \pt-integrated results are shown as a function of multiplicity and centrality, due to a much stronger
nonflow contribution in the higher \pttrig range than that in the \pt-integrated case. \vttsub shows a weak
dependence on the selection of low multiplicity events for both methods. However, the
results from \methodi show a clear \pttrig dependence and large negative values indicating
a significant over-subtraction similar to the case of \pp collisions at \sqs~=~200~GeV shown
in Fig.~\ref{fig:pp200GeV_v22_pT}. \methodii shows a smaller deviation from zero than
\methodi, but the \vttsub values are still larger than 0.001.

Figure~\ref{fig:pAu200GeV_v33_pt} shows the \vthth from the Fourier fit and the subtraction methods.
Unlike the \vtt case, \vthth from the Fourier fit yields negative values.
With the subtraction procedures applied, \vththsub from both methods yields positive values.  Thus
the methods result in an over subtraction, but with a negative sign, hence leading to a strong
\pt-dependent enhancement.

The PHENIX results are measured via multiple detector systems covering a wide
range in pseudorapidity.   However, the midrapidity coverage is limited ($|\eta|<0.35$) and thus the short-range correlations used in \methodii are
not available.   The PHENIX flow measurements generally use the event plane method, but there are comparisons using three sets of two-particle correlations, where
one can then algebraically solve for the anisotropy at midrapidity.    In principle
one can apply \methodi to each of the three sets of two-particle correlations.
We reserve this more detailed study to a future publication.

To summarize the closure test in \pau collisions at \sqsn~=~200~GeV from \hijing:
\begin{enumerate}
    \item There are very substantial jet shape modifications when selecting events based on charged particle multiplicity
    around midrapidity \nch.    These lead to significant distortions in the nonflow subtraction in both \methodb and
    for a larger and smaller acceptance.
    \item Selection of event categories based on forward detectors away from midrapidity, as employed by PHENIX and STAR,
    significantly improve the results of the \hijing closure test.
    \item The \vttsub in integrated \pt as a function of centrality with the two methods have some dependence on the low multiplicity reference selection, but
    are generally within the $|v_{2}| < 0.03$ level in both \methodb.
    \item In contrast the \pt dependent results indicate a significant over-subtraction of nonflow in both \methodb.   The
    over-subtraction is very large in \methodi for $\pt>1~{\rm GeV}/c$.
    \item In case of the \vththsub as a function of \pt, there is an indication of significant bias to increase \vththsub in both \methodb.
\end{enumerate}

\section{Further Tests with \ampt}

\begin{figure*}[htb]
\includegraphics[width=0.49\linewidth]{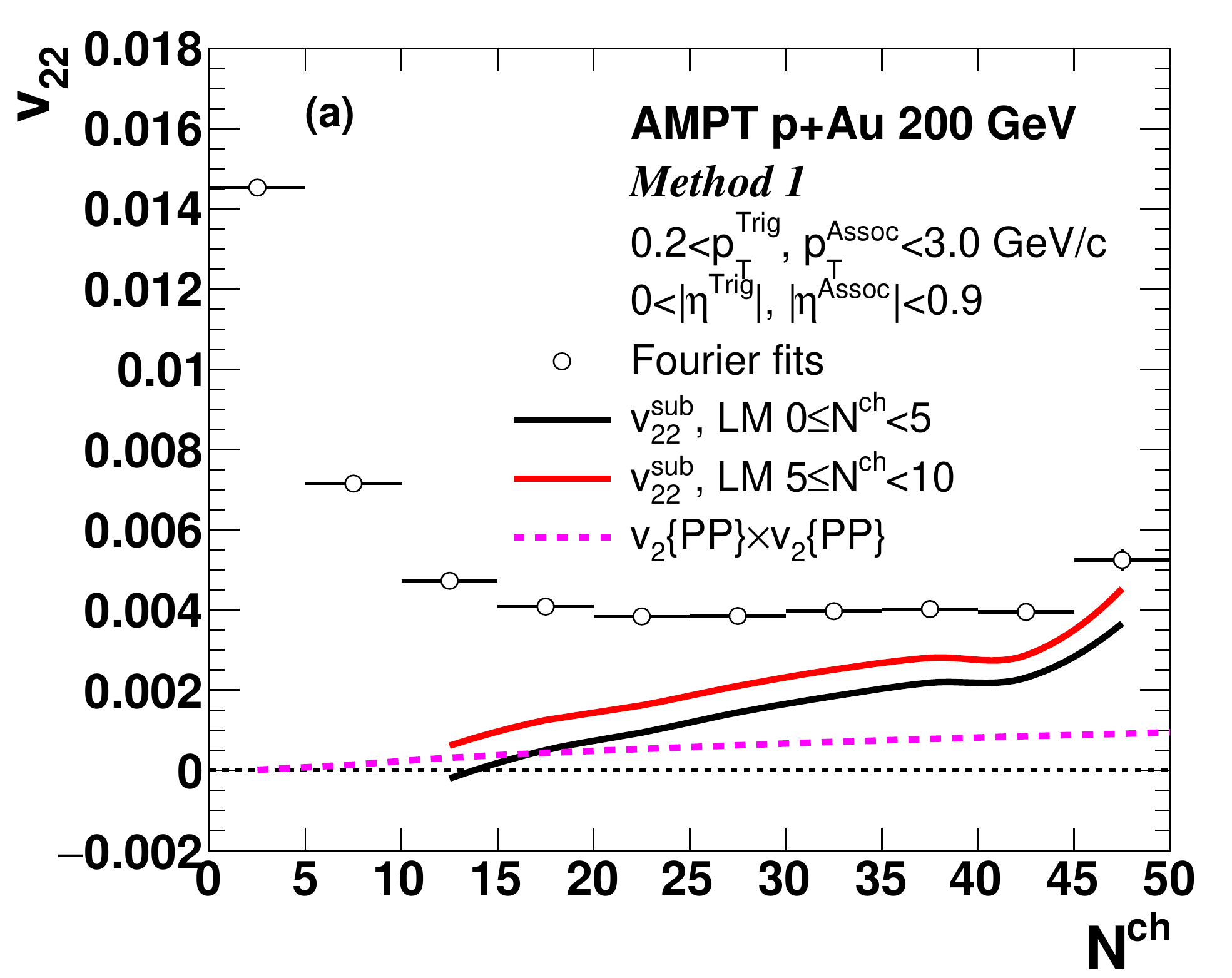}
\includegraphics[width=0.49\linewidth]{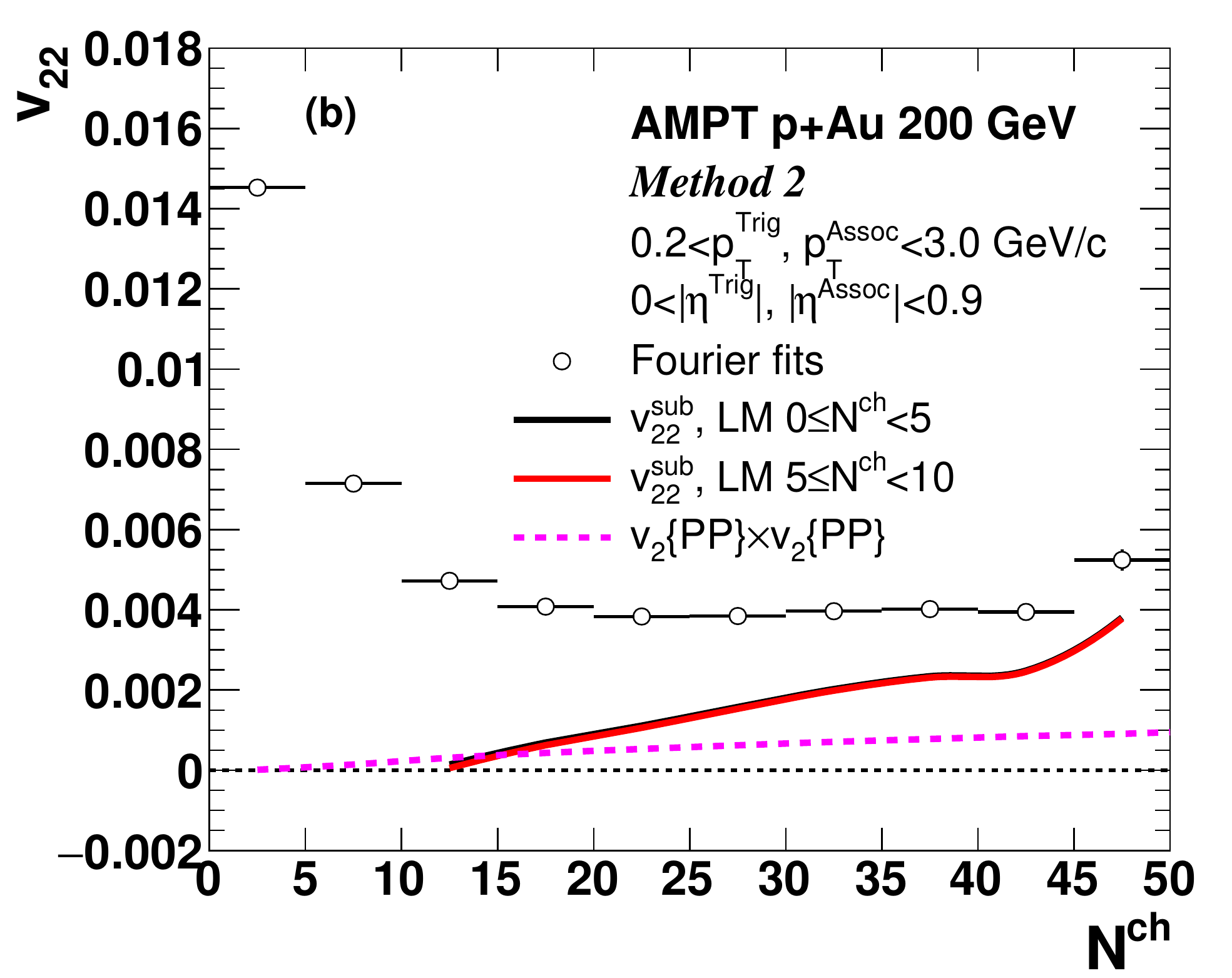}
\caption{\label{fig:ampt_pAu200GeV_v22_mult}
The second-order Fourier coefficient \vtt of long-range ($1<|\deta|<1.8$) two-particle correlation as a function of charged particle multiplicity in \pau collisions at \sqsn~=~200~GeV from \ampt before and after nonflow subtraction. Multiplicity is defined as the number of charged hadrons in $\pt>0.2~{\rm GeV}/c$ and $|\eta|<0.9$.}
\end{figure*}

\begin{figure*}[htb]
\includegraphics[width=0.49\linewidth]{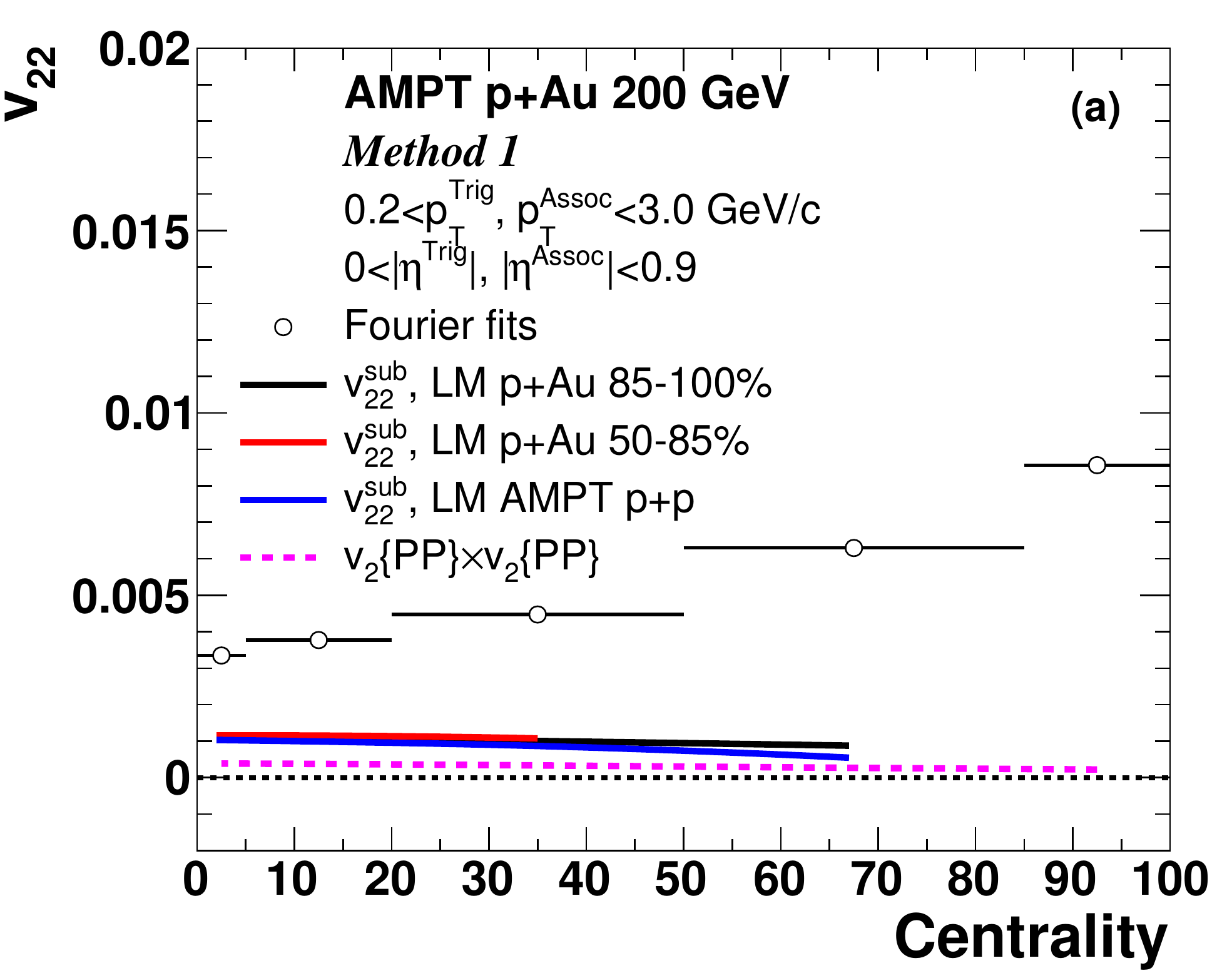}
\includegraphics[width=0.49\linewidth]{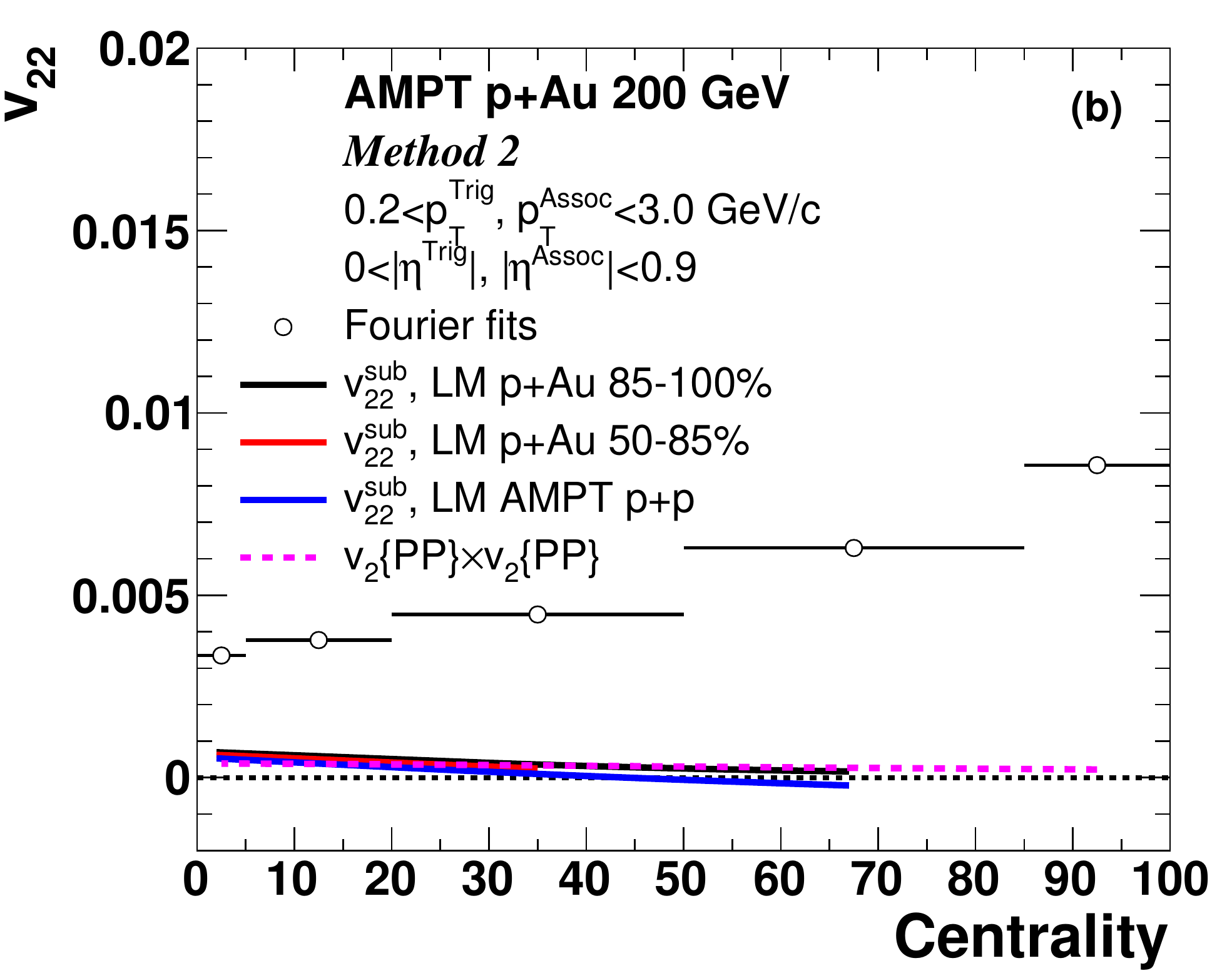}
\caption{\label{fig:ampt_pAu200GeV_v22_cent}
The second-order Fourier coefficient \vtt of long-range ($1<|\deta|<1.8$) two-particle correlation as a function of centrality in \pau collisions at \sqsn~=~200~GeV from \ampt before and after nonflow subtraction. Centrality is defined as the number of charged hadrons in $-5.0<\eta<-3.3$ (Au-going direction).}
\end{figure*}

\begin{figure*}[htb]
\includegraphics[width=0.49\linewidth]{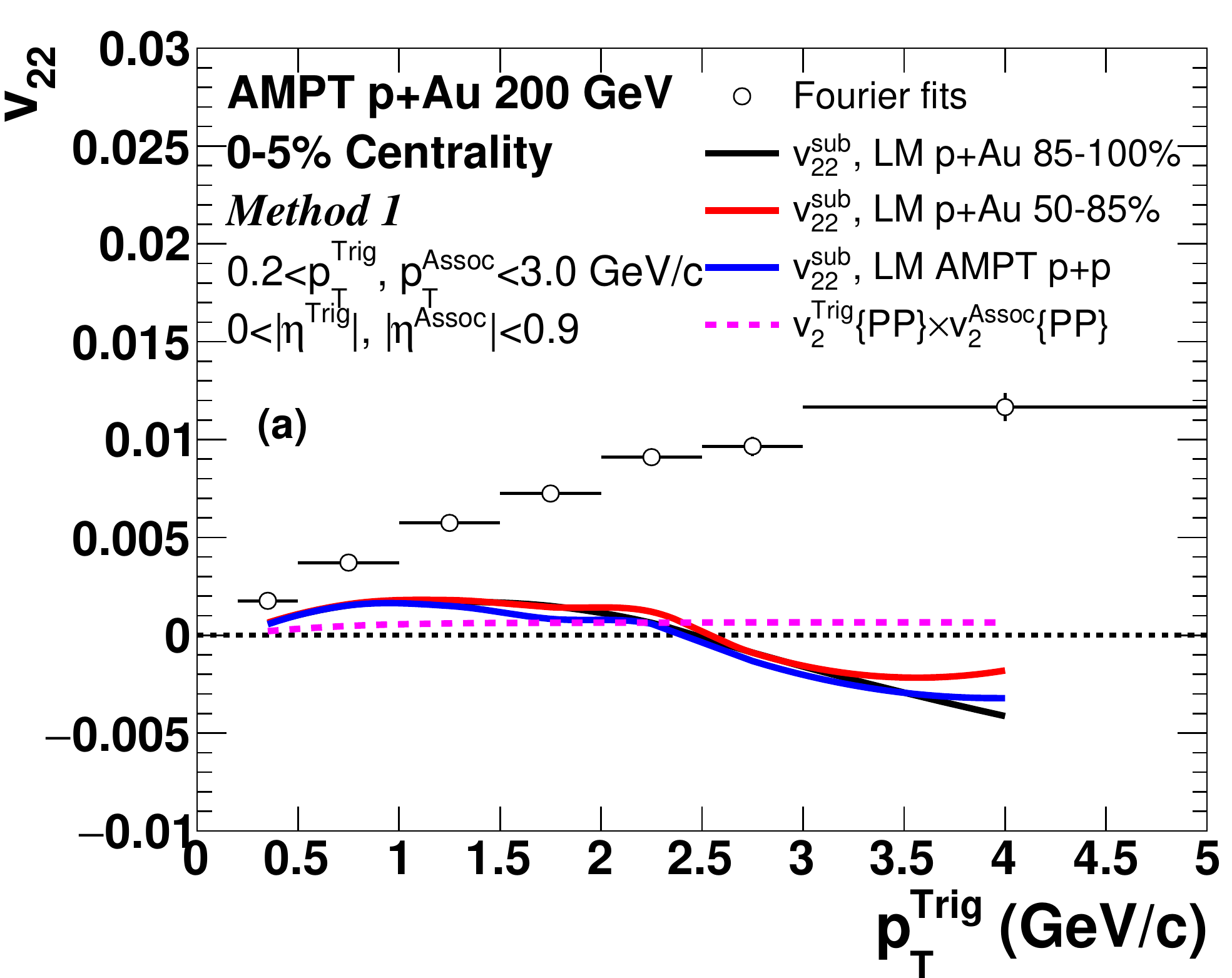}
\includegraphics[width=0.49\linewidth]{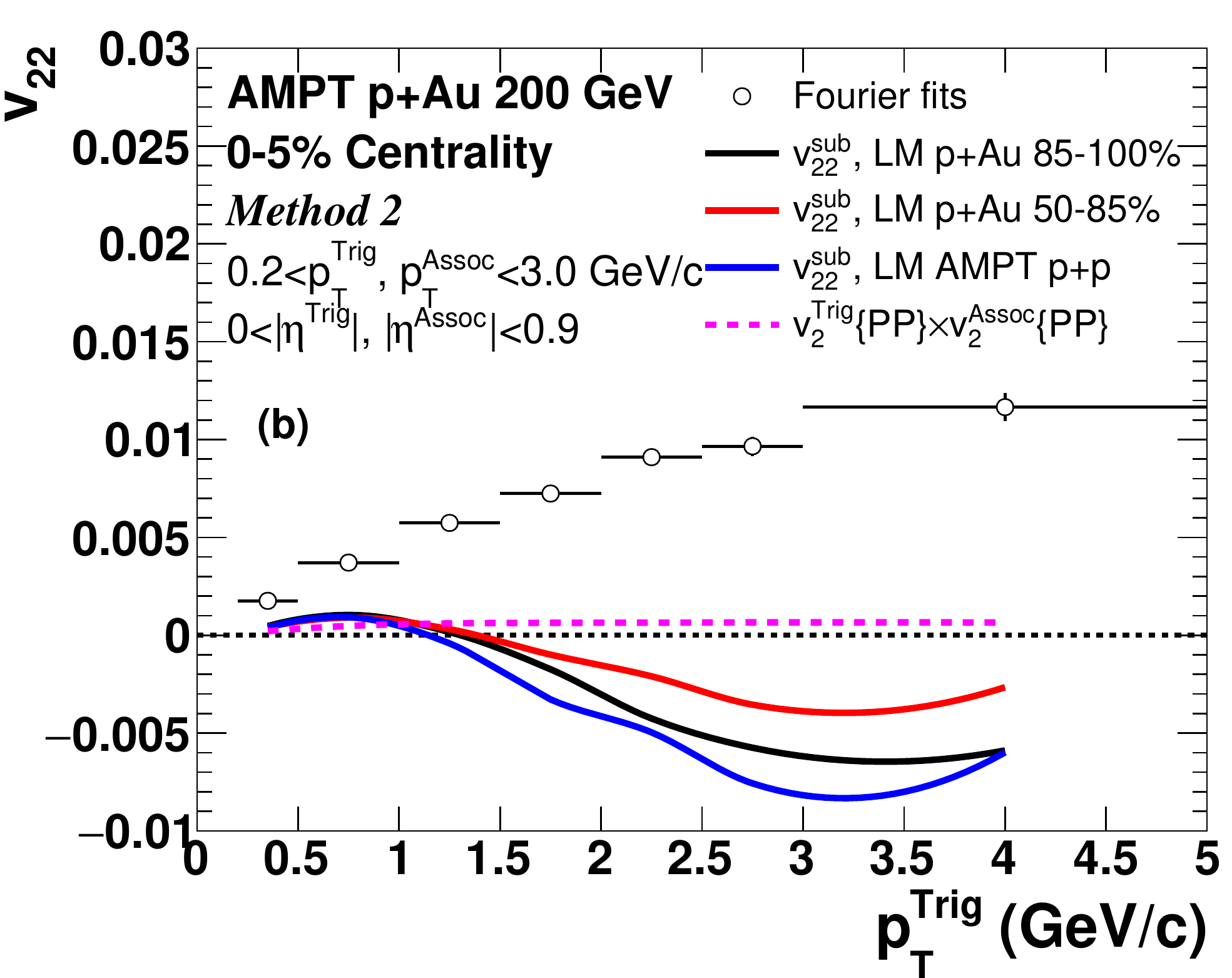}
\caption{\label{fig:ampt_pAu200GeV_v22_pt}
The second-order Fourier coefficient \vtt of long-range ($1<|\deta|<1.8$) two-particle correlation as a function of \pt in 0--5\% of \pau collisions at \sqsn~=~200~GeV from \ampt before and after nonflow subtraction. Centrality is defined as the number of charged particles in $-5.0<\eta<-3.3$ (Au-going direction).}
\end{figure*}

\begin{figure*}[htb]
\includegraphics[width=0.49\linewidth]{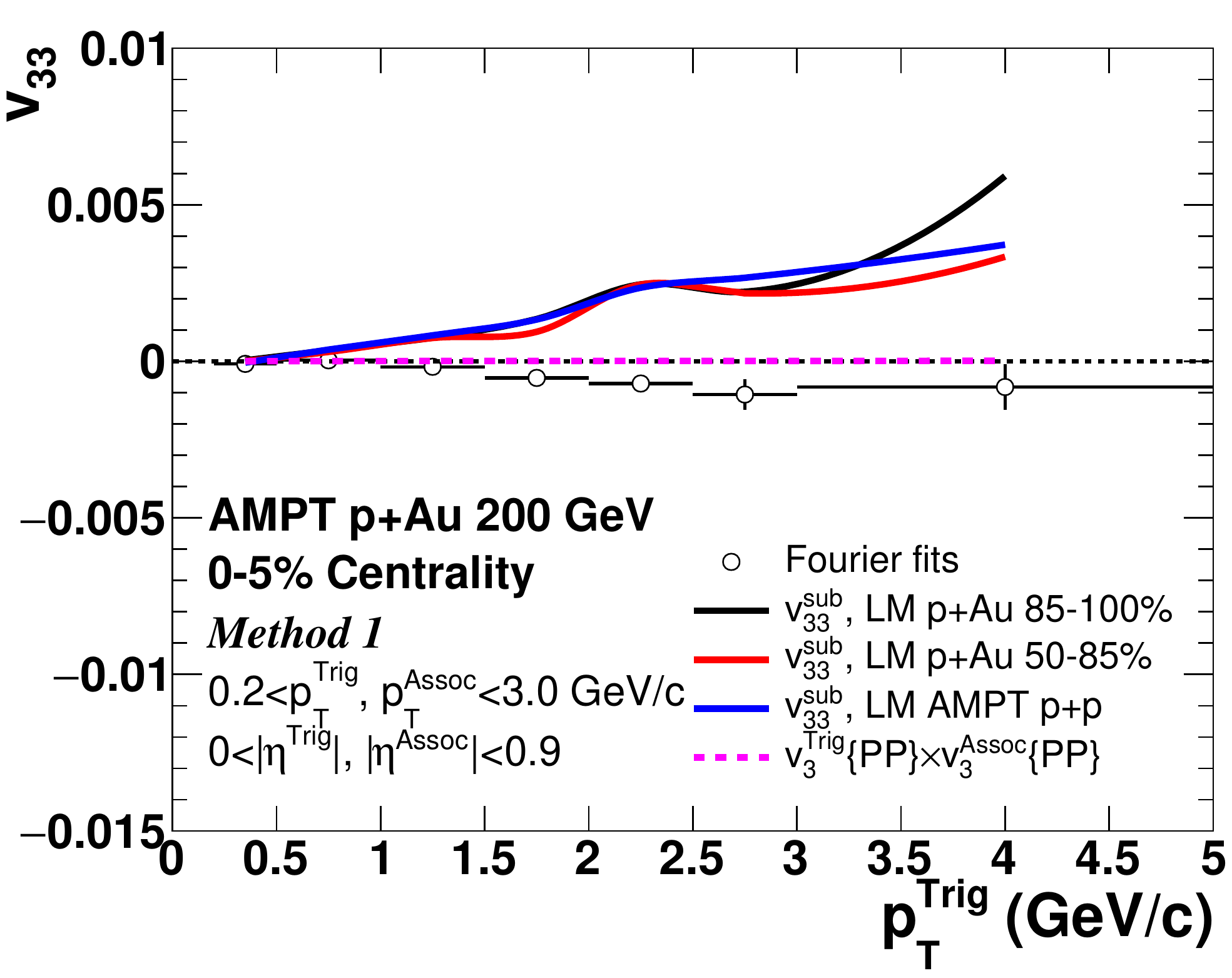}
\includegraphics[width=0.49\linewidth]{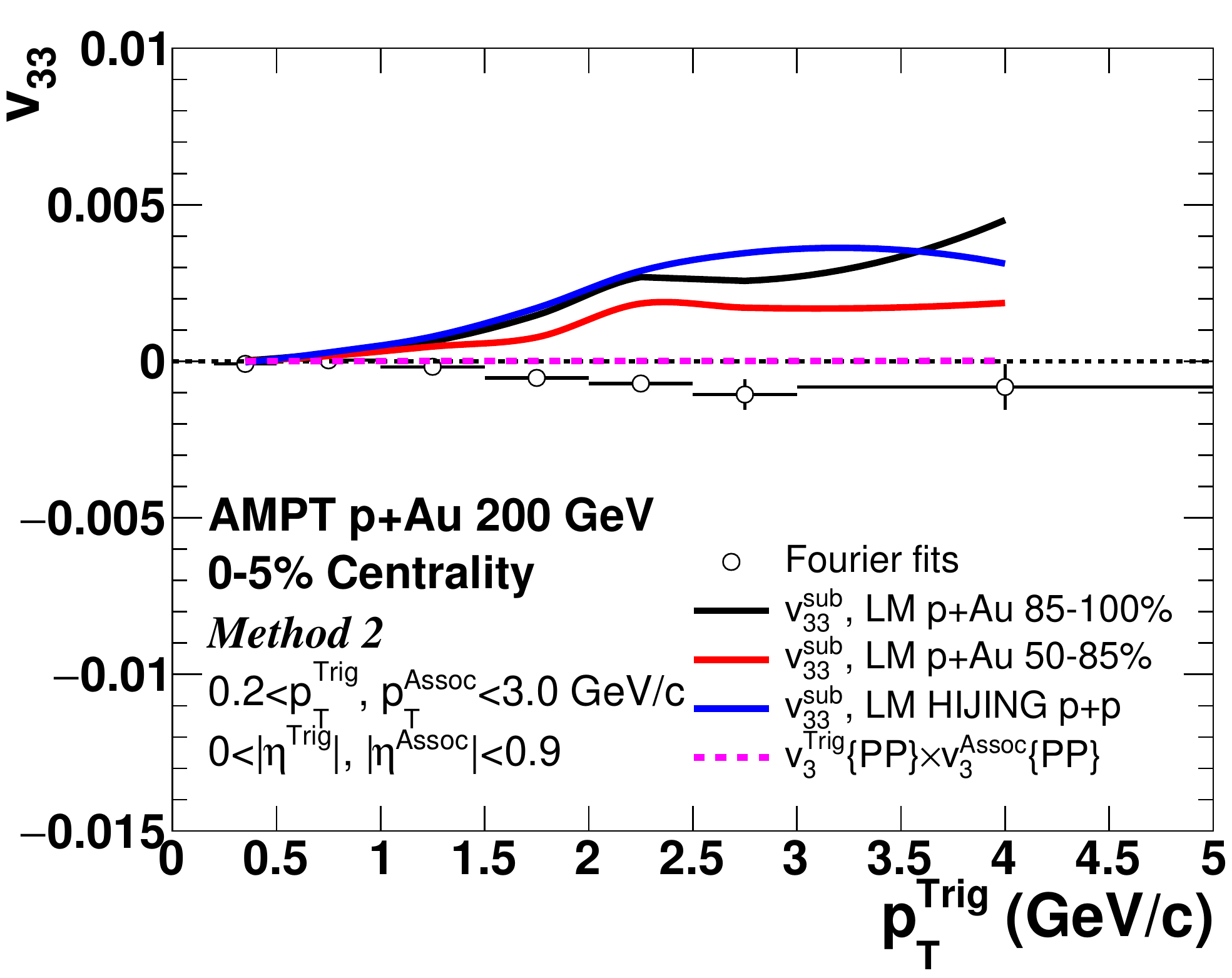}
\caption{\label{fig:ampt_pAu200GeV_v33_pt}
The third-order Fourier coefficient \vthth of long-range ($1<|\deta|<1.8$) two-particle correlation as a function of \pt in 0--5\% of \pau collisions at \sqsn~=~200~GeV from \ampt before and after nonflow subtraction. Centrality is defined as the number of charged particles in $-5.0<\eta<-3.3$ (Au-going direction).}
\end{figure*}

The above tests have a significant limitation in that one is testing procedures to disentangle flow and nonflow on models that have only nonflow.
\ampt~\cite{Lin:2004en} is a model that includes both contributions and can thus be further elucidating.
We follow the previous \ampt studies in Refs.~\cite{Aidala:2017pup} to obtain the ``truth'' flow with respect to the participant plane calculated by using initial-state coordinates of partons resulting from string melting.
More details on the method to calculate the truth flow can be found in Ref.~\cite{Aidala:2017pup}.   We highlight that the finite number of partons used to define the geometry and possible factorization breaking make this a rough
estimate of the truth.
A key item to note is that previous \ampt studies~\cite{Nagle:2018eea} indicate that flow and nonflow do not factorize, in part because partons from jets can rescatter with medium partons.
For both of these reasons, we are not expecting a perfect closure but rather examining possible trends in the subtraction procedure.

For this test, we used \ampt v2.26 with string melting, and a few important parameters are listed in Table~\ref{tab:ampt}.
Note that the updated string parameters, PARJ(41) and PARJ(42), introduced in Ref.~\cite{Lin:2014tya} have been found to be important in recovering the peak structure from jets in near-side short-range correlation functions.

\begin{table}[tbh]
\caption{\label{tab:ampt}
Parameters used in \ampt.}
\begin{ruledtabular}
\begin{tabular}{cc}
 Parameter & Value \\\hline
 ISOFT & 4 \\
 PARJ(41) & 0.55\\
 PARJ(42) & 0.15\\
 Parton screening mass & 6.45d0 (0.75~mb)\\
 $\alpha$ in parton cascade & 0.47d0 \\
\end{tabular}
\end{ruledtabular}
\end{table}

We follow the same procedure applied above with \hijing \pau and \pp events. Two-particle \dphi and
\deta correlations are made from charged hadrons from $6\times 10^{8}$ \pau and $2\times 10^{8}$ \pp \ampt
events. Figure~\ref{fig:ampt_pAu200GeV_per_trig_yield_eta0_9} in the Appendix shows the two-dimensional correlation functions at short ($0<|\deta|<0.5$) and long ($1<|\deta|<1.8$)
range and various multiplicity bins in \pau collisions at \sqsn~=~200~GeV from \ampt.
Charged hadrons in $0.2<\pt<3~{\rm GeV}/c$ and $|\eta|<0.9$ are used for the \dphi
correlation functions, and multiplicity is defined as the number of charged hadrons in
$\pt>0.2~{\rm GeV}/c$ and $|\eta|<0.9$. One thing to point out from the comparison of
correlations between \ampt and \hijing (shown in
Fig.~\ref{fig:pAu200GeV_per_trig_yield_eta0_9}) is that there is a more pronounced peak
structure at the near-side ($\dphi\approx0$) in the long-range correlations which has contributions
from the truth flow in \ampt. Another difference is that the shape of the
short-range \dphi correlation function in the lowest multiplicity bin ($0\leq\nch<5$) is
similar to the shape in higher multiplicity bins in \ampt events so that the nonflow
subtraction results may not depend as much on the selection of the low multiplicity bin.

Figure~\ref{fig:ampt_pAu200GeV_v22_mult} shows the \vtt from Fourier fits to the two
particle \dphi correlation functions in long range as a function of
multiplicity at midrapidity. As with the previous studies, the solid lines are the
\vttsub with two different selections of low multiplicity events, and the dashed line is
\vtt calculated as a product of truth flow ($v_{2}^{2}\{\rm PP\}$) with an assumption of
factorization. The \vttsub with both methods and $v_{2}^{2}\{\rm PP\}$ are close to zero
for low multiplicity events and increase with multiplicity, but the \vttsub is much
larger than $v_{2}^{2}\{\rm PP\}$. One caveat is \vtt, which is a root-mean-square of
$v_{2}$ ($\langle v_{2}^{2} \rangle$), is higher than a product of average $v_{2}$
(${\langle v_{2} \rangle}^2$).  We have estimated the difference via the relation
 $\langle v_{2}^{2} \rangle/ {\langle v_{2}\rangle}^2 \propto \langle \varepsilon_{2}^{2} \rangle/ {\langle \varepsilon_{2}\rangle}^2 $ is $\approx20\%$, which is much smaller than the observed difference.

Another set of two-particle \dphi correlations in short ($|\deta|<0.5$) and long
($1<|\deta|<1.8$) ranges with charged hadrons in the same kinematic range but different
event multiplicity categorization defined at backward rapidity are shown in
Fig.~\ref{fig:ampt_pAu200GeV_per_trig_yield_eta0_9_centrality} in
the Appendix. From \pp collisions to various centrality ranges of \pau
collisions, the shape of the correlations are quite similar except for the near-side peak
structure at long range, possibly due to the truth flow in \ampt.
Figure~\ref{fig:ampt_pAu200GeV_v22_cent} shows the \vtt from Fourier fits in long-range
as a function of centrality. The solid lines are the \vttsub with two different
selections of low multiplicity events, and the dashed lines represent \vtt from the
truth flow, $v_{2}^{2}\{\rm PP\}$. The \vttsub with the \methodi in central \pau
collisions is larger than the $v_{2}^{2}\{\rm PP\}$ like the \vttsub as a function of
multiplicity at midrapidity. The \vttsub using \methodii is smaller than the \vttsub
from \methodi and is consistent with the $v_{2}^{2}\{\rm PP\}$.

An additional test with \ampt events is done for different \pttrig bins in 0--5\%
central \pau collisions. Figure~\ref{fig:ampt_pAu200GeV_per_trig_yield_eta0_9_pt} in
the Appendix shows the two-particle \dphi correlations in short
($|\deta|<0.5$) and long ($1<|\deta|<1.8$) range for various \pttrig bins in \pp and
0--5\% central \pau collisions from \ampt. The correlations are made
from charged hadrons in $|\eta|<0.9$, and the \pt range of associated particles is
$0.2<\ptassoc<3~{\rm GeV}/c$. Centrality is defined with charged hadron multiplicity
in $-5.0<\eta<-3.3$ in the direction of the Au ion. As the case of \hijing \pp and \pau
events, the shape of the two-particle \dphi correlations in the lowest \pttrig bin is
different from that in other \pttrig bins both in \pp and 0--5\% \pau collisions, but the shape
at the same \pttrig bin of \pp events and 0--5\% \pau events looks comparable.

Figure~\ref{fig:ampt_pAu200GeV_v22_pt} shows the \vtt from Fourier fits to the long-range
two-particle \dphi correlation as a function of \pttrig in 0--5\%
central \pau collisions. The solid lines are \vttsub with low
multiplicity selections from \pp and two peripheral \pau centrality bins,
and the dashed line represents the \vtt from the truth flow,
$v_{2}^{\rm Trig}\{{\rm PP}\} \times v_{2}^{\rm Assoc}\{{\rm PP}\}$. Similar to the
case of \hijing \pau events shown in Fig.~\ref{fig:pAu200GeV_v22_pt}, the nonflow
effects increase with \pttrig. The \vttsub with both methods are below zero in
$\pttrig>2.5~{\rm GeV}/c$ using \methodi and $\pttrig>1~{\rm GeV}/c$ using \methodii.
This clearly indicates an over-subtraction in these \pttrig ranges, because the \vtt
from truth flow in \ampt should give positive values even though the truth \vtt from
two-particle correlation may not be exactly same as
$v_{2}^{\rm Trig}\{{\rm PP}\} \times v_{2}^{\rm Assoc}\{{\rm PP}\}$, shown as the dashed
line. The deviation of the \vttsub from zero is larger with \methodii which is opposite
to the closure test results with \hijing events shown in
Fig.~\ref{fig:pAu200GeV_v22_pt}. This is possibly due to the different shape of two
particle \dphi correlation function between \hijing and \ampt both at short and long ranges presented in
Figs.~\ref{fig:pAu200GeV_per_trig_yield_eta0_9_pt}
and~\ref{fig:ampt_pAu200GeV_per_trig_yield_eta0_9_pt}.

Figure~\ref{fig:ampt_pAu200GeV_v33_pt} shows the \vthth from the Fourier fit and the subtraction methods.
Note that the \vthth from the truth flow, $v_{3}^{\rm Trig}\{{\rm PP}\} \times v_{3}^{\rm Assoc}\{{\rm PP}\}$, are non zero but have very small values.
Similar to the \hijing study, \vthth from the Fourier fit yields negative values, but the magnitude is much smaller in \ampt, possibly due to the truth flow contribution in \ampt.
In the subtraction results, \vththsub from both methods yields positive values that are significantly larger than the truth flow.   As was the case in the \hijing study, there is an over-subtraction, again with a
negative sign, and thus a strong \pt-dependent enhancement.

The summary of the \ampt studies is as follows:

\begin{enumerate}
    \item Selection of \pau event categories based on midrapidity multiplicity \nch result in significant undersubtraction
    of nonflow contributions in \ampt.
    \item Selection of \pau event categories based on forward detectors away from midrapidity, as employed by PHENIX and STAR,
    improve the results of the \ampt test.  The results as a function of centrality for the \vttsub values in the two methods
    match the truth extracted \ampt flow qualitatively.
    \item In contrast the \pt dependent results indicate a significant over-subtraction of nonflow in both \methodb.   The
    over-subtraction is very large in \methodii for $\pt>1~{\rm GeV}/c$.
    \item In case of the \vththsub as a function of \pt, there is an indication of significant bias to increase \vththsub in both \methodb.
\end{enumerate}

\section{Conclusions}

We examined closure tests of the nonflow subtraction methods used with two-particle
correlations developed by ATLAS and CMS Collaborations to study the collective behavior
of particle production in small collision systems at the LHC. Monte Carlo event
generators ({\sc pythia} and {\sc hijing}) including no collective flow are use to
quantify a level of closure. In the test for \pp collisions at \sqs~=~13~TeV at the LHC
where a large pseudorapidity gap ($|\deta|>2$) can be applied, the shape of the \dphi
correlation functions in short and long range is relatively stable throughout the event multiplicity
and \pt ranges for which the resulting second-order Fourier coefficients after nonflow
subtraction (\vttsub) are less than 0.001 in these ranges.

We also tested the nonflow
subtraction methods for \pp collisions at \sqs~=~200~GeV, and it is observed that the shape
of the \dphi correlation functions changes significantly with event multiplicity and \pt
ranges. Therefore, the nonflow subtraction results are very sensitive to the selection
of the low multiplicity reference used for the subtraction procedure, indicating that one
should be extremely careful applying the same nonflow subtraction techniques to the
\pp data at RHIC in future analyses.

In the test with \hijing for \pau collisions at \sqsn~=~200~GeV,
the large sensitivity to the low multiplicity event selection that was observed in
\pp collisions at \sqs~=~200~GeV is still apparent when event activity is categorized
at the same rapidity range as that of the two-particle correlations. The test with a
smaller \deta gap ($1<|\deta|<1.8$) shows a significant difference between the methods,
because it starts to be affected by short-range jet correlation contributions. Although the event
activity categorization at backward rapidity helps to reduce the dependence of the low
multiplicity event selection on the results of integrated \pt ($0.2<\pt<3~{\rm GeV}/c$),
the test results as a function of \pt show a clear over-subtraction in
$\pt>1~{\rm GeV}/c$ with both methods for $v_2$ and $v_3$.

We extended this study with \ampt, which includes both flow and nonflow effects, and
the nonflow subtraction results were compared to the truth flow with respect to the
participant plane. The \vttsub and \vththsub are inconsistent with the truth flow in \ampt,
and the \pt dependent results show negative values at higher \pt with both methods.
Thus in \pau collisions, both \hijing and \ampt studies indicate that both \methodb result
in an over-subtraction of nonflow and thus an underestimate of real elliptic flow and overestimate of real triangular flow.

\section*{Acknowledgments}
We acknowledge Jiangyong Jia and Wei Li for useful discussions and a careful reading of the manuscript.
We acknowledge Julia Velkovska for useful discussions and the suggestion to extend this study to
third order flow coefficients.
SHL, KKH, JLN acknowledge support from the U.S. Department of Energy, Office of Science, Office of Nuclear Physics under Contract No. DE-FG02-00ER41152.
QH and DVP acknowledge support from the U.S. Department of Energy, Office of Science, Office of Nuclear
Physics under Contract No. DE-SC0018117.

\appendix

\section{Two-particle \dphi correlation functions}
\label{sec:app:cf}

\begin{figure*}[htb]
\includegraphics[width=1.0\linewidth]{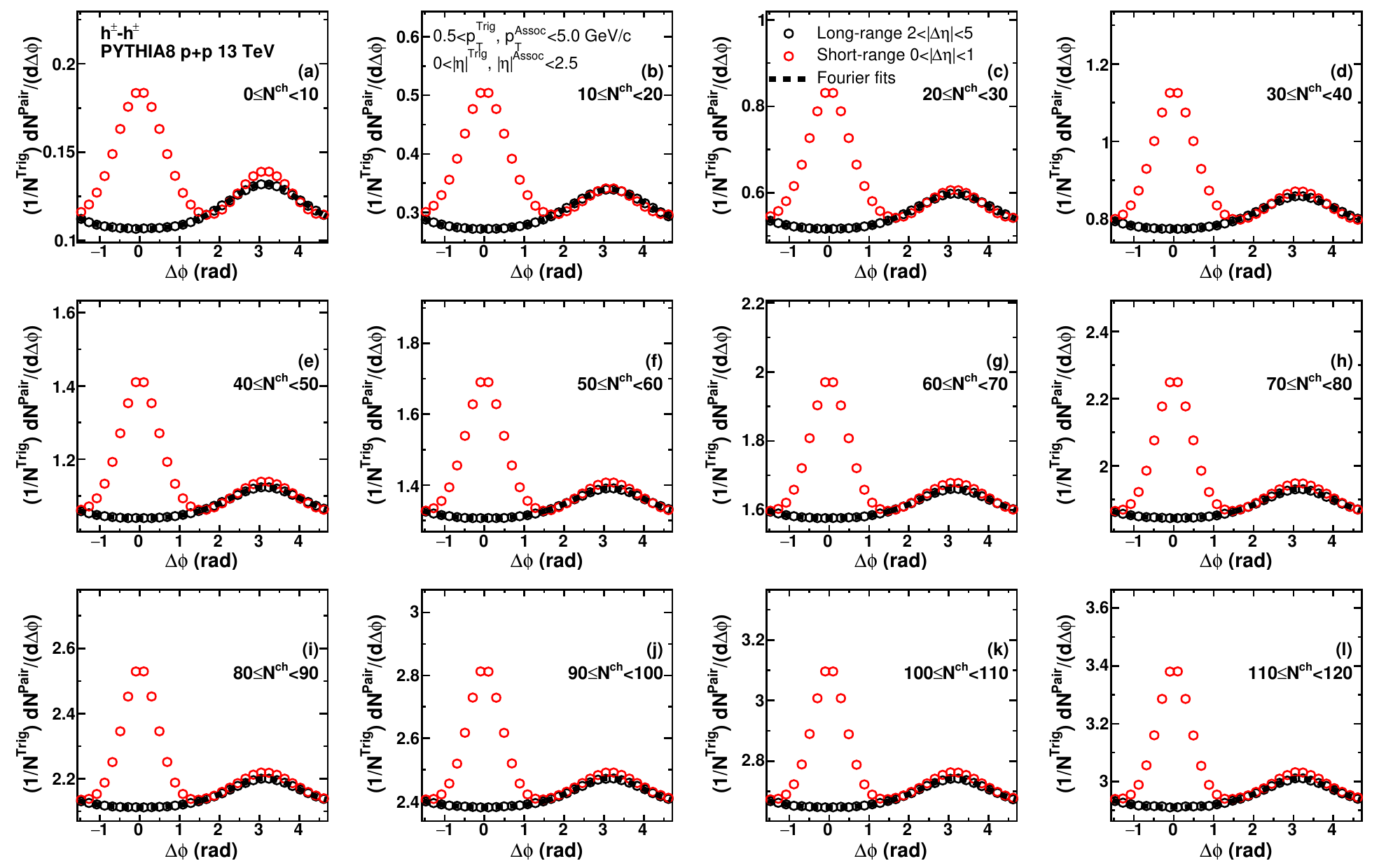}
\caption{\label{fig:pp13TeV_per_trig_yield}
Two-particle \dphi correlation function at short ($|\deta|<1$) and long ($2<|\deta|<5$) ranges in \pp collisions at \sqs~=~13~TeV from \pythia. Charged hadrons in $0.5<\pt<5~{\rm GeV}/c$ and $|\eta|<2.5$ are used for the correlation function. Each panel shows a different multiplicity range, and the multiplicity is defined as the number of charged hadrons in $\pt>0.4~{\rm GeV}/c$ and $|\eta|<2.5$.}
\end{figure*}

\begin{figure*}[htb]
\includegraphics[width=1.00\linewidth]{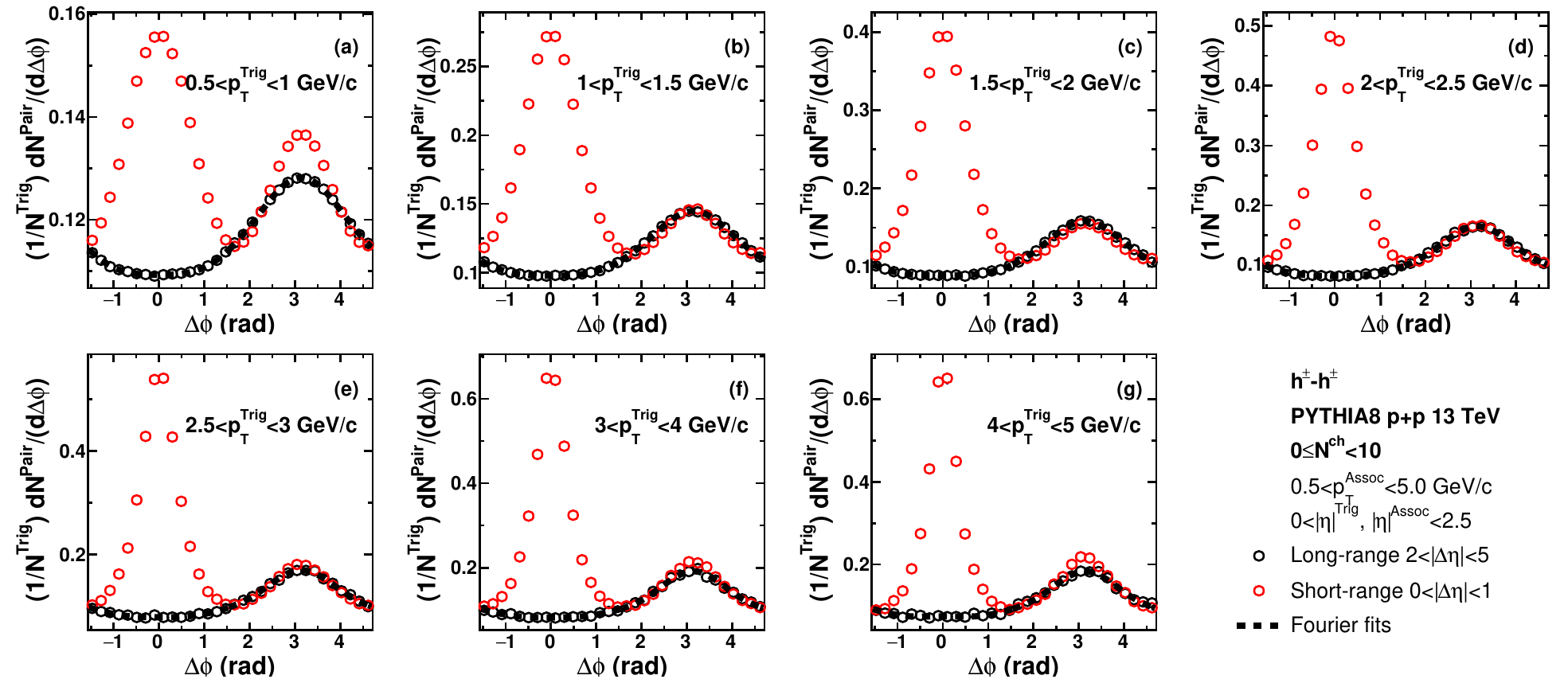}
\includegraphics[width=1.00\linewidth]{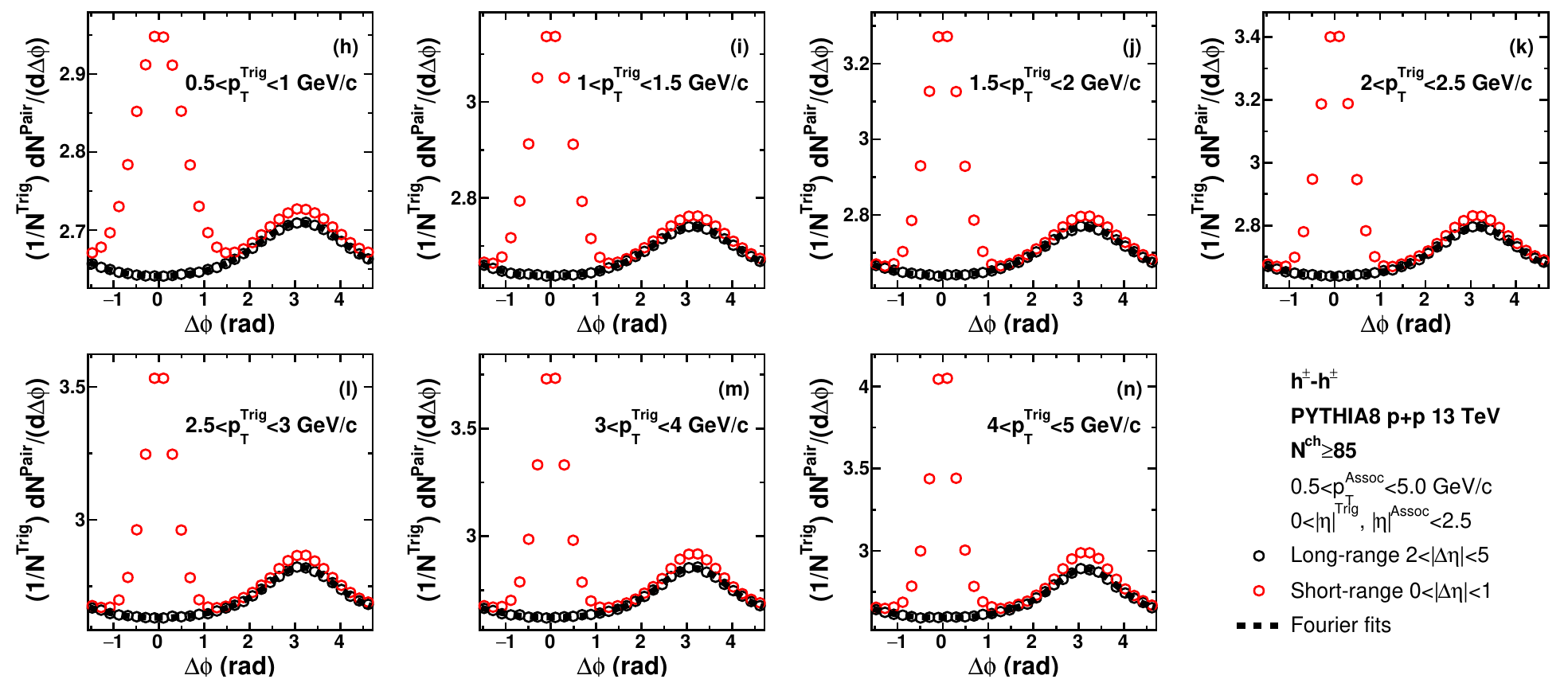}
\caption{\label{fig:pp13TeV_per_trig_yield_pTbin}
Two-particle \dphi correlation function at short ($|\deta|<1$) and long ($2<|\deta|<5$) ranges in \pp collisions at \sqs~=~13~TeV from \pythia. Charged hadrons in $0.5<\pt<5~{\rm GeV}/c$ and $|\eta|<2.5$ are used for the correlation function. Panels in top (a)--(g) [bottom (h)--(n)] two rows are in a multiplicity range of $0\leq\nch<10$ ($\nch\geq85$), and the multiplicity is defined as the number of charged particles in $\pt>0.4~{\rm GeV}/c$ and $|\eta|<2.5$. Each panel represents a different \pt range of trigger particles.}
\end{figure*}

\begin{figure*}[htb]
\includegraphics[width=1.00\linewidth]{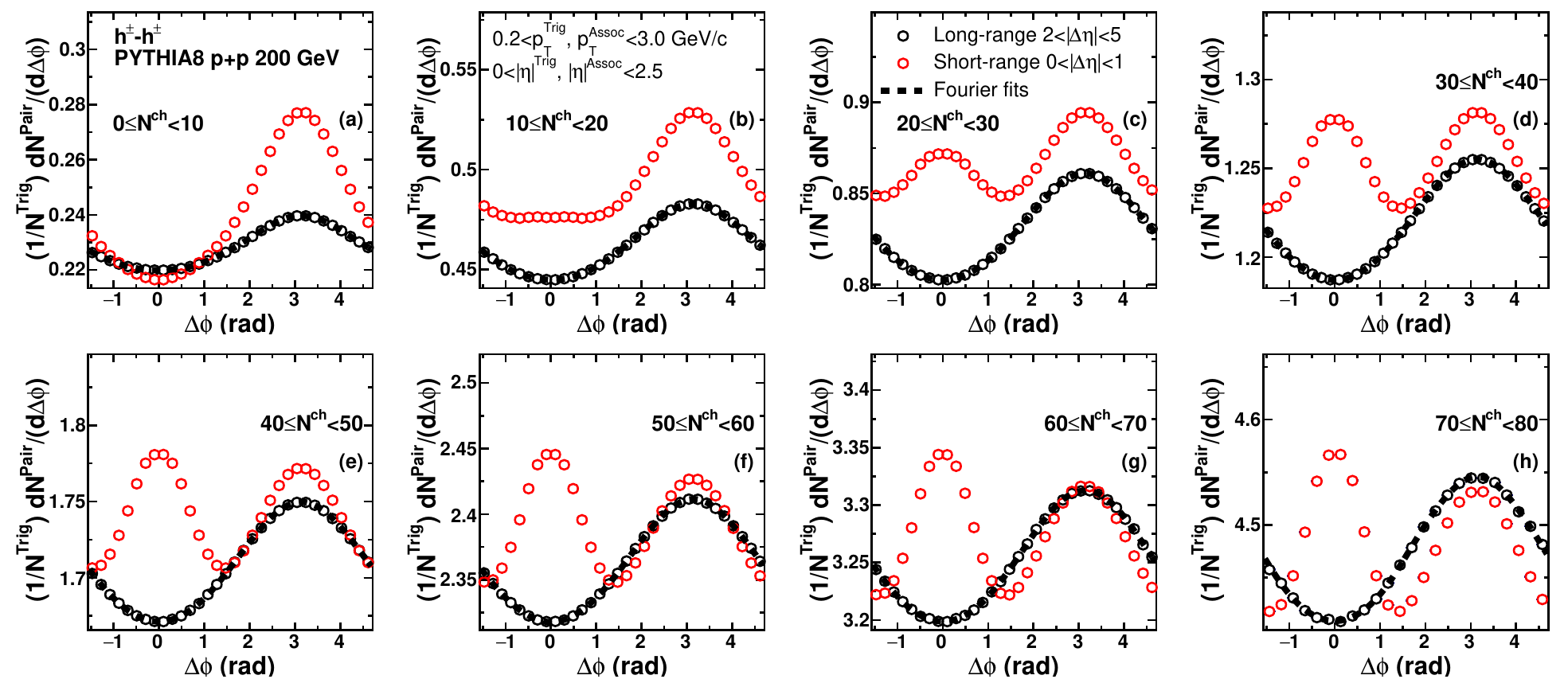}
\caption{\label{fig:pp200GeV_per_trig_yield}
Two-particle \dphi correlation function at short ($|\deta|<1$) and long ($2<|\deta|<5$) ranges in \pp collisions at \sqs~=~200~GeV from \pythia. Charged hadrons in $0.2<\pt<3~{\rm GeV}/c$ and $|\eta|<2.5$ are used for the correlation function. Each panel shows a different multiplicity range, and the multiplicity is defined as the number of charged hadrons in $\pt>0.2~{\rm GeV}/c$ and $|\eta|<2.5$.}
\end{figure*}

\begin{figure*}[htb]
\includegraphics[width=1.00\linewidth]{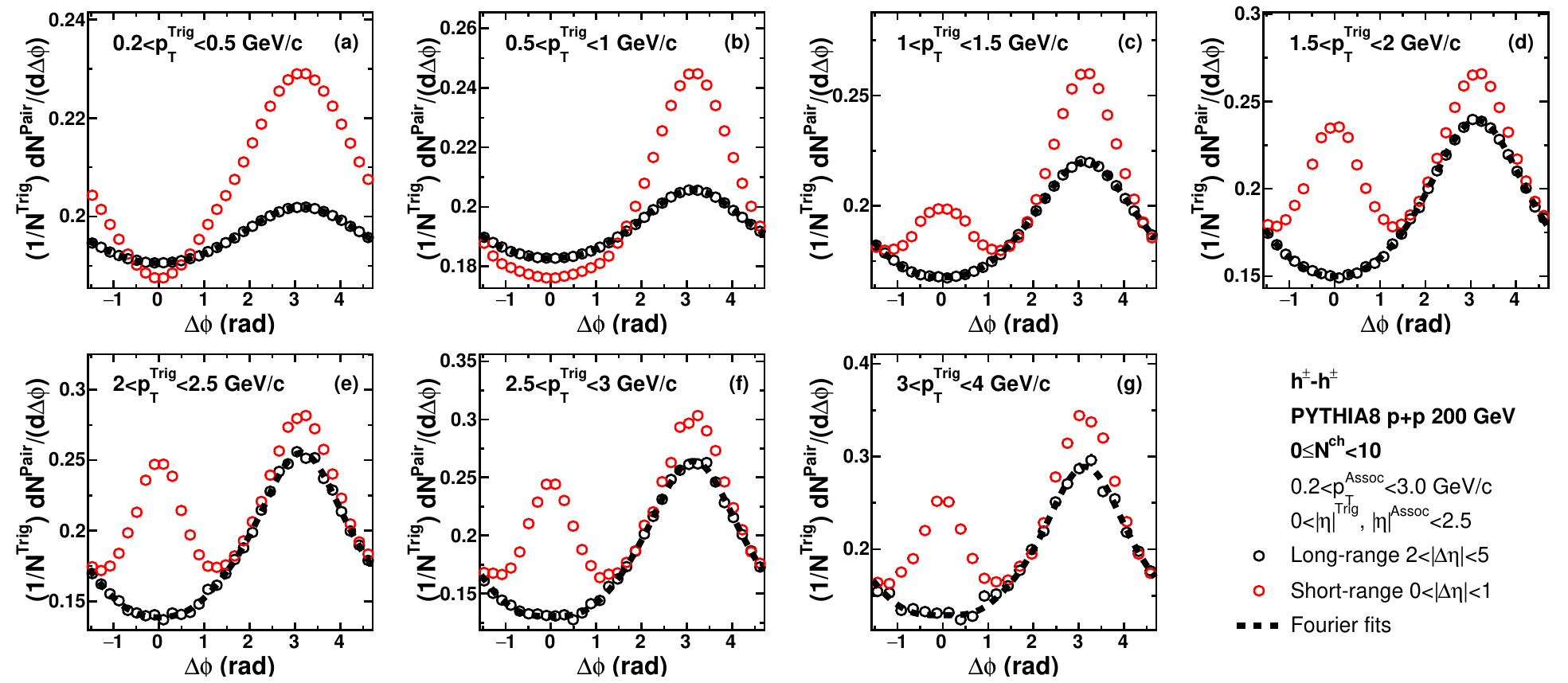}
\includegraphics[width=1.00\linewidth]{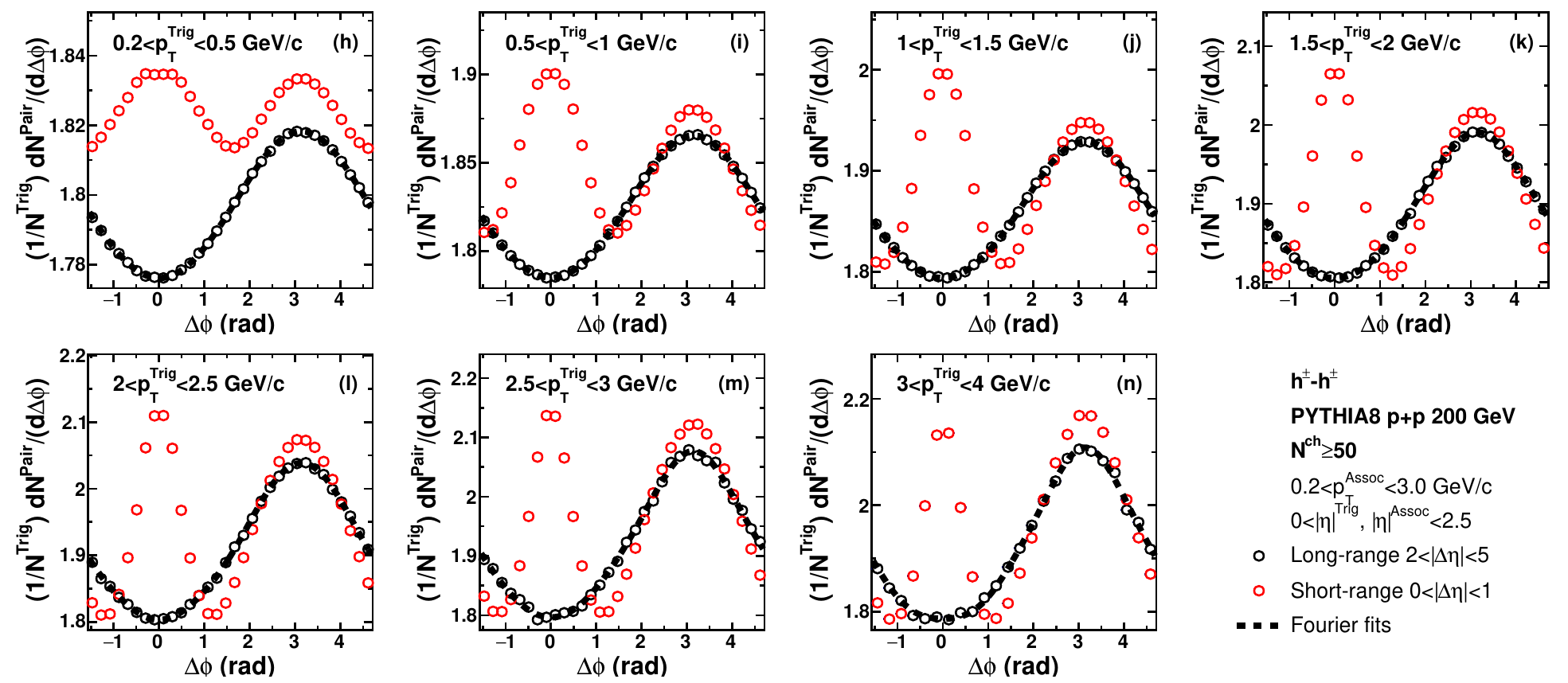}
\caption{\label{fig:pp200GeV_per_trig_yield_pTbin}
Two-particle \dphi correlation function at short ($|\deta|<1$) and long ($2<|\deta|<5$) ranges in \pp collisions at \sqs~=~200~GeV from \pythia. Charged hadrons in $0.2<\pt<3~{\rm GeV}/c$ and $|\eta|<2.5$ are used for the correlation function. Panels in top (a)--(g) [bottom (h)--(n)] two rows are in a multiplicity range of $0\leq\nch<10$ ($\nch\geq50$), and the multiplicity is defined as the number of charged hadrons in $\pt>0.2~{\rm GeV}/c$ and $|\eta|<2.5$. Each panel represents a different \pt range of trigger particles.}
\end{figure*}

\begin{figure*}[htb]
\includegraphics[width=1.00\linewidth]{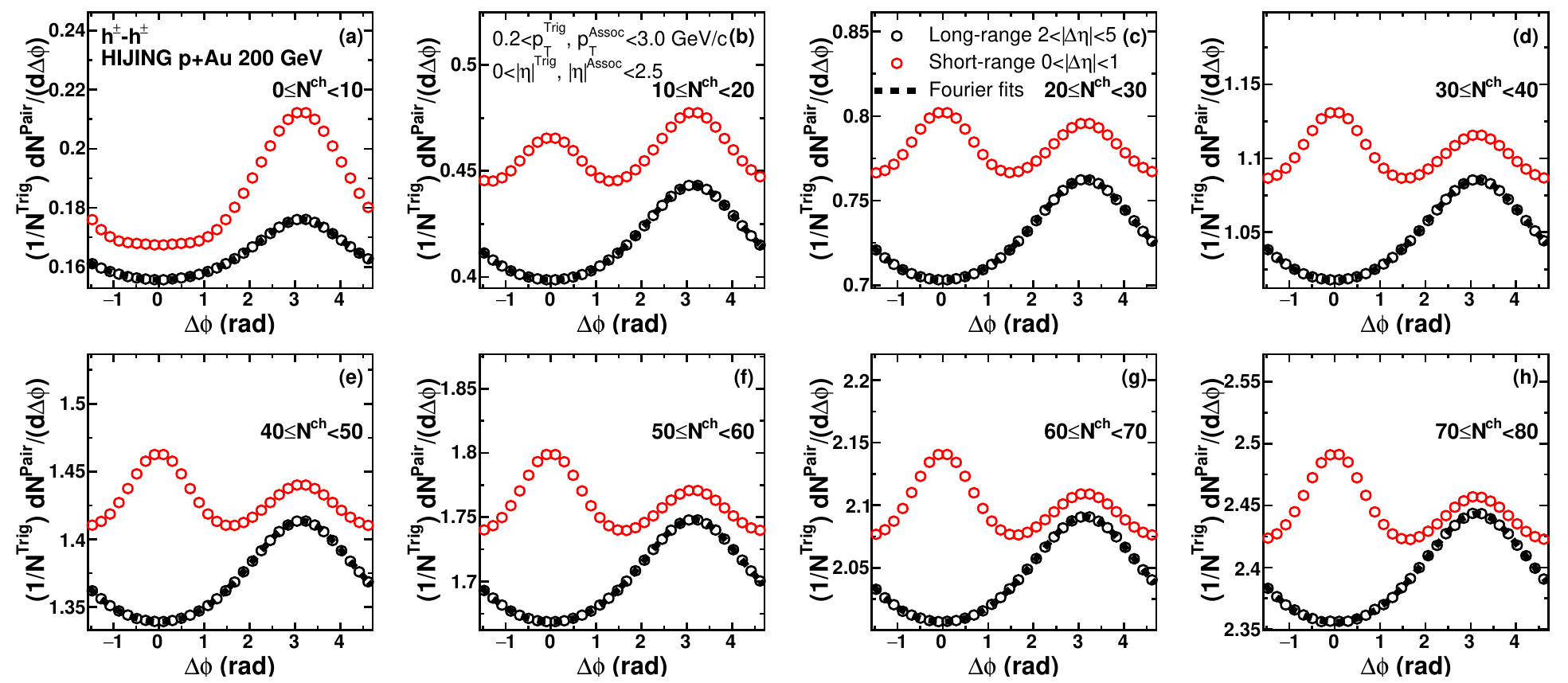}
\caption{\label{fig:pAu200GeV_per_trig_yield}
Two-particle \dphi correlation function at short ($|\deta|<1$) and long ($2<|\deta|<5$) ranges in \pau collisions at \sqsn~=~200~GeV from \hijing. Charged hadrons in $0.2<\pt<3~{\rm GeV}/c$ and $|\eta|<2.5$ are used for the correlation function. Each panel shows a different multiplicity range, and the multiplicity is defined as the number of charged particles in $\pt>0.2~{\rm GeV}/c$ and $|\eta|<2.5$.}
\end{figure*}

\begin{figure*}[htb]
\includegraphics[width=1.00\linewidth]{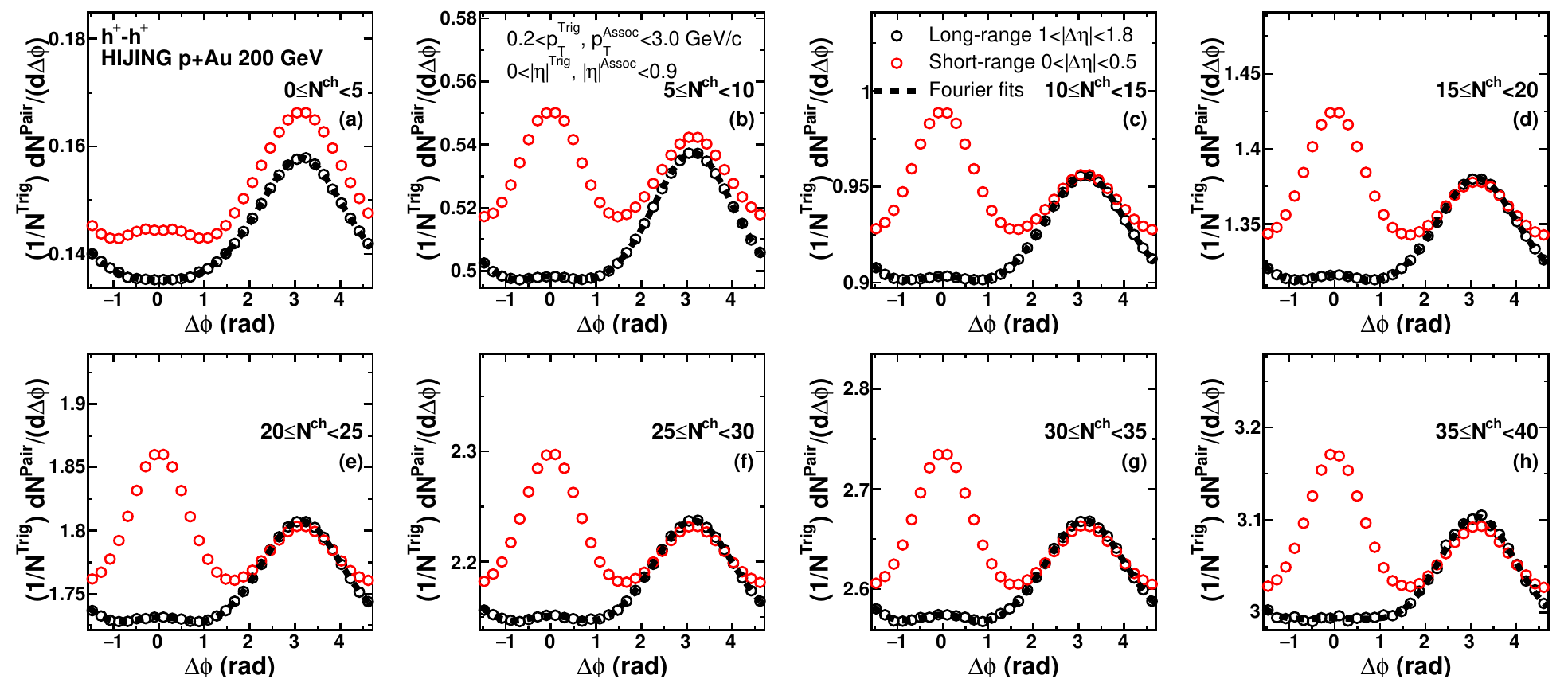}
\caption{\label{fig:pAu200GeV_per_trig_yield_eta0_9}
Two-particle \dphi correlation function at short ($|\deta|<0.5$) and long ($1<|\deta|<1.8$) ranges in \pau collisions at \sqsn~=~200~GeV from \hijing. Charged hadrons in $0.2<\pt<3~{\rm GeV}/c$ and $|\eta|<0.9$ are used for the correlation function. Each panel shows a different multiplicity range, and the multiplicity is defined as the number of charged particles in $\pt>0.2~{\rm GeV}/c$ and $|\eta|<0.9$.}
\end{figure*}

\begin{figure*}[htb]
\includegraphics[width=0.75\linewidth]{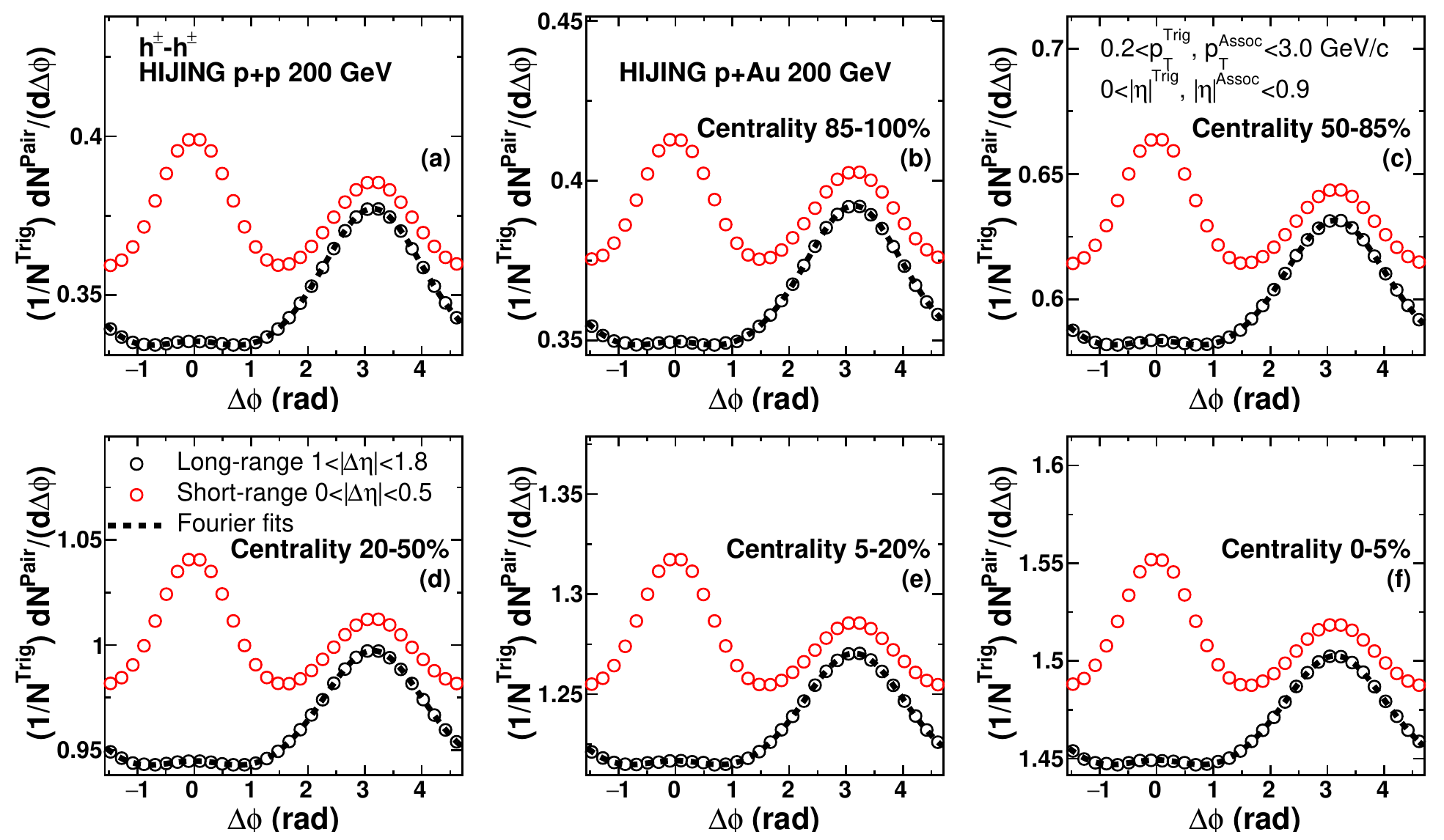}
\caption{\label{fig:pAu200GeV_per_trig_yield_eta0_9_centrality}
Two-particle \dphi correlation function at short ($|\deta|<0.5$) and long ($1<|\deta|<1.8$) ranges in \pp (a) and \pau (b)--(f) collisions at \sqsn~=~200~GeV from \hijing. Charged hadrons in $0.2<\pt<3~{\rm GeV}/c$ and $|\eta|<0.9$ are used for the correlation function. Each panel of \pau collisions shows a different centrality range, and the centrality is defined as the number of charged particles in $-5.0<\eta<-3.3$ (Au-going direction).}
\end{figure*}

\begin{figure*}[htb]
\includegraphics[width=1.0\linewidth]{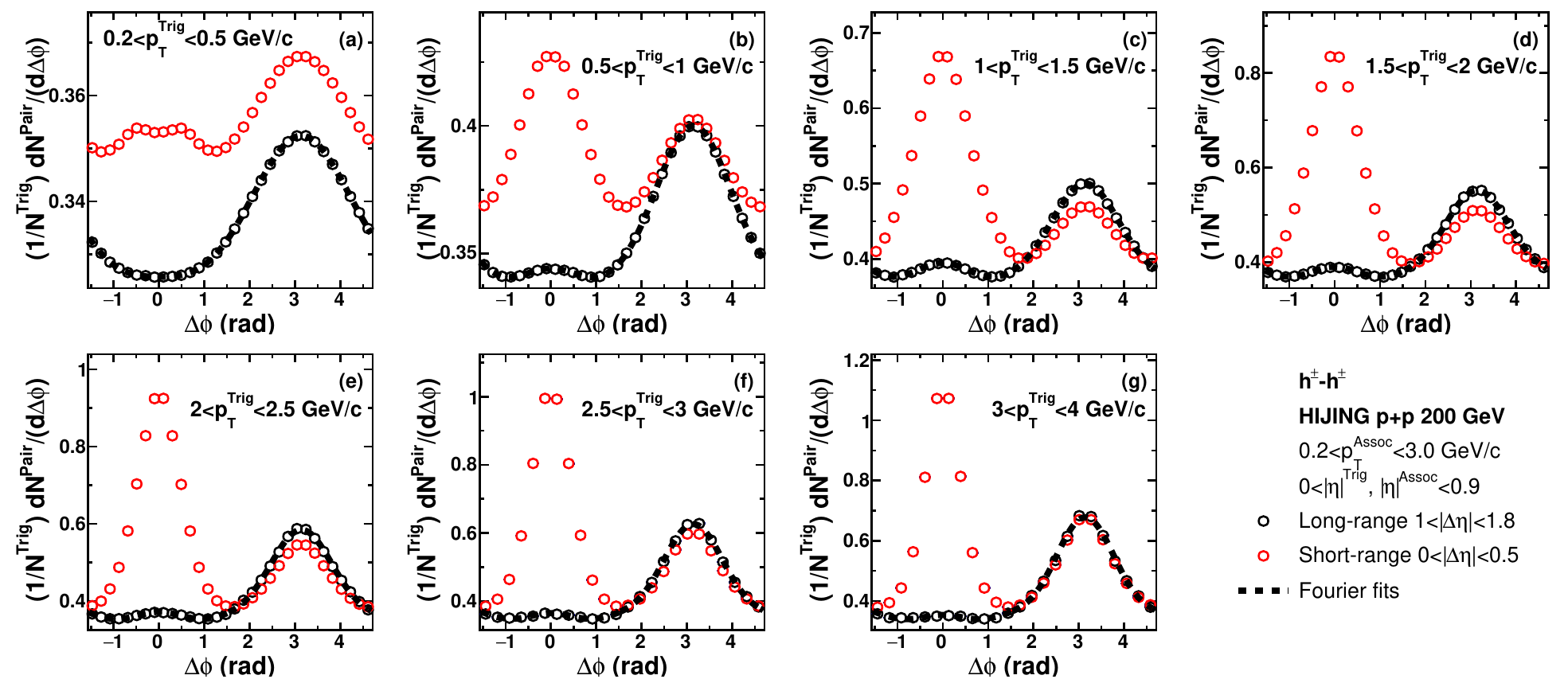}
\includegraphics[width=1.0\linewidth]{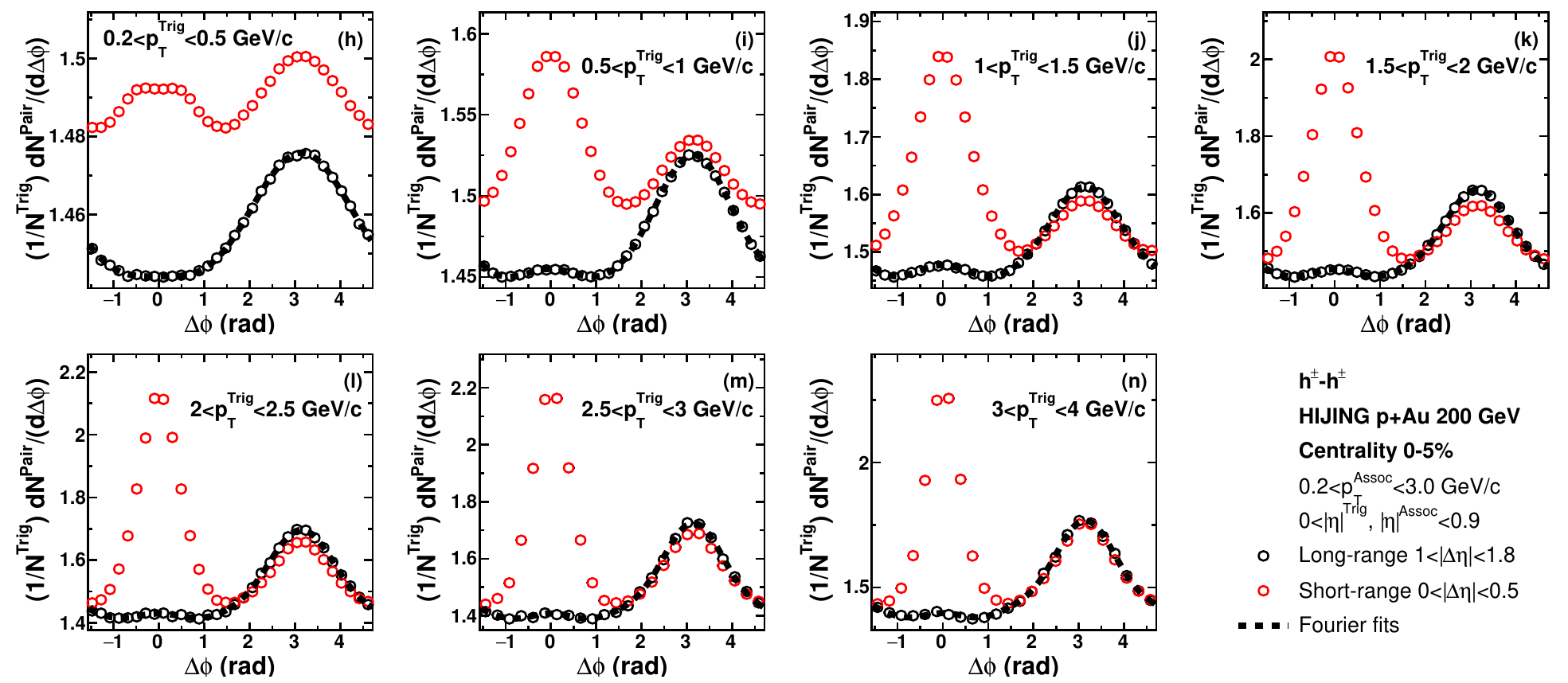}
\caption{\label{fig:pAu200GeV_per_trig_yield_eta0_9_pt}
Two-particle \dphi correlation function at short ($|\deta|<0.5$) and long ($1<|\deta|<1.8$) ranges in \pp (a)--(g) and 0--5\% central \pau (h)--(n) collisions at \sqsn~=~200~GeV from \hijing. Charged hadrons in $|\eta|<0.9$ are used for the correlation function, and the \pt range of associated particle is $0.2<\ptassoc<3~{\rm GeV}/c$. Each panel of \pau collisions shows a different \pt range of trigger particles.}
\end{figure*}

\begin{figure*}[htb]
\includegraphics[width=1.00\linewidth]{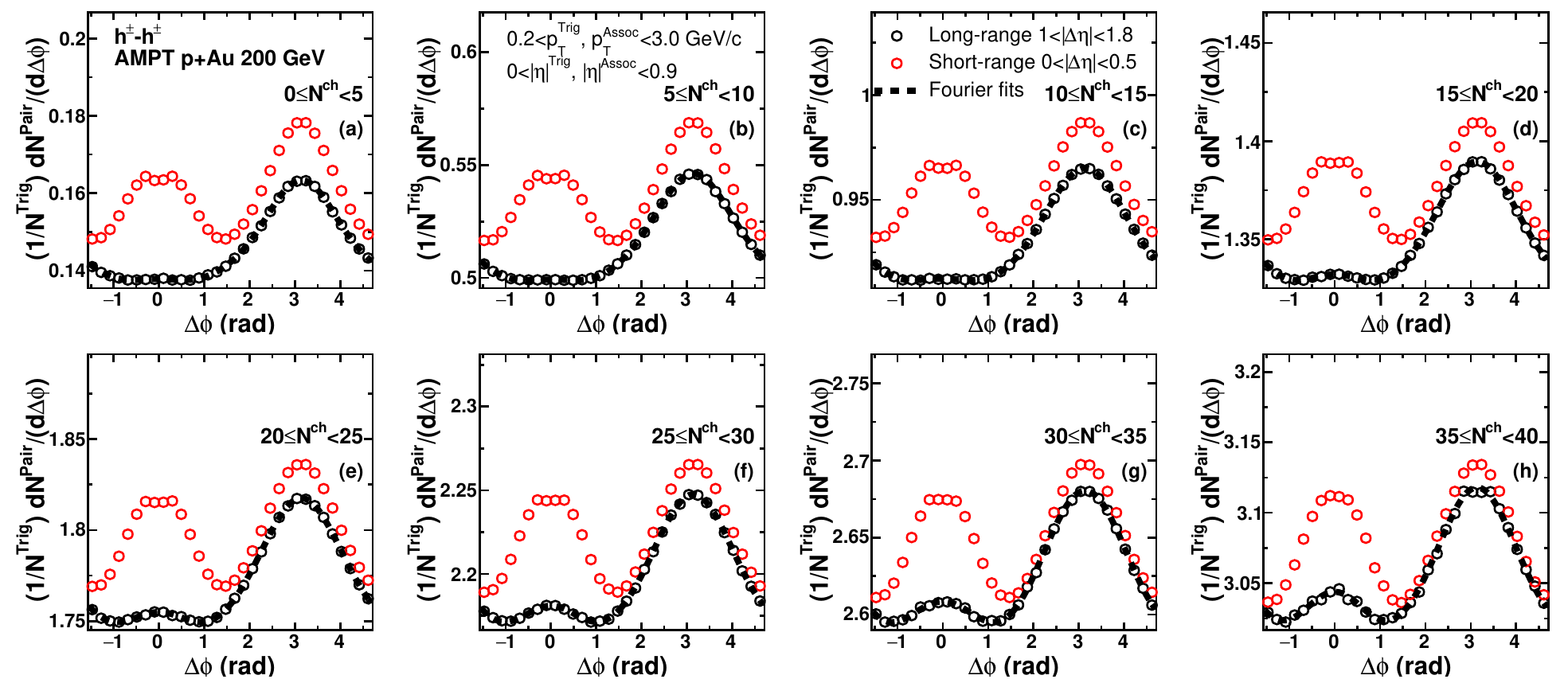}
\caption{\label{fig:ampt_pAu200GeV_per_trig_yield_eta0_9}
Two-particle \dphi correlation function at short ($|\deta|<0.5$) and long ($1.0<|\deta|<1.8$) ranges in \pau collisions at \sqs~=~200~GeV from \ampt. Charged hadrons in $0.2<\pt<3~{\rm GeV}/c$ and $|\eta|<0.9$ are used for the correlation function. Each panel shows a different multiplicity range, and the multiplicity is defined as the number of charged particles in $\pt>0.2~{\rm GeV}/c$ and $|\eta|<0.9$.}
\end{figure*}

\begin{figure*}[htb]
\includegraphics[width=0.75\linewidth]{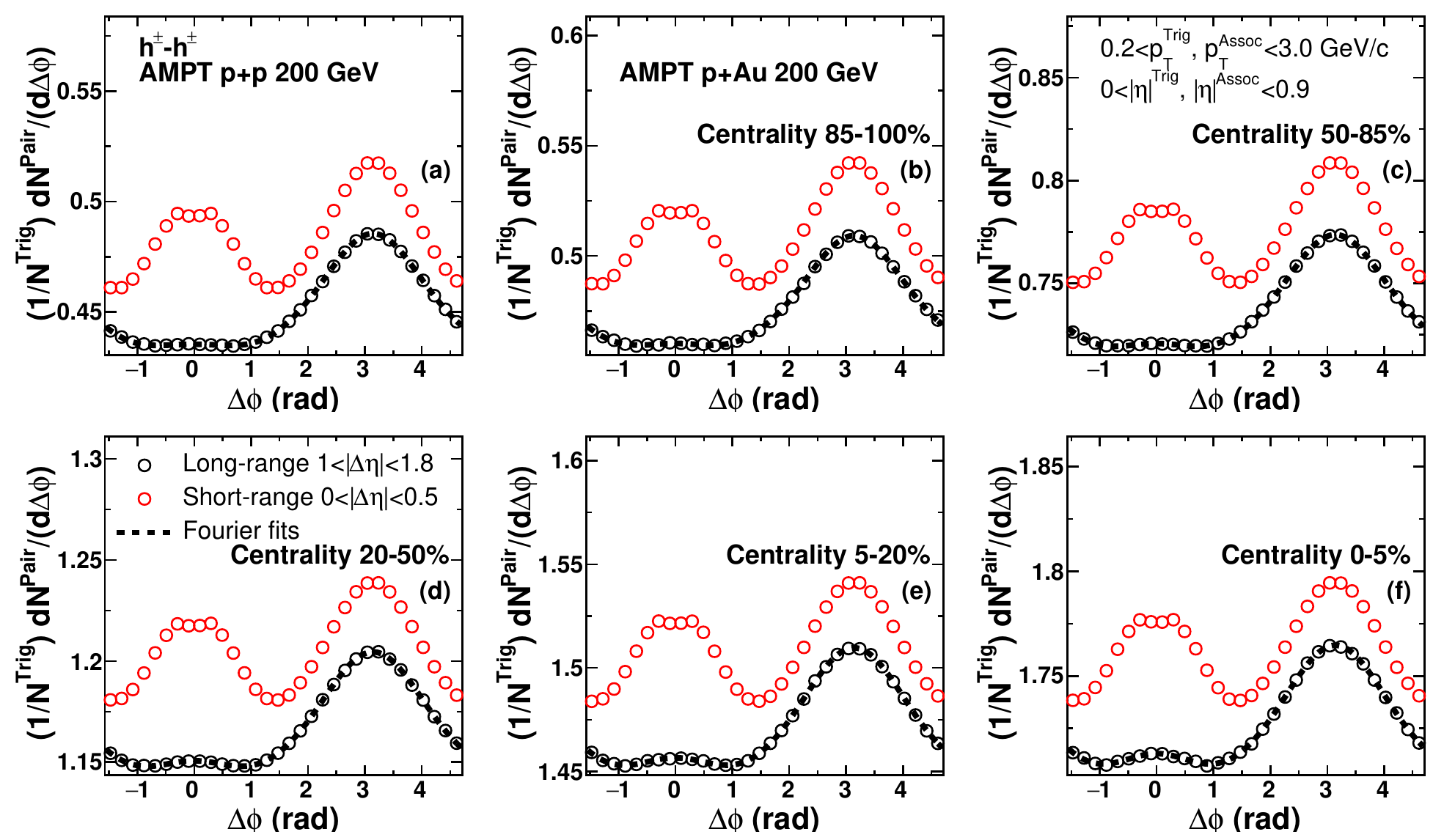}
\caption{\label{fig:ampt_pAu200GeV_per_trig_yield_eta0_9_centrality}
Two-particle \dphi correlation function at short ($|\deta|<0.5$) and long ($1<|\deta|<1.8$) ranges in \pp (a) and \pau (b)--(f) collisions at \sqsn~=~200~GeV from \ampt. Charged hadrons in $0.2<\pt<3~{\rm GeV}/c$ and $|\eta|<0.9$ are used for the correlation function. Each panel of \pau collisions shows a different centrality range, and the centrality is defined as the number of charged particles in $-5.0<\eta<-3.3$ (Au-going direction).}
\end{figure*}

\begin{figure*}[htb]
\includegraphics[width=1.0\linewidth]{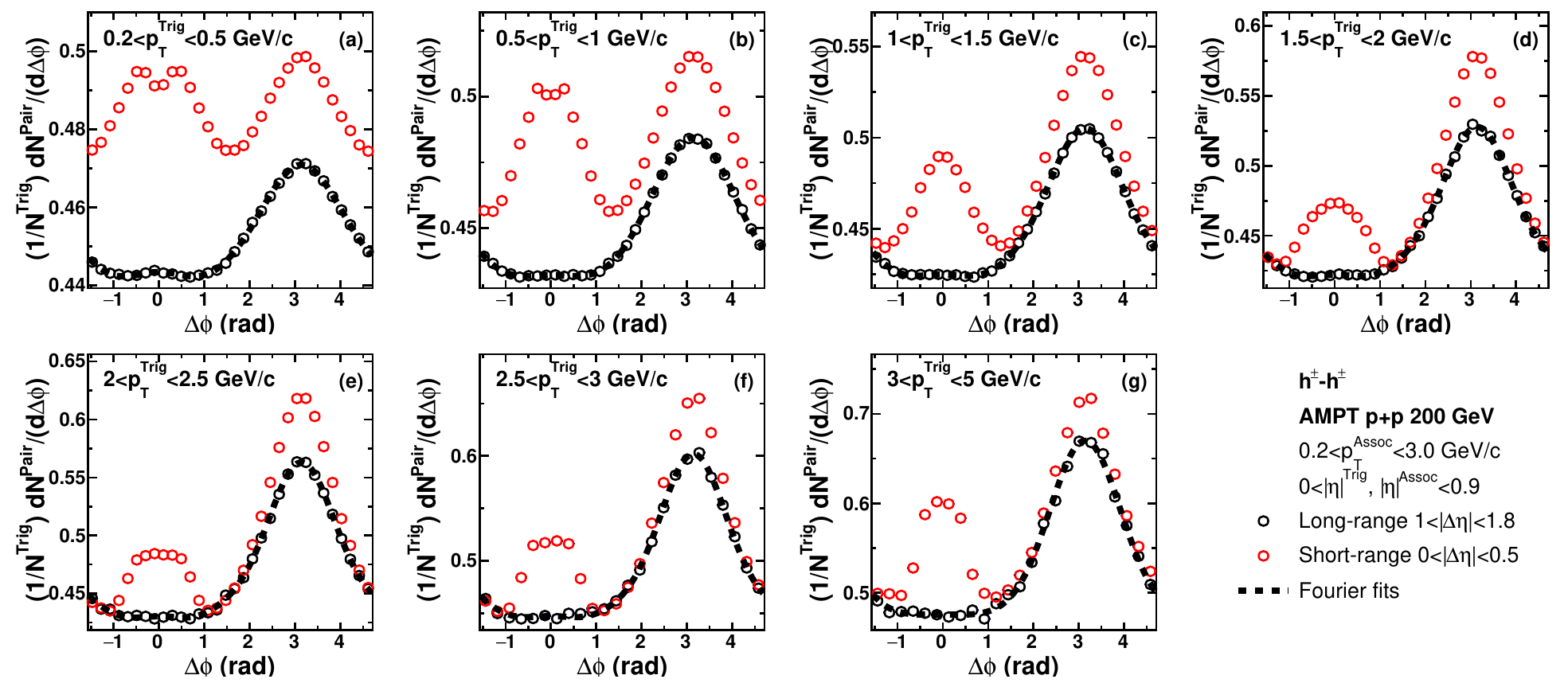}
\includegraphics[width=1.0\linewidth]{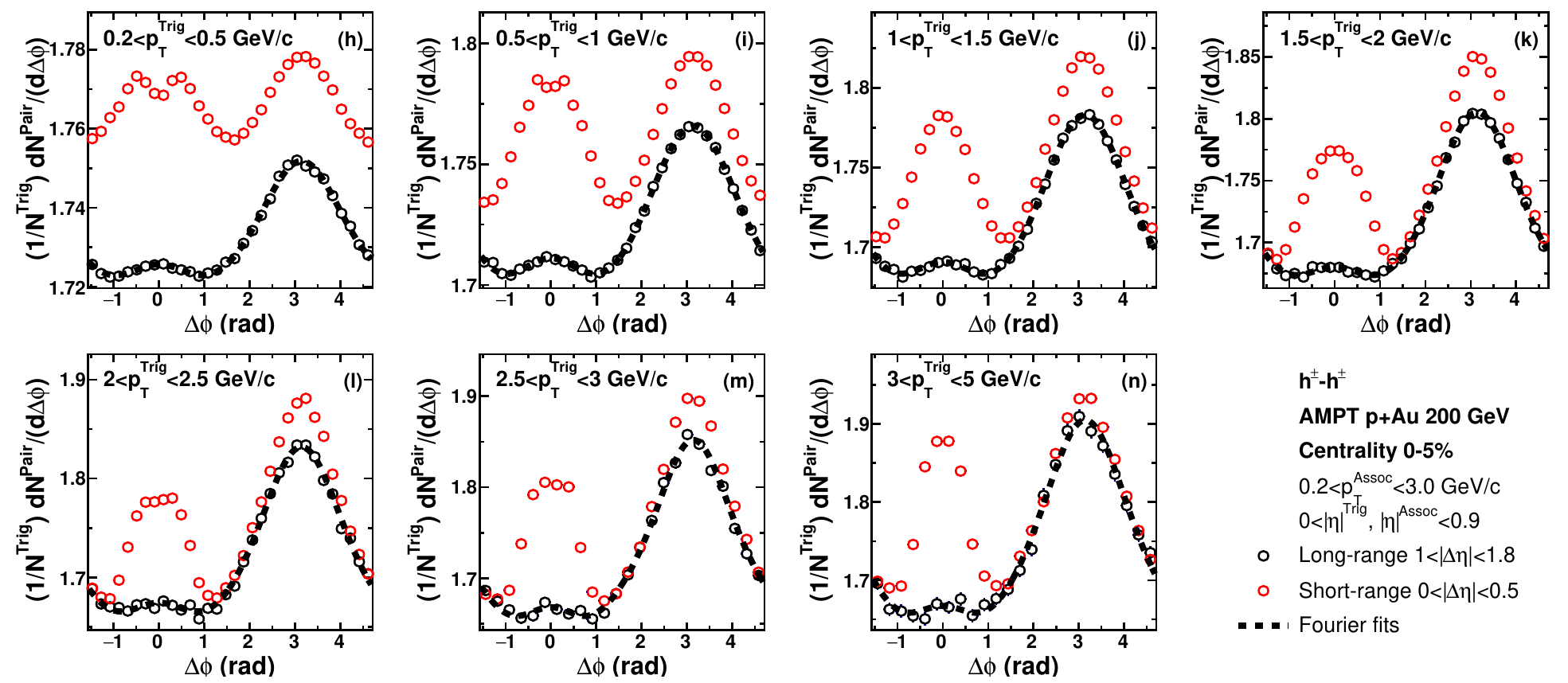}
\caption{\label{fig:ampt_pAu200GeV_per_trig_yield_eta0_9_pt}
Two-particle \dphi correlation function at short ($|\deta|<0.5$) and long ($1<|\deta|<1.8$) ranges in \pp (a)--(g) and 0--5\% central \pau (h)--(n) collisions at \sqsn~=~200~GeV from \ampt. Charged hadrons in $|\eta|<0.9$ are used for the correlation function, and the \pt range of associated particle is $0.2<\ptassoc<3~{\rm GeV}/c$. Each panel of \pau collisions shows a different \pt range of trigger particles.}
\end{figure*}

Here we present two-particle \dphi correlation functions described as Eq.~(\ref{eq:fourier}) in Sec.~\ref{sec:intro} with charged hadrons from event generators, and these correlation functions are used for \vtt and \vttsub calculations (see Figs.~\ref{fig:pp13TeV_per_trig_yield}--\ref{fig:ampt_pAu200GeV_per_trig_yield_eta0_9_pt}). We specifically follow the formula for per-trigger yields as a function of \dphi within a certain \deta range described in Ref.~\cite{Khachatryan:2016txc},

\begin{equation}
\label{eq:per_trig_yield}
\begin{aligned}
    f(\Delta\phi) &= \int \frac{1}{N^{\rm Trig}} \frac{\mathrm{d}^{2}N^{\rm Pair}}{\mathrm{d}\deta \mathrm{d}\dphi} \mathrm{d}\deta\\
    &= \int B(0,0) \frac{S(\deta,\dphi)}{B(\deta,\dphi)} \mathrm{d}\deta,
\end{aligned}
\end{equation}
where \deta and \dphi are the differences of $\eta$ and $\phi$ between trigger and associated particles. $S(\deta,\dphi)$ and $B(\deta,\dphi)$ are the yields of pairs normalized by the number of trigger particles, $N^\mathrm{Trig}$, in the same and mixed events respectively,

\begin{equation}
    S(\deta,\dphi) = \frac{1}{N^{\rm Trig}}\frac{\mathrm{d}^{2}N^{\rm Same}}{\mathrm{d}\deta \mathrm{d}\dphi},
\end{equation}

\begin{equation}
    B(\deta,\dphi) = \frac{1}{N^{\rm Trig}}\frac{\mathrm{d}^{2}N^{\rm Mix}}{\mathrm{d}\deta \mathrm{d}\dphi}.
\end{equation}
The purpose of normalizing by the mixed event pair distributions is to account for the geometric $\deta$-dependent pair acceptance effect.

\clearpage


\bibliography{main}

\end{document}